\documentclass[fleqn,usenatbib]{mnras}

\usepackage{newtxtext,newtxmath}
\usepackage[T1]{fontenc}
\DeclareRobustCommand{\VAN}[3]{#2}
\let\VANthebibliography\thebibliography
\def\thebibliography{\DeclareRobustCommand{\VAN}[3]{##3}\VANthebibliography}

\usepackage{graphicx}	
\usepackage{amsmath}	
\usepackage{xcolor}     
\usepackage{longtable}
\usepackage{tablefootnote}
\usepackage{multirow}
\usepackage{makecell}

\newcommand{\oiii}{[O\,\textsc{iii}]}
\newcommand{\nii}{[N\,\textsc{ii}]}
\newcommand{\sii}{[S\,\textsc{ii}]}
\newcommand{\oi}{[O\,\textsc{i}]}

\newcommand{\siii}{[S\,\textsc{iii}]}
\newcommand{\oii}{[O\,\textsc{ii}]}

\newcommand{\hii}{H\,\textsc{ii}}
\newcommand{\ha}{\mbox{$\rmn{H}\alpha$}}
\newcommand{\hb}{\mbox{$\rmn{H}\beta$}}

\newcommand{\kms} {$\mathrm{km\,s}^{-1}$}

\newcommand{\reff}{r$_{\rm eff}$}
\newcommand{\DAP}{{\tt DAP}}
\newcommand{\MAPS}{{\tt MAPS}}

\title[PHANGS Nebular Catalogue]{The PHANGS--MUSE Nebular Catalogue}

\author[Groves et al.]{B. Groves$^{1}$\thanks{email:brent.groves@uwa.edu.au}, 
K. Kreckel$^{2}$, 
F. Santoro$^{3}$, 
F. Belfiore$^{4}$, 
E. Zavodnik$^{2}$, 
E. Congiu$^{5}$, 
O. V. Egorov$^{2}$, \newauthor
E. Emsellem$^{6}$, 
K. Grasha$^{7}$,
A. Leroy$^{8}$, 
F. Scheuermann$^{2}$, 
E. Schinnerer$^{3}$, 
E. J. Watkins$^{2}$, 
A. T. Barnes$^{9}$, 
\newauthor
F. Bigiel$^{9}$,
D.~A.~Dale$^{10}$, 
S.~C.~O. Glover$^{11}$,
I. Pessa$^{3}$,
P. Sanchez-Blazquez$^{12,13}$, \newauthor
and T. G. Williams$^{3}$ 
\\
$^{1}$International Centre for Radio Astronomy Research, University of Western Australia, 7 Fairway, Crawley, 6009 WA, Australia \\
$^{2}$Astronomisches Rechen-Institut, Zentrum f\"{u}r Astronomie der Universit\"{a}t Heidelberg, M\"{o}nchhofstra{\ss}e 12-14, 69120 Heidelberg, Germany \\
$^{3}$Max Planck Institut f\"{u}r Astronomie, K\"{o}nigstuhl 17, 69117 Heidelberg, Germany\\
$^{4}$INAF -- Osservatorio Astrofisico di Arcetri, Largo E. Fermi 5, I-50157 Firenze, Italy \\
$^{5}$ Departamento de Astronom\'{i}a, Universidad de Chile, Camino del Observatorio 1515, Las Condes, Santiago, Chile\\
$^{6}$European Southern Observatory, Karl-Schwarzschild Stra{\ss}e 2, D-85748 Garching bei M\"{u}nchen, Germany\\
$^{7}$Research School of Astronomy and Astrophysics, Australian National University, Canberra, ACT 2611, Australia\\
$^{8}$Department of Astronomy, The Ohio State University, 140 West 18th Avenue, Columbus, Ohio 43210, USA \\
$^{9}$Argelander-Institut f\"{u}r Astronomie, Universit\"{a}t Bonn, Auf dem H\"{u}gel 71, 53121 Bonn, Germany\\
$^{10}$Department of Physics \& Astronomy, University of Wyoming, Laramie, WY, 82071, USA\\
$^{11}$Universit\"{a}t Heidelberg, Zentrum f\"{u}r Astronomie, Institut f\"{u}r theoretische Astrophysik, Albert-Ueberle-Str. 2, 69120 Heidelberg, Germany \\
$^{12}$Departamento de F\'{i}sica de la Tierra y Astrof\'{i}sica, Universidad Complutense de Madrid, E-28040 Madrid, Spain\\
$^{13}$Instituto de F\'{i}sica de part\'{i}culas y del Cosmos, IPARCOS,  E-28040 Madrid, Spain\\
}

\date{Accepted XXX. Received YYY; in original form ZZZ}

\pubyear{2022}

\begin{document}
\label{firstpage}
\pagerange{\pageref{firstpage}--\pageref{lastpage}}
\maketitle

\begin{abstract}
Ionized nebulae provide critical insights into the conditions of the interstellar medium (ISM). Their bright emission lines enable the measurement of physical properties, such as the gas-phase metallicity, across galaxy disks and in distant galaxies. The PHANGS--MUSE survey has produced optical spectroscopic coverage of the central star-forming discs of 19 nearby main-sequence galaxies. Here, we use the \ha\ morphology from this data to identify 30,790 distinct nebulae, finding thousands of nebulae per galaxy. For each nebula, we extract emission line fluxes and, using diagnostic line ratios, identify the dominant excitation mechanism. A total of 
23,244 nebulae (75\%) are classified as \hii\ regions. The dust attenuation of every nebulae is characterised via the Balmer decrement and we use existing environmental masks to identify their large scale galactic environment (centre, bar, arm, interarm and disc). Using strong-line prescriptions, we measure the gas-phase oxygen abundances (metallicity) and ionization parameter for all \hii\ regions.  With this new catalogue, we measure the radial metallicity gradients and explore second order metallicity variations within each galaxy. By quantifying the global scatter in metallicity per galaxy, we find a weak negative correlation with global star formation rate and stronger negative correlation with global gas velocity dispersion (in both ionized and molecular gas). With this paper we release the full catalogue of strong line fluxes and derived properties, providing a rich database for a broad variety of ISM studies. 
\end{abstract}

\begin{keywords}
galaxies:ISM -- \hii\ Regions -- galaxies: abundances
\end{keywords}



\section{Introduction}

Emission lines from ionized nebulae play a fundamental role in our understanding of galaxy evolution. Apart from their use in determining spectroscopic redshifts, emission lines have been used to determine star-formation rates, the presence of active galactic nuclei (AGN), galaxy dynamics, gas-phase metal abundances and more \citep[see e.g.][]{Kewley2019}. However, in most extragalactic studies, the emission lines have been typically measured from a single spectrum of the entire galaxy (e.g.\ SDSS, \citealt{Abazajian2009}; GAMA, \citealt{Driver2009}; VVDS, \citealt{LeFevre2003}). Resolved maps of nearby emission line galaxies in strong spectral lines has been available 
with narrow filters, slit spectra on bright \hii\ regions \citep[e.g.][]{Croxall2009, Moustakas2010} and even spectrally resolved with Fabry-Perot surveys \citep[e.g.][]{Veilleux2002, Epinat2008, Moiseev2015, Silchenko2019}, but it is only with the advent of integral field spectrographs (IFS) that maps of multiple emission lines across thousands of galaxies have become common, 
with most achieving kpc scale sampling (e.g.\ CALIFA, \citealt{Sanchez2012}; VENGA, \citealt{Blanc2013}; MaNGA, \citealt{Bundy2015}; SAMI, \citealt{Croom2021}). 

However, the individual ionized nebulae that the emission lines originate from (e.g.~\hii\ regions, supernova remnants, planetary nebulae) are typically $<$100 pc in size. This means that in these large surveys, the nebulae cannot be spatially distinguished from each other or the faint surrounding diffuse ionized gas \citep[DIG;][]{Reynolds1990}.  While there have been efforts to account for the DIG \citep[e.g.][]{Zhang2017,EspinosaPonce2020}, only in nearby galaxies (or rare lensed systems) are we able to achieve the 10--100 pc resolution required to separate out the individual ionized nebula \citep[i.e.~\hii\ regions;][]{Kennicutt1989}. Once imaged in multiple emission lines, it is then possible to use line ratios to distinguish nebular emission in \hii\ regions from other nebular sources \citep[e.g.~supernovae (SNe), planetary nebulae (PNe)][]{Ciardullo2002, Smith2005, Long2010}.  
Long-slit studies have compiled growing samples of emission line spectroscopy for \hii\ regions \citep[e.g.][]{Pilyugin2014, Berg2020}, 
yet generally these studies pre-select for the brightest regions, which introduces biases in the sampling across galaxy disks. 

With the advent of new instruments and approaches, 
wide-area and high angular resolution spectral maps are now available. This is made possible through long-slit spectral stepping \citep[e.g.~the TYPHOON survey][]{Ho2017}, imaging Fourier transform spectrographs \citep[e.g.~the SIGNALS survey with SITELLE on the CFHT;][]{Rousseau-Nepton2019} or IFS  \citep[e.g.~MAD or TIMER on the MUSE/VLT;][]{Erroz-Ferrer2019, Gadotti2019}. Such maps not only make it possible for individual nebulae to be identified, but the integrated emission line fluxes within each nebula mean that they can be classified and their key physical properties measured. Furthermore, sensitive data can also examine the properties of the intervening diffuse ionized gas. Full spectral maps can also be used to understand the properties of the underlying stellar populations, including the stars that are the potential sources of ionizing photons. Surveys using spectral maps have already identified leaking radiation from \hii\ regions as the dominant 
ionizing source of the DIG \citep{DellaBruna2020, Belfiore2022}, quantified the impact of the DIG on the measurement of the gas-phase metallicity \citep{Poetrodjojo2019}, and determined that metallicity variations exist within galaxies on top of the well known radial gradients \citep{Ho2017, Kreckel2019, Metha2021, Sanchez-Menguiano2020, Williams2022}.

It is within this context that we describe here the PHANGS--MUSE Nebulae Catalogue of over 30,000 nebulae across 19 galaxies, the largest catalogue of high-resolution ($<100$\,pc) extragalactic nebulae with homogeneous optical spectroscopic coverage currently available. 
First introduced by \citet{Santoro2022} when fitting for the \hii\ region luminosity functions, the nebula catalogue has already been used in a number of papers for a range of science topics, including; quantifying the pre-supernova feedback within 6000 of the \hii\ regions \citep{Barnes2021}, modelling of line ratios for diffuse ionized gas surrounding these \hii\ regions \citep{Belfiore2022}, and interpolating metallicities measured at each \hii\ region to construct full coverage metallicity maps \citep{Williams2022}. With this paper we release a full catalogue of emission line properties and derived physical properties associated with the objects.

This catalogue is based on the mosaicked MUSE IFS observations from the Physics at High-Angular resolution in Nearby GalaxieS program \citep[PHANGS--MUSE survey][]{Emsellem2022}, which are drawn from the larger PHANGS survey \citep[PHANGS]{Leroy2021}\footnote{\url{http://www.phangs.org}}.  We summarise this survey and the specific subsample targeted with MUSE in Section \ref{sec:sample}. We present our methods of constructing our catalogue of nebular emitting objects in Section \ref{sec:methods}. We describe how derived properties are obtained and included in the catalogue as value-added products in Section \ref{sec:VAC}. We present results focused on the metallicity measurements in those objects classified as \hii\ regions in Section \ref{sec:results}. We discuss the interpretation of the metallicity variation we observe in Section \ref{sec:discussion}, as well as other technical aspects of our catalogue, and conclude in Section \ref{sec:conclusion}. 

\section{The PHANGS--MUSE Survey}\label{sec:sample}

The PHANGS survey was designed specifically to resolve galaxies into the individual elements of the star-formation process: molecular clouds, \hii\ regions, and stellar clusters. Driven by this aim, the full PHANGS sample was originally determined by selecting southern-sky accessible ($-75^{\circ}\le\delta\le+25^{\circ}$, for ALMA \& MUSE), low inclination ($i<75^{\circ}$), massive star-forming galaxies ($\log(M_*/M_{\odot}) > 9.75$ and $\log(sSFR/{\rm yr}^{-1}) > -11$) within $\sim23$\,Mpc, such that $1"<100$\,pc \citep[described in full detail][as the PHANGS--ALMA sample]{Leroy2021}. In addition to ALMA data, there is a wealth of data from other telescopes such as the Hubble Space Telescope \citep[the PHANGS-HST survey][]{Lee2022}, and the MUSE IFS on ESO's VLT \citep{Bacon2010}, known as the PHANGS--MUSE survey \citep{Emsellem2022}.

The PHANGS--MUSE survey is an ESO large program ($\sim 170$\,h, PI Schinnerer) aimed at spectroscopically mapping the discs of 19 nearby star-forming galaxies. 
This subsample of PHANGS (originally selected to align with the PHANGS--ALMA pilot surveys) covers a broad range in stellar mass, but is biased somewhat to main-sequence massive galaxies (Table \ref{tab:galaxies}). Further details on the sample, observations, reduction and MUSE data products are described in \citet{Emsellem2022}. Here we summarise the sample and data, and refer the reader to \citet{Emsellem2022} for the full details. Public data products, including data cubes and line maps, are available at the ESO archive \footnote{\url{https://archive.eso.org/scienceportal/home?data_collection=PHANGS}} and CADC\footnote{\url{https://www.canfar.net/storage/vault/list/phangs/RELEASES/PHANGS-MUSE}}.

The physical properties of these galaxies are listed in Table \ref{tab:galaxies}. As all our galaxies are within $\sim$20\,Mpc, the typical seeing (0\farcs91) of our MUSE observations means that all structures down to 100\,pc can be isolated within the disk environment, with a median physical resolution of $\sim$70\,pc. The galaxy distances we use are the latest compilation from \citet{Anand2021}, including new tip of the red giant branch distances from the PHANGS--HST observations and new planetary nebula luminosity function distances measured from the PHANGS--MUSE data itself \citep{Scheuermann2022}. The inclination and position angle of the galaxies were determined by \citet{Lang2020} from the PHANGS ALMA CO rotation curves analysis, or near-IR imaging when the CO data was not available or the fit to the CO velocity field was deemed unreliable. The position angle and inclinations were used to determine the deprojected radial distances used in this paper.  The listed stellar masses and star formation rates are global measures from UV and IR photometry \citep{Leroy2019}. Representative disc scale lengths are provided, both the 25th magnitude B-band isophotal radius (R25) from RC3 (\citealt{RC3} via HyperLEDA; \citealt{Makarov2014}) and the effective radius containing half of the stellar mass of the galaxy (\reff). These quantities are compiled and computed in \cite{Leroy2021}. 

Due to their proximity, the stellar structures of the galaxies, such as spiral arms and bars, are clearly resolved. Within this work we use the structural morphology masks created by \citet{Querejeta2021} using \emph{Spitzer} $3.6\,\mu$m imaging. 
\citet{Querejeta2021} used photometric fitting that decomposes the galaxies into bulges and discs \citep{Salo2015}, then further divided the structures visually and through fitting into centres, bars (with the bar mask defined by a fitted ellipse), rings and lenses, and spiral arms. Spiral arms were fitted as logarithmic spiral curves with widths fitted to both the stellar and molecular gas surface density, and the inter-arm region was considered to be any region in the disc outside of these. We follow \citet{Querejeta2021} in distinguishing between inner bar and outer disc regions when considering the arm and inter-arms and use their notations (their Figure 2 and Table 1, respectively). These morphological masks allow us to determine the influence of local environment on the nebulae. Given our bias towards massive main sequence galaxies, spiral or disk features are seen in all, and only four out of the 19 galaxies are not barred.

\begin{table*}
\caption{General properties of the PHANGS-MUSE galaxies.   }
\label{tab:galaxies}
\centering
\begin{tabular}{lrrrrrrrrr}
\hline \hline
Name & Distance$^{a}$ & $v_\mathrm{sys}^{b}$ & $PA^{c}$ & $i^{c}$ & log$_{10}$ $M_*^{d}$ & $R_{25}^{b}$ & $r_{\rm eff}$ & $E(B-V)_{\rm MW}^{e}$ & resolution \\
 & Mpc & km s$^{-1}$ & deg & deg & M$_\odot$ & arcmin & arcmin & mag & pc \\
 \hline
\hline
IC5332 & 9.0 & 699 & 74.4 & 26.9 & 9.67 & 3.0 & 1.4 & 0.014 & 45 \\
NGC0628 & 9.8 & 651 & 20.7 & 8.9 & 10.34 & 4.9 & 1.4 & 0.061 & 42 \\
NGC1087 & 15.9 & 1502 & 359.1 & 42.9 & 9.93 & 1.5 & 0.7 & 0.030 & 71 \\
NGC1300 & 19.0 & 1545 & 278.0 & 31.8 & 10.62 & 3.0 & 1.2 & 0.026 & 62 \\
NGC1365$^{*}$ & 19.6 & 1613 & 201.1 & 55.4 & 10.99 & 6.0 & 3.3$^{f}$ & 0.018 & 84 \\
NGC1385 & 17.2 & 1477 & 181.3 & 44.0 & 9.98 & 1.7 & 0.7 & 0.017 & 96 \\
NGC1433$^{*}$ & 18.6 & 1057 & 199.7 & 28.6 & 10.87 & 3.1 & 0.8 & 0.008 & 83 \\
NGC1512 & 18.8 & 871 & 261.9 & 42.5 & 10.71 & 4.2 & 0.9 & 0.009 & 96 \\
NGC1566$^{*}$ & 17.7 & 1483 & 214.7 & 29.5 & 10.78 & 3.6 & 0.6 & 0.008 & 76 \\
NGC1672$^{*}$ & 19.4 & 1318 & 134.3 & 42.6 & 10.73 & 3.1 & 0.6 & 0.020 & 73 \\
NGC2835 & 12.2 & 867 & 1.0 & 41.3 & 10.00 & 3.2 & 0.9 & 0.086 & 33 \\
NGC3351 & 10.0 & 775 & 193.2 & 45.1 & 10.36 & 3.6 & 1.0 & 0.024 & 43 \\
NGC3627$^{*}$ & 11.3 & 715 & 173.1 & 57.3 & 10.83 & 5.1 & 1.1 & 0.029 & 69 \\
NGC4254 & 13.1 & 2388 & 68.1 & 34.4 & 10.42 & 2.5 & 0.6 & 0.033 & 61 \\
NGC4303$^{*}$ & 17.0 & 1560 & 312.4 & 23.5 & 10.52 & 3.4 & 0.7 & 0.019 & 96 \\
NGC4321 & 15.2 & 1572 & 156.2 & 38.5 & 10.75 & 3.0 & 1.2 & 0.023 & 59 \\
NGC4535 & 15.8 & 1954 & 179.7 & 44.7 & 10.53 & 4.1 & 1.4 & 0.017 & 80 \\
NGC5068 & 5.2 & 667 & 342.4 & 35.7 & 9.40 & 3.7 & 1.3 & 0.090 & 23 \\
NGC7496$^{*}$ & 18.7 & 1639 & 193.7 & 35.9 & 10.00 & 1.7 & 0.7 & 0.008 & 104 \\

\hline
	\multicolumn{10}{p{.8\textwidth}}{ 
		$^{a}${From the compilation of \citet{Anand2021}.}
		$^{b}${From LEDA \citep{Makarov2014}.}
		$^{c}${From \cite{Lang2020}, based on \mbox{CO(2--1)} kinematics.}
		$^{d}${Derived by \citet{Leroy2021}, using \textit{GALEX} UV and \textit{WISE} IR photometry.}
		$^{e}${From \cite{Schlafly2011}.}
		$^{f}${Due to AGN bias, derived from the  scale length (l$_*$) as r$_{\rm eff}$ = 1.41 l$_*$ following Equation 5 in \citet{Leroy2021}}.
		$^{*}${Classified as an AGN by \citet{Veron-Cetty2010}}
			}
\end{tabular}
\end{table*}

The PHANGS--MUSE large program was observed over several semesters and includes data from other programs and includes MUSE observations in both ground layer AO (adaptive optics) and non AO mode. Combined with variations in seeing, this means that the point spread function (PSF) varied between galaxies and among pointings within the same galaxy. To account for this variation between pointings in the same galaxy, we created mosaicked datacubes with a consistent PSF, where all pointings in a single galaxy were convolved to a single Gaussian PSF, whose size was determined by the pointing with the worst (largest) PSF. We used this optimised convolution data (copt) to identify the nebulae for the catalogue. The consistent PSF across the mosaic avoids issues with variable nebulae sizes across a single galaxy.  

The mosaicked MUSE datacubes were then passed through a data analysis pipeline (DAP) to provide maps of value-added products such as emission lines, mean stellar properties, gas and stellar kinematics and more \citep[as detailed in][]{Emsellem2022}. 
As the emission lines form the key data for this paper, we briefly describe the analysis here. The DAP uses the penalised pixel fitting method \citep[pPXF][]{Cappellari2017} to derive both the stellar continuum and emission lines properties within the spectral range 4850--7000\,\AA. F Before any fitting, the MUSE data is corrected for foreground Galactic extinction, using the \citet{CCM1981} extinction law and the $E(B - V)$ attributed to the Milky Way foreground from \citet{Schlafly2011}. 

To fit the stellar continuum and derive the stellar properties, the datacubes are first spatially Voronoi-binned \citep[using the {\sc vorbin} package][]{Cappellari2003} to achieve a minimum S/N of 35 in the $5300-5500$\,\AA\ wavelength range. 
The continuum between $4850-7000$\,\AA\ in each bin is then fit with a combination of E-MILES simple stellar population model templates \citep{Vazdekis2016} generated with a \citet{Chabrier2003} initial mass function and BaSTI isochrones \citep{Pietrinferni2004}. The Na I D absorption doublet (already removed in AO observations) are masked. The higher spectral resolution templates are convolved to the resolution of the MUSE data before fitting. The spectra is first fit to determine the stellar kinematics using a smaller set of model templates sampled at eight ages ($0.15-14$ Gyr, logarithmically sampled in steps of 0.22 dex), and four metallicities ([Z/H] = $-1.5$; $-0.35$; $-0.06$; $0.4$]). To the fit for the stellar population parameters we fix the kinematics and use a larger set of templates sampled at ages = [0.03, 0.05, 0.08, 0.15, 0.25, 0.40, 0.60, 1.0, 1.75, 3.0, 5.0, 8.5, 13.5] Gyr and [Z/H] = [$-1.49$, $-0.96$, $-0.35$, $+0.06$, $+0.26$, $+0.4$]. When fitting for the stellar population properties we also constrain the average attenuation of the stellar continuum, parametrized by the \citet{Calzetti2001} curve. 
 
To fit the emission lines we rerun pPXF on the mosaicked cubes at an individual spaxel level, with the emission lines treated as additional Gaussian components. The underlying stellar continuum is fit using the smaller set of E-MILES templates and the derived kinematics of the Voronoi bin that contained the individual spaxel, with the inclusion of an 8th-order multiplicative polynomial. We fit all strong emission lines and tie the kinematics (velocity and velocity dispersion) in three groups; Hydrogen lines (\ha, \hb), low ionization lines (e.g., \nii$\lambda6583$, \sii$\lambda \lambda6716,6731$), and high ionization lines (e.g., \oiii$\lambda5007$, \siii$\lambda6312$). These maps of stellar kinematics and mean properties, emission line fluxes, and gas kinematics form the key part of the data analysis pipeline and the PHANGS--MUSE release, as described in \citet{Emsellem2022}.

\section{Methods} 
\label{sec:methods}

\subsection{Nebular catalogue construction}\label{sec:HIIphot}
The PHANGS--MUSE galaxies are replete with emission lines, with more than $95$\% of our 0\farcs2 spaxels within $0.5R_{25}$ containing \ha\ emission at a $>3\sigma$ level \citep[see figure 20 in ][]{Emsellem2022}. With such filled maps, distinguishing individual nebulae from each other and the diffuse ionized gas is difficult, even with a median physical resolution of 70\,pc. Therefore, to identify the nebulae we require an unbiased and robust region identifier. While several such methods exist and have been applied previously (e.g., Clumpfind; \citet{Williams1994,Kreckel2016} or pyHIIExtractor; \citet{LugoAranda2022}), we chose to use HIIphot, a code specifically built to identify and characterise \hii\ regions with their irregular morphology \citep{Thilker2000}. We use a slightly altered version HIIphot to work on the \ha\ maps created from IFS data, first used in \citet{Kreckel2019}. 

Originally designed to be applied to narrow band imaging data centred on the \ha\ line, HIIphot used the associated broad band data used for continuum subtraction from the narrow band data to determine the significance of the \ha\ detection. However, IFS can spectrally resolve any underlying stellar continuum and subtract this as done within the data analysis pipeline. Therefore HIIphot was altered to work on \ha\ maps alone with the associated fitting error map to identify the nebulae and determine their boundaries. However the main algorithm in nebulae identification is still as described in \citet{Thilker2000}.

The key to nebulae identification is to first distinguish individual nebulae, then grow these up to a given termination criterion defining the edges of the nebulae. While a classical photoionized nebula has a clear boundary defined by the edge of the Str\"omgren sphere, real nebulae may be centrally concentrated or appear as rings, or have several peaks and a diffuse boundary due to density variations within the ISM. The angular resolution of our observations means that we only resolve the largest of \hii\ region complexes. In most \hii\ regions our resolution smooths any features and boundaries and, a potentially larger problem we discuss in Section \ref{sec:blending}, merge proximate nebulae. Therefore the choice of controlling parameters is driven by both the dataset and the physics of nebulae. 

As described in \citet{Santoro2022}, to identify the nebulae we first require to detect the peaks in \ha\ emission, or `seed regions', above the diffuse background. We set this background for each galaxy to be the median of all \ha\ pixels within the MUSE FoV with $\Sigma_{\ha} < 1\times 10^{-17}$ erg\, s$^{-1}$\, cm$^{-2}$\, arcsec$^{-2}$. This ranges from $\Sigma_{\ha}=1$ to $3\times10^{-18}$ erg\, s$^{-1}$\, cm$^{-2}$\, arcsec$^{-2}$ across our sample. The detection threshold within HIIphot was set to 3$\sigma$ above this background, where $\sigma$ is the standard deviation of the background pixels and typically around the same level as the background. 

Given the diverse morphologies of \hii\ regions (and other ionized nebulae), HIIphot performs iterative Gaussian smoothing on the \ha\ maps, merging connecting features to create the nebulae `footprints'. To avoid the detection of regions with unphysical sizes, we limit the spatial smoothing to three iterations, each time increasing the smoothing kernel (starting from the original resolution image) by 10\%. These footprints are then further trimmed to a `seed' with a consistent isophotal boundary defined by 50\% of the median within the footprint. Once detected, we further cleaned the seed sample to avoid artefacts due to noise by imposing a S/N cut of 50 above the \ha\ error maps for the integrated flux values.

Defining the boundary edges of nebulae is challenging, with many criteria existing in the literature (e.g.~\ha\ surface brightness, line ratios, \ha\ equivalent widths). By using HIIphot we chose to use the spatial gradient of the \ha\ surface brightness to define the boundaries of our nebulae. As discussed in \citet{Thilker2000}, the choice of terminal gradient is ambiguous, with flatter values leading to larger \hii\ regions that can include the diffuse ionized gas directly associated with the \hii\ region \citep[see, e.g.][for the level of association]{Belfiore2022} but also lead to a more contiguous map of nebulae \citep[Figure 5 in][]{Thilker2000}. The spatial resolution of the data also impacts the exact boundaries, smoothing edges and potentially merging adjoining nebulae. 
We chose to use a single termination gradient of 5.0 EM pc$^{-1}$ (where the emission measure, EM, is in cm$^{-6}$\,pc) for all galaxies (corresponding to $2.43\times10^{-16}$ erg\, s$^{-1}$\, arcsec$^{-2}$\, pc$^{-1}$). This value is similar to that used in other nearby galaxy studies \citep[e.g.]{Oey2007, Zhang2017}, and visually provided the best balance in terms of capturing the total \ha\ flux for each nebula, while limiting the size growth. We chose to use a single termination gradient rather than one for each galaxy for consistency in nebulae identification, even given the factor of $\sim$4 difference in physical resolution across the PHANGS--MUSE sample. 

\begin{figure*}
    \centering
    \includegraphics[width=7in]{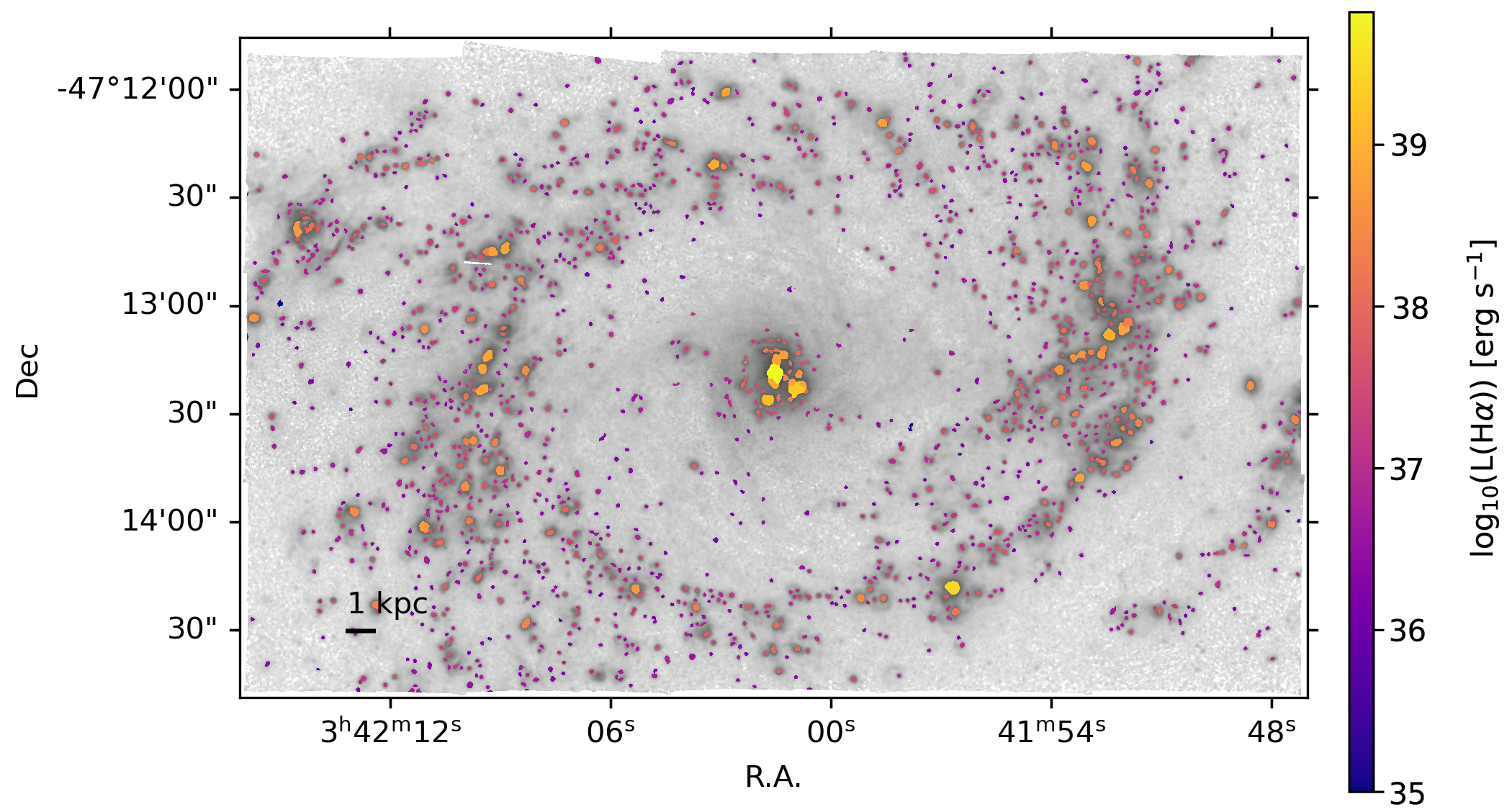}
    \caption{A visualisation of the two dimensional spatial extent and distribution of all nebulae in the galaxy NGC 1433. The background greyscale image shows the \ha\ emission in log scale, and the colour of the nebulae indicates their intrinsic (dust corrected) \ha\ luminosity. A full atlas of each galaxy is available in Appendix \ref{appendix:atlas}.}
    \label{fig:allnebulae}
\end{figure*}

These steps lead to 31,497 identified nebulae with defined boundaries across our sample of 19 galaxies. For each nebula we report the central position in both RA and Dec, weighted by \ha\ intensity, and their position relative to the galaxy centres. We quantify the area encompassed by the nebulae in pixels, however as most of our regions are unresolved or only marginally resolved (see also Section \ref{sec:emission_lines}), we do not provide size measurements (though see Section 4.2 where we do present 10\%, 50\%, and 90\% circularized radial sizes for the overall distributions in each galaxy). As described in \citet[][particularly \S 5.3]{Emsellem2022}, foreground star masks were generated for all PHANGS--MUSE galaxies based on the \emph{Gaia} DR2 catalogue \citep{Gaia2018}. We exclude 98 sources whose footprint falls within the star masks and are likely impacted by artefacts from incorrect stellar continuum subtraction.
We also flag 609 nebulae with centres within 1 PSF FWHM of the edges of our PHANGS--MUSE galaxy footprints. While the emission lines in these regions are likely correctly measured, their proximity to the edges mean that their boundaries are potentially incorrectly defined and that their integrated line fluxes may not represent the total emission of the nebulae. For larger nebulae where the distinct ionized zones can be distinguished (i.e.~the S$^{++}$ and S$^{+}$ zones are resolved), the emission line ratios measured for these regions are potentially incorrect. While these flagged nebulae are included in the full catalogue, we exclude them from our further analysis. For the results presented in this paper, we focus on the remaining 30,790 nebulae (Figure \ref{fig:allnebulae}).

Spatial masks corresponding to the locations of each identified nebula are released as data products accompanying this paper, and can also be found via the PHANGS webpage\footnote{\url{https://sites.google.com/view/phangs/home/data}}. An image atlas showing the footprints of all nebulae in each galaxy is included in Appendix \ref{appendix:atlas}.

\subsection{Emission line measurements}
\label{sec:emission_lines}

With the footprints of all nebulae defined, we integrated the original MUSE spectra within each nebula and re-fit using the same data analysis pipeline (DAP) described in \cite{Emsellem2022} used to create the original \ha\ maps. 
We do this to increase the signal-to-noise of our emission lines and to detect faint spectral features, such as the temperature sensitive auroral lines. The only changes to the pipeline are to extend the wavelength range fitted to include the \siii $\lambda 9069$\,\AA\ emission line and, when integrating the spectra, we use the unconvolved, \emph{native} resolution mosaicked data cubes to minimise the impact of PSF smearing at nebula boundaries. The latter only has a small impact due to the consistency in seeing between observations, but for some galaxies \citep[e.g.~NGC1365, as seen in Table A1 in][]{Emsellem2022} the PSF variation can be a factor of two within the different pointings of the mosaic.

\begin{table*}
	
	\centering

    \caption{Wavelengths and ionisation potential of the relevant ion for each emission line included in the public catalog. All lines are corrected for the Milky Way foreground dust extinction. Wavelengths are taken from the
    National Institute of Standards and Technology 
    (NIST; \url{https://physics.nist.gov/PhysRefData/ASD/lines_form.html}), and are Ritz wavelengths in air (consist with wavelengths in the public data release) except for the H~Balmer lines, in which case we use the `observed ' wavelength in air as reported in NIST. The
    \DAP\ string name is used to identify the correct extension in the PHANGS-MUSE \MAPS\ files. Ionisation
    potentials are taken from \citet{Draine2011}.  }
	
	\begin{tabular}{l l l c c}
		\hline
		Line name & Wavelength & String ID & Ionisation potential & Fixed ratio \\
		 & (air) [\AA] & & [eV] & \\
		\hline
		\multicolumn{5}{c}{Hydrogen Balmer lines} \\
		\hline
		\hb\ & 4861.35 & \texttt{HB4861} & 13.60 & no \\
		\ha\ & 6562.79 & \texttt{HA6562} & 13.60  & no\\
		\hline
		\multicolumn{5}{c}{Low ionisation lines} \\
		\hline
		\oi$\lambda$6300   & 6300.30  & \texttt{OI6300} & --- & no\\
		\nii$\lambda$6548  &  6548.05  & \texttt{NII6548} & 14.53 & 0.34 \nii$\lambda$6584\\
		\nii$\lambda$6584  &  6583.45 & \texttt{NII6583}  & 14.53 & no \\
		\sii$\lambda$6717  & 6716.44   & \texttt{SII6716} & 10.36 & no \\
		\sii$\lambda$6731  & 6730.82  & \texttt{SII6730} & 10.36 & no\\
		\hline
		\multicolumn{5}{c}{High ionization lines} \\
		\hline
		\oiii$\lambda$4959 & 4958.91  & \texttt{OIII4958} & 35.12 & 0.35 \oiii$\lambda$5007 \\ 
		\oiii$\lambda$5007 & 5006.84   & \texttt{OIII5006} &35.12 & no\\ 
		\siii$\lambda$9068 & 	9068.6  & \texttt{SIII9068} & 23.34 & no\\
		\hline
	\end{tabular}
	
	\label{tab:lines}
\end{table*}

As with the global PHANGS--MUSE DAP, we fit all emission lines simultaneously, but also include lines that are fainter (e.g.~\nii$\lambda5754$) and at longer wavelengths (i.e.~\siii$\lambda 9069$). The full list of lines released in this catalogue is given in Table \ref{tab:lines}. Similarly, when fitting lines we assume single Gaussian profiles and tie the kinematics (velocity and velocity dispersion) in three groups: hydrogen lines, low ionization lines, and high ionization lines. Line velocities are reported relative to the systemic velocity of the galaxy, provided in Table \ref{tab:galaxies}. In our fit we account for a Milky Way foreground extinction, assuming the $E(B-V)$ values provided by \cite{Schlafly2011} (also listed in Table \ref{tab:galaxies}) and an \cite{ODonnell1994} extinction law. 

\begin{figure*}
    \centering
    \includegraphics[width=\hsize]{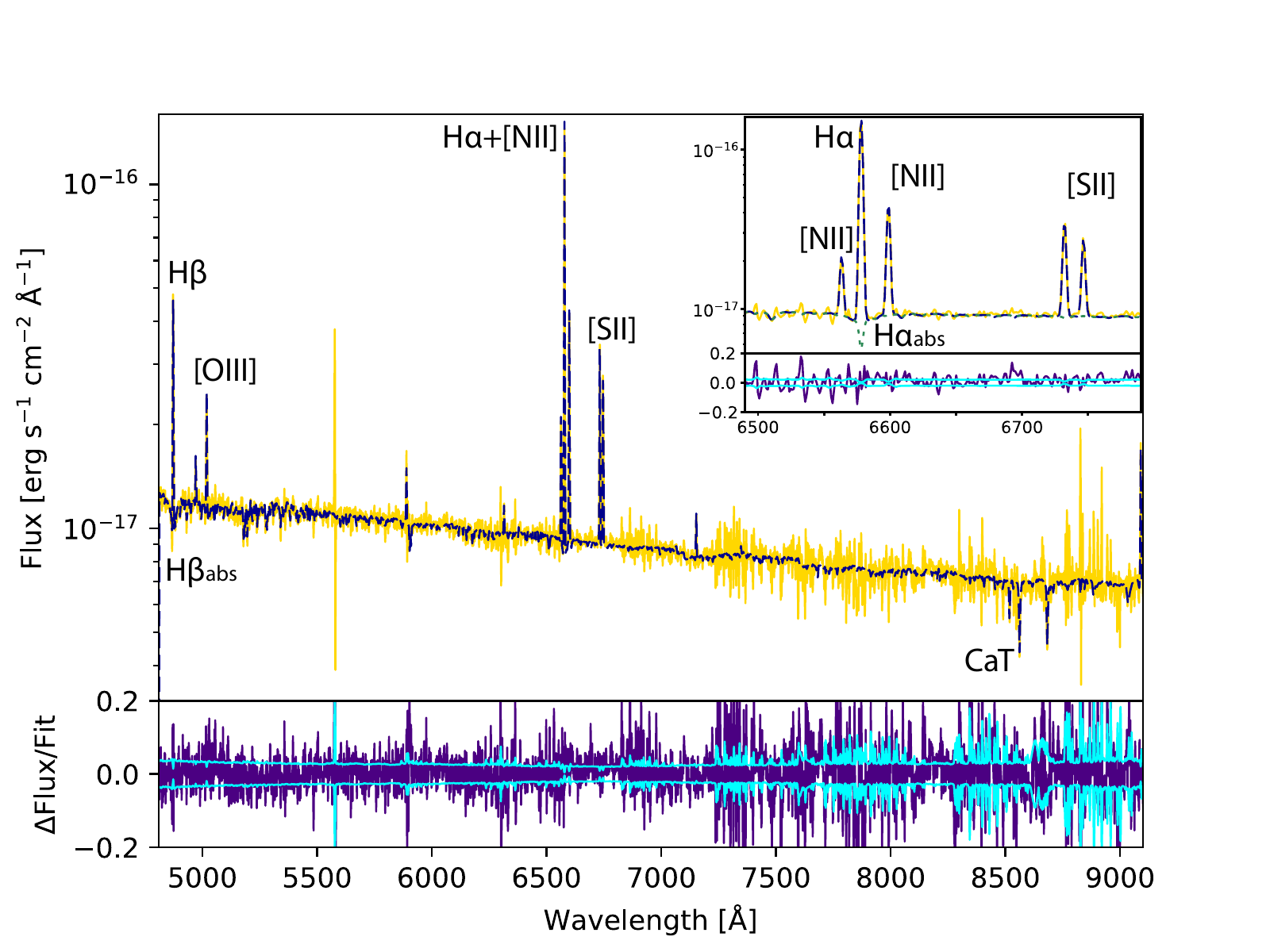}
    \caption{A nebula spectrum from IC5332 (region ID 38, $L_{\ha}=10^{36.8}$\,erg\,s$^{-1}$), showing a typical integrated spectrum (gold), including the underlying stellar (and weak nebula) continuum with clear strong emission and absorption lines (with the strong lines labelled) and visible residuals from the sky background subtraction. Overlaid is the resulting analysis pipeline fit from pPXF (blue dashed line), from which we extract the emission line fluxes and kinematics. The lower panel shows the relative residual of the spectral fit in indigo and relative spectral error in cyan ((Nebula-Fit)/Fit and error/fit, respectively). In the insert we show a zoom in of the \ha\ and \nii\ region, where the spectral lines can be seen more clearly. We also show the underlying continuum fit (dotted green line) revealing the underlying \ha $_{\rm abs}$ feature.}
    \label{fig:specfit}
\end{figure*}

In Figure~\ref{fig:specfit}, we show a typical nebular spectrum from IC5332 (ID:38, $L_{\ha}=10^{36.8}$\,erg\,s$^{-1}$), along with the best fitted spectrum from the data analysis pipeline overlaid, and the relative residual from the fit at the bottom. What is clear from this spectrum are the emission lines, and how well we reproduce these. Also clear are the strong sky line residuals (especially beyond 6800\AA). The underlying continuum is dominated by stellar light, although at the scaling in this figure, only certain stellar absorption features are visible. 

As an insert we show a zoom-in of the \ha\ region of the spectrum with the strong \nii\ and \sii\ lines. Also shown is the stellar continuum fit in green, clearly demonstrating the \ha\ absorption feature (\ha $_{\rm}abs$). In the brightest nebulae, weaker lines such as the \oii $\lambda 7319,7330$\AA\ doublet are clearly visible. In some nebulae, faint residuals around bright lines are visible, suggesting more complex kinematics than can be modelled by a single Gaussian component. However typically these residuals are still at the level of the spectral uncertainty propagated from the MUSE spectral cubes (cyan lines in residual plot). In the example shown in Figure ~\ref{fig:specfit}, 89\% of all pixels shown have residuals within $3\sigma$ of the errors, with sky residuals dominating the outlying pixels.  

We also determine the Balmer emission line equivalent width (EW), as both an indication of the relative brightness of the nebula to the underlying stellar population, and a proxy for the local specific star formation rate surface density. We calculate EW(\ha) and EW(\hb) following the procedure described in \cite{Westfall2019}, as applied within the MaNGA survey for calculation of emission-line moments. 
For EW(\ha), we integrate the flux over a central band from 6557.6 -- 6571.35\AA\ in the rest frame. We calculate the continuum flux using determining the median over a blue (6483.0--6513.0\AA) and red (6623.0--6653.0\AA) channel and then determine the mean between these. For EW(\hb), we use the same approach, with the \hb\ line determined over the 4847.9 -- 4876.6\AA\ band and the continuum flux using 4827.9 -- 4847.9\AA\ and 4876.6 -- 4891.6\AA\ for the blue and red channels.
We calculate the equivalent widths in two ways. The first approach is to use the direct nebulae spectrum, with the line flux simply the integral over the central band minus the determined continuum, what we call here the `raw' EW. However, this ignores the impact of the underlying stellar absorption feature visible in Figure \ref{fig:specfit}. The second method is to integrate over the spectrum once the best-fitting continuum fit from pPXF has been subtracted or the `fit' EW. This accounts for the underlying absorption feature, however requires sufficient S/N in the data to get a good fit to the continuum and is poorly determined in low spectral resolution data, such as narrow-band imaging. In both cases the EW is then the determined line flux over the mean continuum.

The median value across our full catalogue is EW(\ha)$_{\rm raw}\sim20$\AA\ and EW(\hb)$_{\rm raw}\sim1$\AA, with $\sim43$\% of nebulae having EW(\hb)$_{\rm raw}<0$ due to the underlying stellar absorption feature. While the median difference between EW$(\ha)_{\rm raw}-$EW(\ha)$_{\rm fit}$ is typically only $-2.2$\AA, the difference is relatively stronger for the weaker \hb\ line at $-4.1$\AA.

While this presents the EW in a standard format, suitable for comparison with previous work, it is clear that our nebulae are sitting within the central stellar disk of each galaxy. Due to this, our stellar continuum band naturally suffers from a significant contribution of light from old stellar population, which is not associated with the young nebulae. The impact of this background contribution is explored in Scheuermann et al. (submitted). 

Given that nebular objects can be marginally resolved in our data, with \hii\ regions displaying a variety of \ha\ morphologies, determination of the completeness of our catalogue by quantifying the recovery rate of artificial source injection is not straightforward.  In \cite{Santoro2022}, the completeness for our catalogue was estimated in an empirical way by considering the \ha\ line emission outside of the region masks and measuring the \ha\ surface brightness at the 90th percentile level of the surface brightness distribution. This surface brightness was then converted to a luminosity assuming an unresolved point source. By this metric, typical completeness limits are 10$^{36}$ -- 10$^{37}$ erg s$^{-1}$, which are roughly equivalent to the ionizing flux of a single O7V star \citep{Vacca1994}.  We refer the reader to \cite{Santoro2022} for further details and a complete table. For objects classified as \hii\ regions (see Section \ref{sec:bpt}), in Table~\ref{tab:lum_size_dist} we quantify the 10th, 50th and 90th percentile extinction corrected \ha\ luminosities and physical sizes.  

\begin{table}
\caption{Distribution of \hii\ region attenuation-corrected \ha\ luminosities (L$_{H\alpha}$) and sizes ($r_{\rm circ}$) for each galaxy, listing the 10\%, 50\% and 90\% values.     }
\label{tab:lum_size_dist}
\centering
\begin{tabular}{l | rrr | rrr}
\hline \hline
Galaxy & \multicolumn{3}{c}{log(L$_{H\alpha}$ [erg s$^{-1}$])} & \multicolumn{3}{c}{$r_{\rm circ}$ [pc] } \\
 & 10\% & 50\% & 90\% & 10\% & 50\% & 90\% \\
\hline
\hline
IC5332 & 35.9 & 36.4 & 37.2 &   27 &    33 &   46 \\ 
 NGC0628 & 36.2 & 36.7 & 37.8 &   31 &    35 &   58 \\ 
 NGC1087 & 36.7 & 37.5 & 38.5 &   51 &    55 &   98 \\ 
 NGC1300 & 36.6 & 37.2 & 38.1 &   58 &    62 &   88 \\ 
 NGC1365 & 37.0 & 37.7 & 38.9 &   76 &    82 &  126 \\ 
 NGC1385 & 36.7 & 37.5 & 38.8 &   45 &    51 &  103 \\ 
 NGC1433 & 36.5 & 37.1 & 37.9 &   58 &    62 &   85 \\ 
 NGC1512 & 36.8 & 37.4 & 38.1 &   79 &    83 &  115 \\ 
 NGC1566 & 36.6 & 37.3 & 38.6 &   48 &    53 &  100 \\ 
 NGC1672 & 37.0 & 37.6 & 38.9 &   64 &    70 &  121 \\ 
 NGC2835 & 36.5 & 37.1 & 38.1 &   48 &    52 &   84 \\ 
 NGC3351 & 36.3 & 36.8 & 37.7 &   36 &    39 &   58 \\ 
 NGC3627 & 37.0 & 37.7 & 38.9 &   41 &    49 &   96 \\ 
 NGC4254 & 36.7 & 37.5 & 38.7 &   41 &    48 &   91 \\ 
 NGC4303 & 36.9 & 37.6 & 38.7 &   46 &    56 &  107 \\ 
 NGC4321 & 36.9 & 37.6 & 38.5 &   59 &    64 &   96 \\ 
 NGC4535 & 36.4 & 37.0 & 38.1 &   32 &    40 &   71 \\ 
 NGC5068 & 35.8 & 36.3 & 37.5 &   19 &    23 &   45 \\ 
 NGC7496 & 36.6 & 37.2 & 38.3 &   56 &    61 &   96 \\ 
 \hline
\end{tabular}
\end{table}

\section{Value-Added Products}
\label{sec:VAC}
Given the large suite of emission lines measured within our nebulae, there are multiple physical properties that can be determined. We include these value-added properties in the nebulae catalogue, though note that different calibrations can be used for many of these properties resulting in systematically different results. For all properties below we only present the results where the relevant lines have a S/N greater than 5, and when the nebular classification is appropriate (e.g.\ metallicities are only calculated for \hii\ regions).  A complete list of all columns contained in our nebular catalogue is provided in Table \ref{tab:catalog_data}. 

\begin{table*}
	
	\centering

    \caption{Columns in the catalogue} \label{tab:catalog_data}
	
	\begin{tabular}{l c c}
		\hline
		Column & Unit & Description \\
		\hline
		\hline
		gal\_name & & galaxy name \\
		region\_ID & & region ID \\
        cen\_ra & deg & RA (J2000) center, weighted by \ha\ intensity \\ 
        cen\_dec & deg & Dec (J2000), weighted by \ha\ intensity \\
        flag\_edge & & flag set to 1 if within one PSF of the field edge \\
        flag\_star & & flag set to 1 if overlapping with a star \\
        deproj\_r\_R25 & R25 & Deprojected distance from galaxy center in units of R25 \\
        deproj\_r\_reff & r$_{\rm eff}$ & Deprojected distance from galaxy center in units of r$_{\rm eff}$\\
        deproj\_phi & deg & Deprojected position angle within galaxy disk \\
        region\_area & pixels & \hii\ region area \\
        emline*\_FLUX$^{\dag}$ & 1e-20 erg/s/cm$^2$  & emission line fluxes (see Table \ref{tab:lines})\\     
        emline*\_FLUX\_CORR$^{\dag}$ & 1e-20 erg/s/cm$^2$  & attenuation-corrected emission line fluxes (see Table \ref{tab:lines})\\
        & & assuming an \cite{ODonnell1994} extinction curve and R$_V$ = 3.1 \\
        emline*\_VEL$^{\dag}$ & km/s & line velocity relative to v$_{sys}$ (Table \ref{tab:galaxies})\\
        emline*\_SIGMA$^{\dag}$ & km/s & line velocity dispersion, corrected for instrumental broadening\\
        AV$^{\dag}$ & mag & A$_V$, V-band attenuation derived from the Balmer decrement \\
        & & assuming an \cite{ODonnell1994} extinction curve and R$_V$ = 3.1\\
        EW\_HA6562\_raw$^{\dag}$ & \AA & Equivalent width of H$\alpha$, measured directly\\
        EW\_HB4861\_raw$^{\dag}$ & \AA & Equivalent width of H$\beta$, measured directly\\
	EW\_HA6562\_fit$^{\dag}$ & \AA & Equivalent width of H$\alpha$, measured after stellar continuum subtracted\\
        EW\_HB4861\_fit$^{\dag}$ & \AA & Equivalent width of H$\beta$, measured after stellar continuum subtracted\\
        HA6562\_LUM\_CORR & erg/s & attenuation corrected \ha\ luminosity  \\
        BPT\_NII & & BPT flag, see Table \ref{tab:bpt_flags}\\
        BPT\_SII & & BPT flag, see Table \ref{tab:bpt_flags}\\
        BPT\_OI & & BPT flag, see Table \ref{tab:bpt_flags}\\
        met\_scal$^{\dag}$ & & Metallicities determined using the Scal prescription \citep{Pilyugin2016}\\
        Delta\_met\_scal & & Offset in metallicity relative to the radial gradient (Table \ref{tab:metal_grad_scal})\\
        logU$^{\dag}$ & & Ionization parameter derived from \siii/\sii\ using the prescription in \citealt{Diaz1991}\\
        Environment & & Environment flag, as in Table \ref{tab:env_flags}\\
		\hline
		\multicolumn{3}{l}{$^{*}$emission lines are listed in Table \ref{tab:lines}}\\
		\multicolumn{3}{l}{$^{\dag}$Note that corresponding errors are included as *\textunderscore ERR}
	\end{tabular}
\end{table*}

\subsection{Dust Attenuation}\label{sec:dust}

All line fluxes are provided in our catalogue as observed values, yet the derivations of physical quantities (e.g., metallicity, ionization parameter) are typically based on intrinsic line fluxes. Therefore, before deriving any quantities, the measured fluxes need to be corrected for reddening due to dust. 
We assume here an \citet{ODonnell1994} 
extinction curve with an $R_{V}=3.1$, that represents a small modification of the \citet{CCM1981} extinction curve. We then derive the reddening, $E(B-V)$, based on this curve and assume an intrinsic Balmer ratio of \ha/\hb$=2.86$.  In practice the choice of extinction curve has little impact on the corrections, as extinction curves do not deviate significantly across the MUSE wavelength range and the \ha\ and \hb\ emission lines are bracketing most of the emission lines of interest. However the derived $A_V$, and hence line luminosity, is directly dependent upon our assumed value of $R_{V}$.
By using the \cite{ODonnell1994} extinction curve we are assuming that the nebula itself only experiences attenuation from a uniform foreground dust layer. In reality, complex dust geometries within the nebulae \citep[as seen in nearby \hii\ regions like the Tarantula;][]{DeMarchi2016} as well as blending of multiple regions along our line of sight might bias our inferred extinction. However, we believe at our $<$100~pc scales with distinguished nebulae and the thin star-forming disk the foreground screen assumption is more justified than a mixed-media model for the majority of our nebulae. 

\begin{figure*}
    \centering
\includegraphics[width=0.95\hsize]{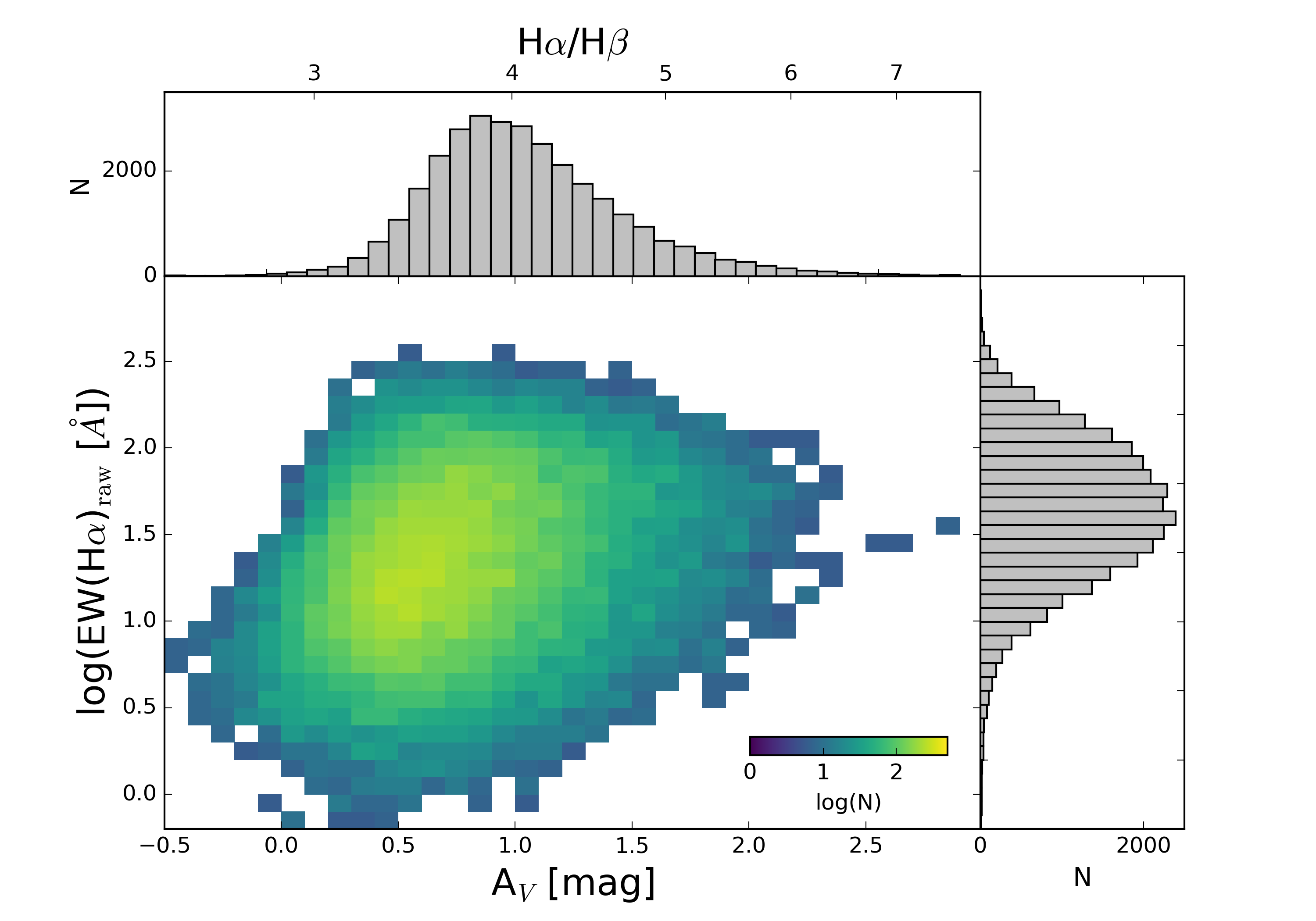}
\caption{The distribution of nebulae attenuations, $A_V$, derived from the Balmer decrement and the \citet{ODonnell1994} attenuation curve with an $R_{V}=3.1$, compared to the \ha\ emission line equivalent width determined from the integrated spectra (EW(\ha)$_{\rm raw}$). 
The central plot shows the 2D histogram of $A_V$ with the measured EW(\ha)$_{\rm raw}$, while the outer histograms show the 1D distributions of both properties. The Balmer decrement (\ha/\hb) is also shown directly on the top axis.}
    \label{fig:Av_EW}
\end{figure*}

In Figure \ref{fig:Av_EW}, we show the distribution of V-band attenuations, $A_V$, experienced by the nebulae with significant detections of \hb\ (S/N $>$ 5; 31,377 objects; 99.9\% of the sample).
We find a median $A_V=0.72$ mag (16\%--84\% range is 0.34--1.2 mag) and a tail of highly attenuated nebulae (5\% of objects have $A_V$ $>$ 1.7 mag). When we weight the $A_V$'s linearly by the intrinsic \ha\ luminosity (attenuation corrected), we find a median $A_V=1.4$ mag (16\%--84\% range is 0.8--2.5 mag), a significant increase. While it does appear in the sample (and as suggested by Figure \ref{fig:Av_EW}) that the brightest \hii\ regions are more attenuated, this difference in median $A_V$ is also caused by the highly obscured faint \hii\ regions being undetected in \hb.  Also visible in this figure are a small subset ($<$3\%) with unphysical attenuation ($A(V) <0$, meaning $\ha/\hb < 2.86$) even with a S/N $>$ 5 in both \ha\ and \hb. 80\% of these are consistent within the 3$\sigma$ line flux uncertainties with a value of $\ha/\hb = 2.86$.  The remainder are typically found in nebulae with low \ha\ equivalent widths, as can be seen in the central plot in Figure \ref{fig:Av_EW}, suggesting that for most of these nebulae the underlying Balmer absorption features are incorrectly subtracted leading to an overestimated \hb\ flux. However, it may also be that the intrinsic Balmer ratio for some of these nebulae is less than our fiducial value of 2.86 due to physical reasons associated with the nebula itself (e.g.~Planetary Nebulae are both faint and typically several thousand Kelvin hotter than \hii\ regions and therefore have an intrinsically lower ratio). 
For these unphysical attenuations we set $A_V=0$ mag when considering the reddening correction of the line ratios. All emission lines included in our catalogue are also provided as corrected values ('*\_CORR'; see Table \ref{tab:catalog_data}) by applying our determined $A_V$ and chosen extinction curve. 

\subsection[Emission line diagnostic classifications and HII region catalogue construction]{Emission line diagnostic classifications and \hii\ region catalogue construction}
\label{sec:bpt}

In defining the nebulae catalogue, we have used the HIIphot code \citep{Thilker2000}. However, within the PHANGS--MUSE data we also clearly see \ha\ emission associated with supernova remnants, planetary nebulae, and regions ionized by active galactic nuclei (AGN). As a first pass at separating these regions we use emission line ratio diagrams \citep[commonly called BPT diagrams after their introduction in][]{Baldwin1981} and use the diagnostic curves described in \citet{Kewley2001} and \citet{Kauffmann2003} to classify the nebulae. We note that while the \citet{Kauffmann2003} curve is derived empirically from global spectra, it still provides a useful constraint on whether ionization by processes other than UV photons from OB-stars are playing a role in the nebulae (e.g.~shocks, AGN, etc.; \citealt{Law2021, Belfiore2022})

\begin{figure*}
    \centering
    \includegraphics[width=7in]{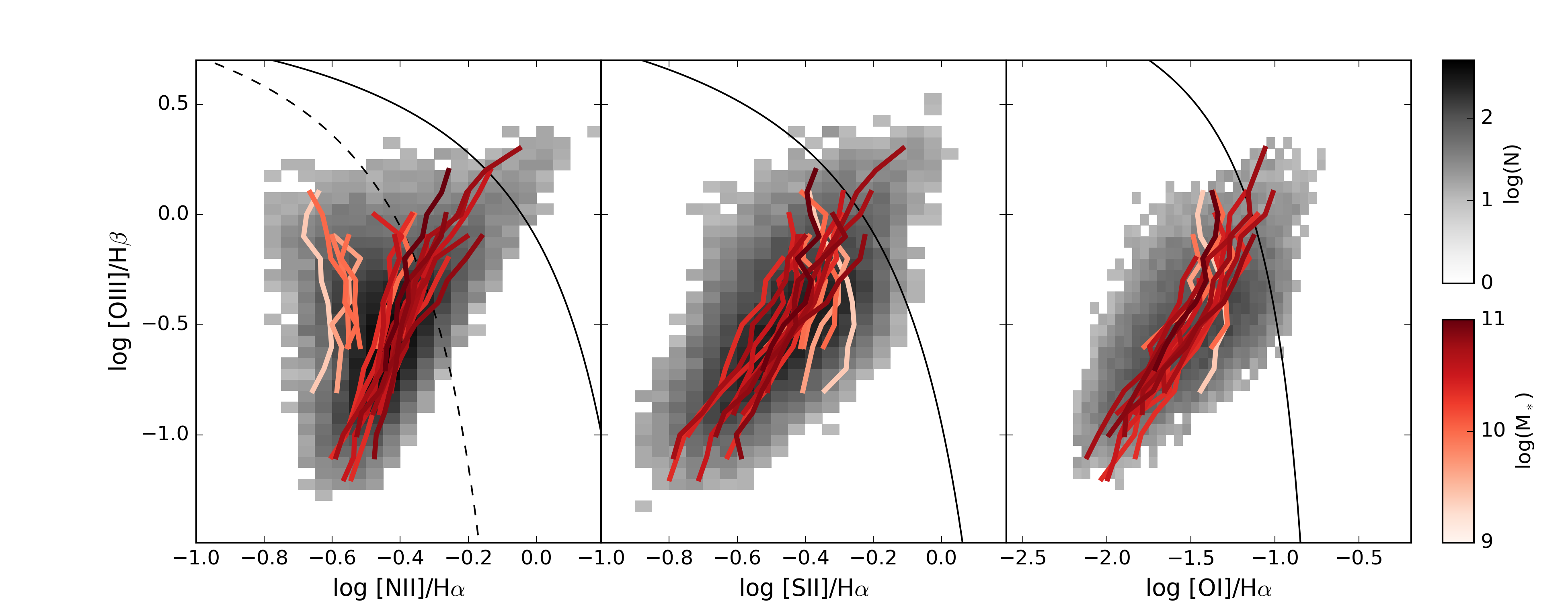}
    \caption{The distribution of our nebulae on the three BPT \citep{Baldwin1981} diagnostic diagrams considered when flagging for ionization mechanism (Table \ref{tab:bpt_flags}). From left to right, we show  \oiii$\lambda 5007/\hb$ versus \nii$\lambda 6583/\ha$, \sii$\lambda\lambda$6716,6730/\ha, and \oi$\lambda$6300/\ha. Overlaid are the \citet{Kauffmann2003} diagnostic curve (dashed line) and \citet{Kewley2006} diagnostic curves (solid lines). Non-stellar ionizing sources are typically found above these lines, and objects between the two demarcations are considered composites. Greyscale demonstrates the distribution of nebulae in these diagrams, and 79\% of nebulae are consistent with photoionization. Coloured lines (scaled to the galaxy stellar mass) indicate galaxy trends (binned along the y-axis). \nii/\ha\ and \sii/\ha\ show clear secondary correlations with stellar mass (corresponding to trends in metallicity). 
    }
    \label{fig:bpt}
\end{figure*}

We use the three strong line diagnostic diagrams (Figure \ref{fig:bpt}) to classify our nebulae; \oiii/\hb\ versus \nii$\lambda6584$/\ha, \oiii/\hb\ versus \sii$\lambda6717,6731$/\ha, and \oiii/\hb\ vs \oi$\lambda6300$/\ha. We note that different galaxies follow tracks that are slightly offset and correlate with the total stellar mass of the galaxy (and presumably its metallicity), with all galaxies shown individually in Appendix \ref{appendix:bpt}. For each diagnostic, we flag the nebulae with S/N $<$ 5 in any of the lines used in the diagnostic, then mark the remaining as \hii\ regions, composites, clear AGN impact or LINER-like (indicative of shocks or strong contributions from more diffuse ionized gas) spectra (Table \ref{tab:bpt_flags}).  
We construct an \hii\ region sample from those objects classified as \hii\ regions by all three diagnostics (20,577 objects), as well as objects where \oi\ is not detected with S/N $>$ 5 but they are otherwise consistent with the \hii\ region BPT diagnostics (2,667 objects). 
This results in a total of 23,244 (74.0\%)  
of objects that are consistent with photoionization by massive stars, and we consider this to be our full \hii\ region catalogue. This sample would increase by about 1500 objects if we reduced our S/N requirement to 3, and it would only increase by about $\sim$100 objects if we included objects that fall below the BPT demarcations when accounting the line flux uncertainties.  

Given the factor of two variation in distance and moderate variation in sensitivity between galaxies, we do not achieve uniform detection thresholds across the sample. In addition, the various \hii\ region morphologies considered by HIIphot do not lead to homogeneous  luminosity thresholds in our source identification.  
To provide a general quantification of our typical \hii\ region properties per galaxy, we quote the 10th, 50th and 90th percentiles in both attenuation-corrected \ha\ luminosity and \hii\ region size in Table \ref{tab:lum_size_dist}. Here, the size is taken as the circularized radius that results in an equal area to the area of the \hii\ region mask. We note that the vast majority of our regions are unresolved, as reflected by the close agreement in 10\% and 50\% sizes, along with the clear correlation with galaxy distance. Because of this, we purposefully do not include size measurements in our catalogue. The 10th percentile luminosities provide a general idea of the completeness limits for each galaxy, and we refer to \cite{Santoro2022} for more detailed discussion. Future work will aim to map out of the \hii\ region selection function more completely and provide homogenised 150~pc scale catalogues.

\begin{table}
\caption{BPT flags included in the catalog   }
\label{tab:bpt_flags}
\centering
\begin{tabular}{lcr}
\hline \hline
Column name & Value & Meaning\\
\hline
\hline
BPT\_NII & 0 & star formation \\
 & $-1$ & low S/N $<$ 5 \\
 & 1 & composite \\
 &  3 & AGN \\
 \hline
BPT\_SII & 0 & star formation \\
 & $-1$ & low S/N $<$ 5 \\
 & 2 & LI(N)ER  \\
 & 3  & AGN \\
 \hline
 BPT\_OI & 0 & star formation \\
 & $-1$ & low S/N $<$ 5 \\
 & 3  & AGN \\
\hline
\end{tabular}
\end{table}

\subsection{Gas-phase metallicities}
\label{sec:metallicities}

To derive the gas-phase metallicity there are a wide range of prescriptions in the literature that can be applied to the nebulae classified as \hii\ regions. 
Systematic differences between these prescriptions are well known in the literature, and routinely produce absolute measurements that differ by 0.2 dex, and even up to 0.7 dex, in $12+log({\rm O/H})$ for the same \hii\ regions \citep[see, e.g.,][for reviews on this problem]{Peimbert2017, Kewley2019}. While qualitatively the difference between individual \hii\ regions is typically maintained (metal-poor remain poor), the scale in these differences can also be markedly different, as shown by \citet{Kewley2008} in SDSS galaxies. 

\begin{figure*}
    \centering
    \includegraphics[width=7.5in]{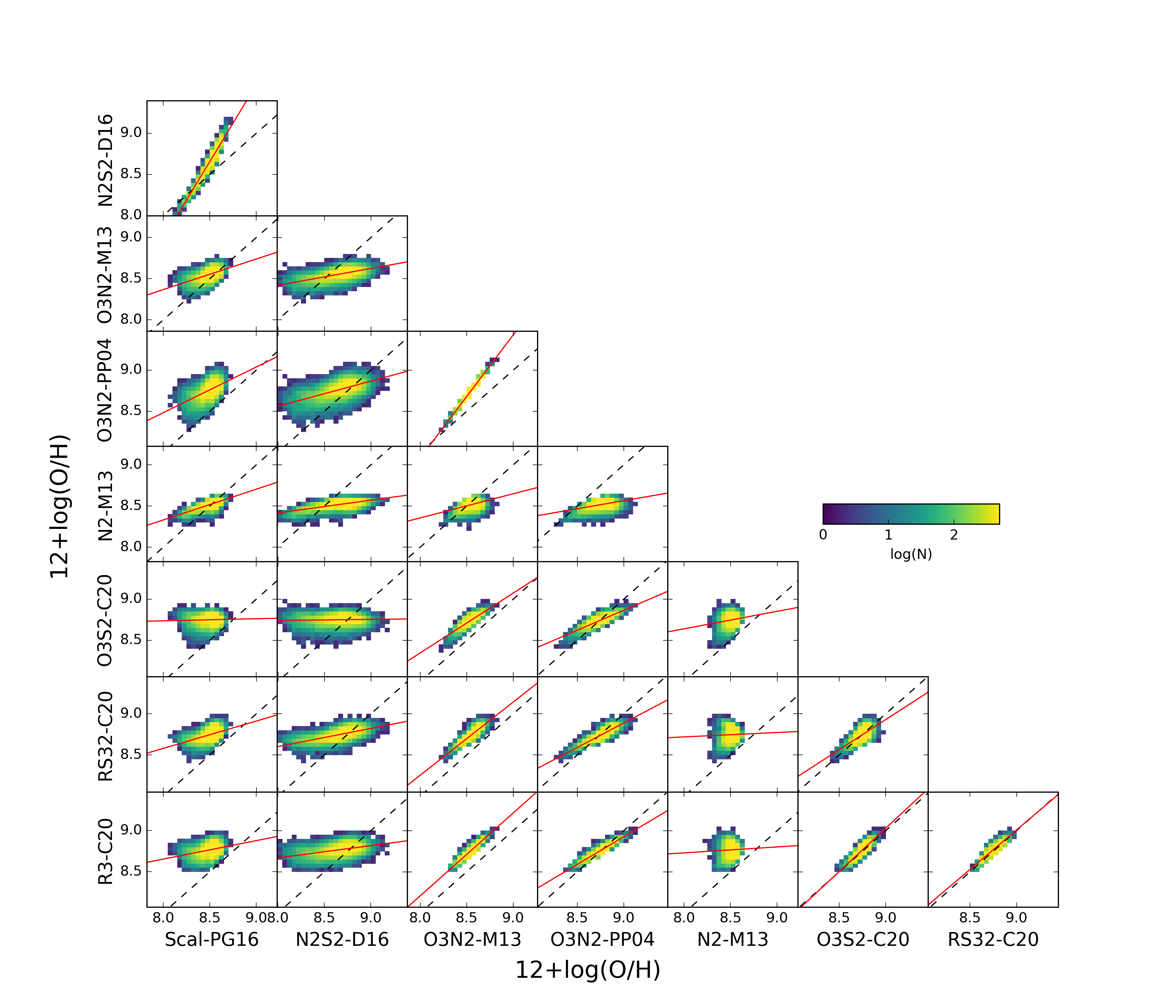}
    \caption{A comparison of 12+log(O/H) metallicities across eight different prescriptions (listed in Table \ref{tab:metal_prescriptions}). The corresponding 1-to-1 line (dashed) is shown in each panel. A linear fit is given for each (red line), with fit parameters listed in Table \ref{tab:metal_conversions}. Systematic offsets of 0.2 dex and scatter of 0.2 dex is apparent between prescriptions. Note that the O3N2-PP04 and O3N2-M31 rely on exactly the same line ratios and hence show a perfect correlation.  Our favoured prescription is the \citet{Pilyugin2016} Scal, which uses 3 different line ratios (and implicitly the Balmer decrement) to remove degeneracies due to ionization parameter variations, and shows a strong correlation with N2S2-D16.   }
    \label{fig:metallicity_corner_plot}
\end{figure*}

We demonstrate these differences in Figure \ref{fig:metallicity_corner_plot}, where we apply eight different prescriptions from the literature (Table \ref{tab:metal_prescriptions}) to our 23,244 \hii\ regions  
and compare the resulting metallicity measurements. We note that, as the wavelength coverage of our galaxies by MUSE excludes emission lines below 4800~\AA, some of the standard metallicity prescriptions using lines such as the \oii $\lambda 3727$ doublet cannot be applied here \citep[e.g.][]{Kobulnicky2004, PT05, Kewley2019}. 
In a pair-wise comparison, we compute a linear conversion between prescriptions, and tabulate the fits (shown in red in Figure \ref{fig:metallicity_corner_plot}) in Table \ref{tab:metal_conversions}. Note that the number of \hii\ regions in each panel differs slightly depending upon the detection ($S/N > 5$) of the lines involved. Most values show a positive correlation, though the strength of the correlation and the scatter between measurements vary wildly, with offsets of up to 0.2 dex and scatter of up to 0.2 dex apparent. This should serve as a warning that when comparing metallicity measurements in the literature, it is important to ensure consistent prescriptions are applied. The conversions we provide are only applicable over the metallicity range covered by our sample, defined as the 5--95 percentiles of the distribution. In some cases (e.g.~N2-M13 versus R3-C20) no clear correlation between the prescriptions is observed over the narrow metallicity range covered.

\begin{table}
\caption{Metallicity prescriptions considered in Figure \ref{fig:metallicity_corner_plot}    }
\label{tab:metal_prescriptions}
\centering
\begin{tabular}{lrrr}
\hline \hline
Abreviation & Lines used & Reference\\
\hline
\hline
Scal-PG16 & \hb, \oiii, \nii, \sii & \cite{Pilyugin2016} \\
O3N2-M13 & \ha, \oiii, \ha, \nii & \cite{Marino2013} \\
O3N2-PP04 & \hb, \oiii, \ha, \nii & \cite{PP04} \\
N2-M13 &  \ha, \nii & \cite{Marino2013} \\
N2S2-D16 & \ha, \nii, \sii & \cite{Dopita2016} \\
O3S2-C20 & \hb, \oiii, \ha, \sii & \cite{curti2020} \\
RS32-C20 & \hb, \oiii, \ha, \sii & \cite{curti2020}  \\
R3-C20 & \hb, \oiii  & \cite{curti2020} \\
\hline
\end{tabular}
\end{table}

As described in \cite{Kreckel2019}, we favour the S calibration (Scal) prescription defined in \cite{Pilyugin2016}, hereafter Scal-PG16, and include these calculated metallicities in our value-added catalogue. The Scal-PG16 prescription was empirically calibrated against a sample of 313 \hii\ regions where direct auroral line detections provided measurements of the electron temperature, and hence more robust determination of 12+log(O/H). As it relies on a larger number of emission lines than other prescriptions (Table~\ref{tab:metal_prescriptions}), it is less biased by ionization parameter variations, which can cause line ratio variations and results in degeneracies in the metallicity determination when only one or two line ratios are considered for the prescription \citep{Kewley2002}. However, note that for the range of metallicities encountered in our sample the calibration is based only on a small fraction of \hii\ regions. 

The Scal-PG16 prescription relies on three standard diagnostic line ratios: 
\begin{equation}
\begin{array}{l}
{\rm N}_2  = ({\rm \nii \lambda 6548+ \lambda 6584}) /{{\rm H}\beta },  \\
{\rm S}_2  = ({\rm \sii \lambda 6717+ \lambda 6731}) /{{\rm H}\beta },  \\
{\rm R}_3  = ({{\rm \oiii} \lambda 4959+ \lambda 5007}) /{{\rm H}\beta },
\end{array}
\end{equation}
where attenuation corrected line fluxes are used (and therefore implicitly includes the ratio of Balmer lines). 
It is defined separately over the upper and lower branches in log~${\rm N}_{2}$. The upper branch
(log~${\rm N}_{2} \ge -0.6$) is calculated as
\begin{eqnarray}
\footnotesize
       \begin{array}{lll}
     {\rm 12+log(O/H)}  & = & \rm  8.424 + 0.030 \, \log (R_{3}/S_{2}) + 0.751 \, \log N_{2}   \\  
                          & + & \rm (-0.349 + 0.182 \, \log (R_{3}/S_{2}) + 0.508 \log N_{2})   \\ 
                          & \rm \times & \log S_{2}   \\ 
     \end{array}
\label{equation:ohsu}
\end{eqnarray}
and the lower branch
(log~$N_{2} < -0.6$) is calculated as
\begin{eqnarray}
\footnotesize
       \begin{array}{lll}
     {\rm 12+log(O/H)}  & = &  \rm 8.072 + 0.789 \, \log (R_{3}/S_{2}) + 0.726 \, \log N_{2}   \\  
                          & + & \rm  (1.069 - 0.170 \, \log (R_{3}/S_{2}) + 0.022 \log N_{2})    \\ 
                          & \times & \rm \log S_{2}   \\ 
     \end{array}
\label{equation:ohsl}
\end{eqnarray}

\begin{figure}
    \centering
    \includegraphics[width=3in]{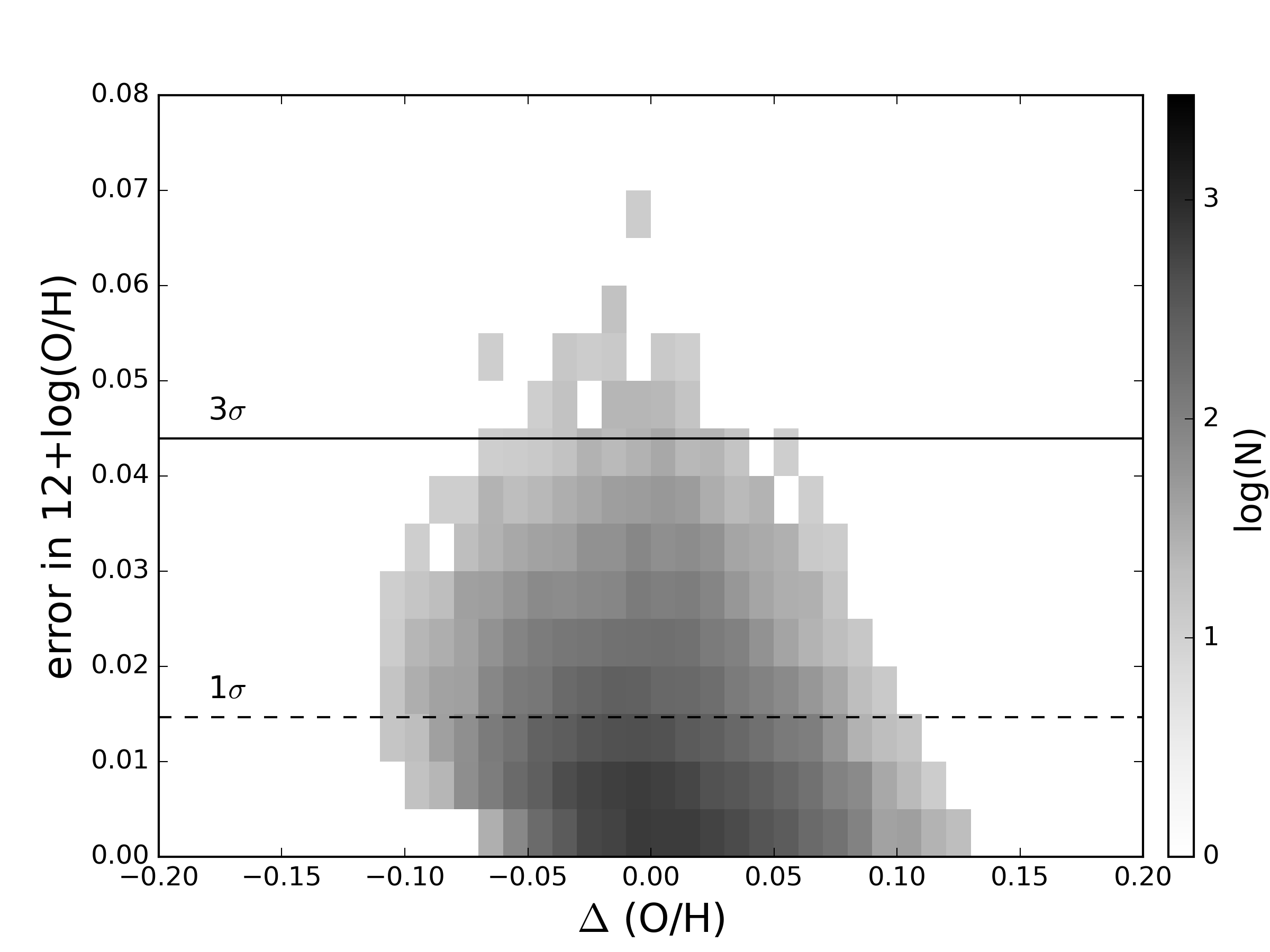}
    \caption{Distribution of errors in 12+log(O/H) as a function of offset from the radial gradient in each galaxy ($\Delta$(O/H)) based on the Scal-PG16
    prescription. These errors are based on Monte Carlo resampling of the associated line flux uncertainties, and do not take into account systematic uncertainties related to the calibration of the metallicity prescription.  Typical 1$\sigma$ uncertainties are 0.01 dex.  }
    \label{fig:scal_errs}
\end{figure}

The Scal-PG16 prescription is highly correlated with the \cite{Dopita2016} N2S2-D16 prescription, which is calibrated against photoionization models but has similarly been designed to minimise degeneracies with ionization parameter.  We calculate the uncertainties associated with the Scal-PG16 metallicity by Monte Carlo error propagation of the emission line flux errors, with 1000 samples used to determine the 1$\sigma$ distribution corresponding to each measured metallicity. Our \hii\ regions cover a range of metallicities from 8.1 $<$ 12+log(O/H) $<$ 8.7 with typical statistical uncertainties of 0.01 dex (Figure \ref{fig:scal_errs}). 

In an independently constructed catalogue, also identified using the HIIphot package, metallicity measurements from 8,914 \hii\ regions in eight of these galaxies have been published previously \citep{Kreckel2019}. Line flux measurements were made using a different data reduction pipeline and different analysis approach, such that the physical extent of the regions may differ and the fitting of the underlying stellar continuum has changed. Comparing our new catalogue with the previous one, we find that $\sim$7000 \hii\ regions cross-match within 1\arcsec, and for these objects the median difference in metallicity is negligible (0.0003 dex) and the standard deviation is small ($<$ 0.02 dex).  

\begin{table*}
\caption{Linear fit parameters for converting metallicities, see red lines in Figure \ref{fig:metallicity_corner_plot}. The range (5--95 percentiles) over which this conversion holds is shown in square brackets. For each conversion we provide in parenthesis the following quantities: intercept, slope, scatter about the conversion.    }
\label{tab:metal_conversions}
\centering
\begin{tabular}{p{1.5cm} | rrrrrrr}
\hline \hline
N2S2-D16 \newline [8.3, 8.9] & (-7.2, 1.9, 0.06)  &  &  &  &  &  \\ 
 O3N2-M13 \newline [8.4, 8.6] & (5.4, 0.4, 0.05)  & (6.8, 0.2, 0.05)  &  &  &  &  &  \\ 
 O3N2-PP04 \newline [8.6, 8.9] & (4.0, 0.6, 0.08)  & (6.1, 0.3, 0.08)  & (-4.0, 1.5, 0.00)  &  &  &  &  \\ 
 N2-M13 \newline [8.4, 8.6] & (5.3, 0.4, 0.03)  & (7.2, 0.2, 0.04)  & (6.0, 0.3, 0.04)  & (6.8, 0.2, 0.04)  &  &  &  \\ 
 O3S2-C20 \newline [8.6, 8.8] & (8.5, 0.0, 0.06)  & (8.6, 0.0, 0.06)  & (2.5, 0.7, 0.03)  & (4.5, 0.5, 0.03)  & (6.9, 0.2, 0.06)  &  &  \\ 
 RS32-C20 \newline [8.5, 8.9] & (5.9, 0.3, 0.06)  & (6.8, 0.2, 0.05)  & (1.1, 0.9, 0.03)  & (3.5, 0.6, 0.03)  & (8.3, 0.1, 0.07)  & (2.3, 0.7, 0.05)  &  \\ 
 R3-C20 \newline [8.7, 8.9] & (6.9, 0.2, 0.06)  & (7.5, 0.1, 0.06)  & (0.2, 1.0, 0.02)  & (2.9, 0.7, 0.02)  & (8.2, 0.1, 0.07)  & (-0.4, 1.1, 0.04)  & (0.4, 1.0, 0.03)  \\ 
 \hline 
 & Scal-PG16 & N2S2-D16 & O3N2-M13 & O3N2-PP04 & N2-M13 & O3S2-C20 & RS32-C20 \\ 
 & [8.3, 8.6] &  [8.3, 8.9] &  [8.4, 8.6] &  [8.6, 8.9] &  [8.4, 8.6] &  [8.6, 8.8] &  [8.5, 8.9] \\ 
  \hline
  \hline
\end{tabular}
\end{table*}

\subsection{Ionization Parameter}

In addition to the actual metal abundance, another parameter of the ISM that regulates line emissivity is the ionization parameter, the ratio of the density of ionizing photons to the number density of hydrogen atoms in the gas ($U$). Empirically, this is typically parametrized by line ratios of different ions arising from the same element. Based on theoretical photoionization models \citep{Kewley2002}, \oiii/\oii\ is sensitive to changes in $U$ but has a strong secondary dependence on metallicity. \siii/\sii\ shows very little dependence on metallicity but is less commonly used to trace $U$ due to the difficulty in observing the red \siii $\lambda\lambda$ 9068,9532 lines. MUSE now gives us coverage of the \siii $\lambda$9068 emission line, enabling us to explore ionization parameter variations more directly using the \siii/\sii\ line ratio.  

However, while the \siii/\sii\ line ratio appears to be a robust tracer of $U$ in theoretical models, comparisons have uncovered large offsets between model predictions for the ratio and empirical results \citep{Mingozzi2020}. We apply the prescription of \cite{Diaz1991} to determine $U$ from this diagnostic ratio as 

\begin{eqnarray}
\footnotesize
       \begin{array}{lll}
    \log (U) & = & -1.684 \log(\mathrm{\siii} \lambda\lambda 9068, 9532/\mathrm{\sii} \lambda\lambda 6716,6730) \\
    & - &  2.986. 
    \end{array}
\end{eqnarray}

Here we measure only the shorter wavelength \siii$\lambda$9068 emission line but assume a ratio of \siii $\lambda$9532 = $\rm 2.5 \times \siii \lambda$9068 fixed by atomic physics \citep{Osterbrock2006, Tayal2019}. We require a S/N $>$ 5 in all lines, and are able to compute $U$ for 20,781 objects (66.2\% of the sample), nearly all of which (20,083 objects; 97\% of objects with measured $U$) are classified as \hii\ regions (see Section \ref{sec:bpt}).

\begin{figure*}
    \centering
    \includegraphics[width=7in]{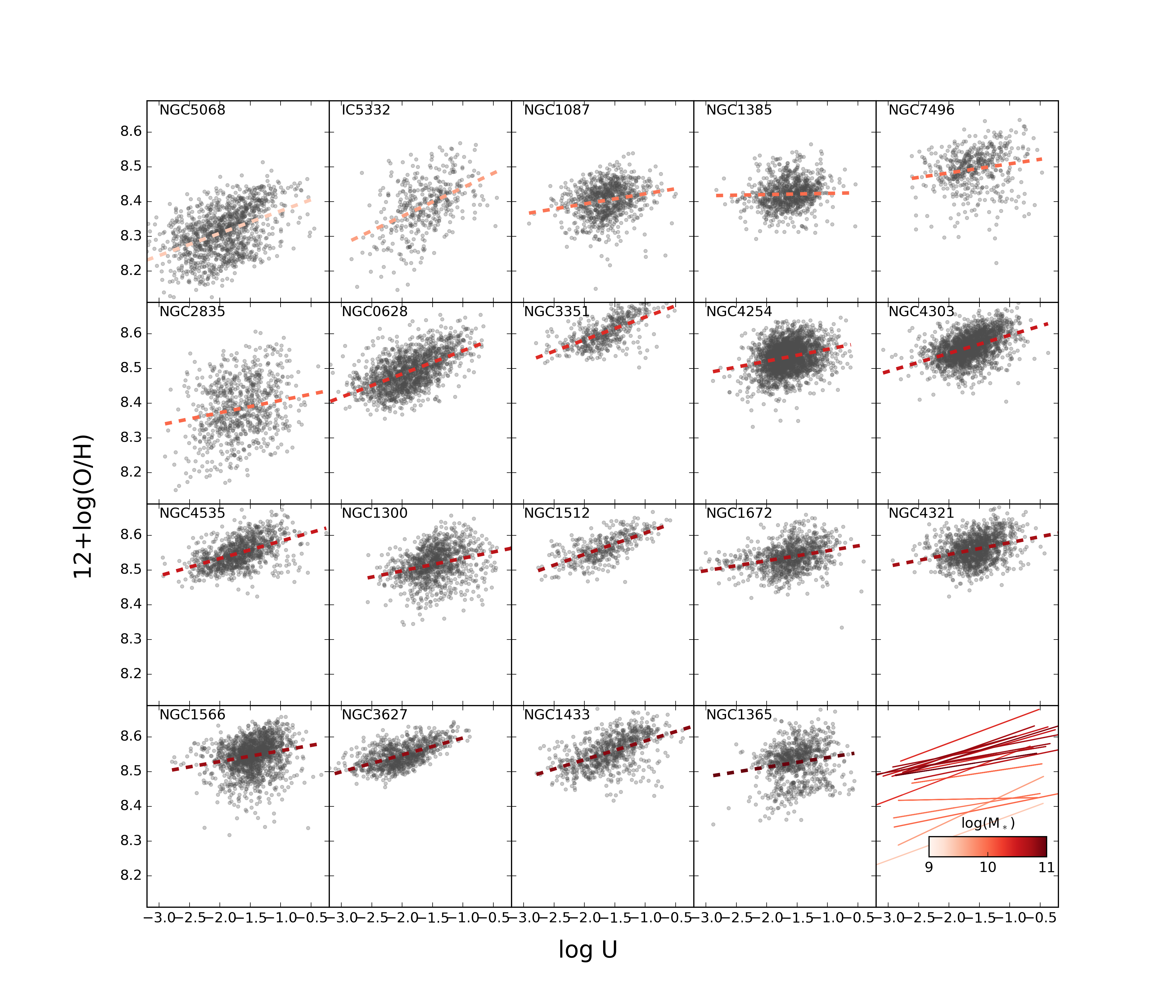}
    \caption{Metallicity (12+log(O/H)) as a function of ionization parameter (U) for \hii\ regions within each galaxy. Galaxies are sorted by stellar mass from low (top left) to high (bottom right). Linear fits are performed in each galaxy, demonstrating the prevalence of positive correlations across the sample. Apart from the well-established mass-metallicity relation, no obvious trends are seen as a function of total galaxy stellar mass (red colour scale, bottom right). }
    \label{fig:U_vs_metal}
\end{figure*}

In Figure \ref{fig:U_vs_metal} we see that our \hii\ regions cover a range of ionization parameters from $-3.0 < \log U < -0.5$, with no systematic differences by galaxy stellar mass. Typical uncertainties are 0.04 dex. Apparent within each galaxy is a positive correlation between $U$ and 12+log(O/H), first reported in luminous star-forming galaxies \citep{Dopita2014}, and recently identified as a robust trend across \hii\ region samples \citep{Kreckel2019, Grasha2022}. This is contrary to theoretical predictions by \citet{Massey2005}, and is also not well reflected in photoionization models \citep{Ji2022}, indicating a need for additional model development. No clear correlation of $U$ with stellar mass of the galaxy due to the scatter within each galaxy and the differing slopes of $U$ versus (O/H) found between galaxies.

\subsection{Nebulae Environments}

The nebulae we identify do not exist in isolation, but rather are part of the larger scale structure of our galaxies. Therefore we also include in our catalogue parameters that trace the different galactic environments in which they occur. 

\begin{figure*}
    \centering
    \includegraphics[width=3in]{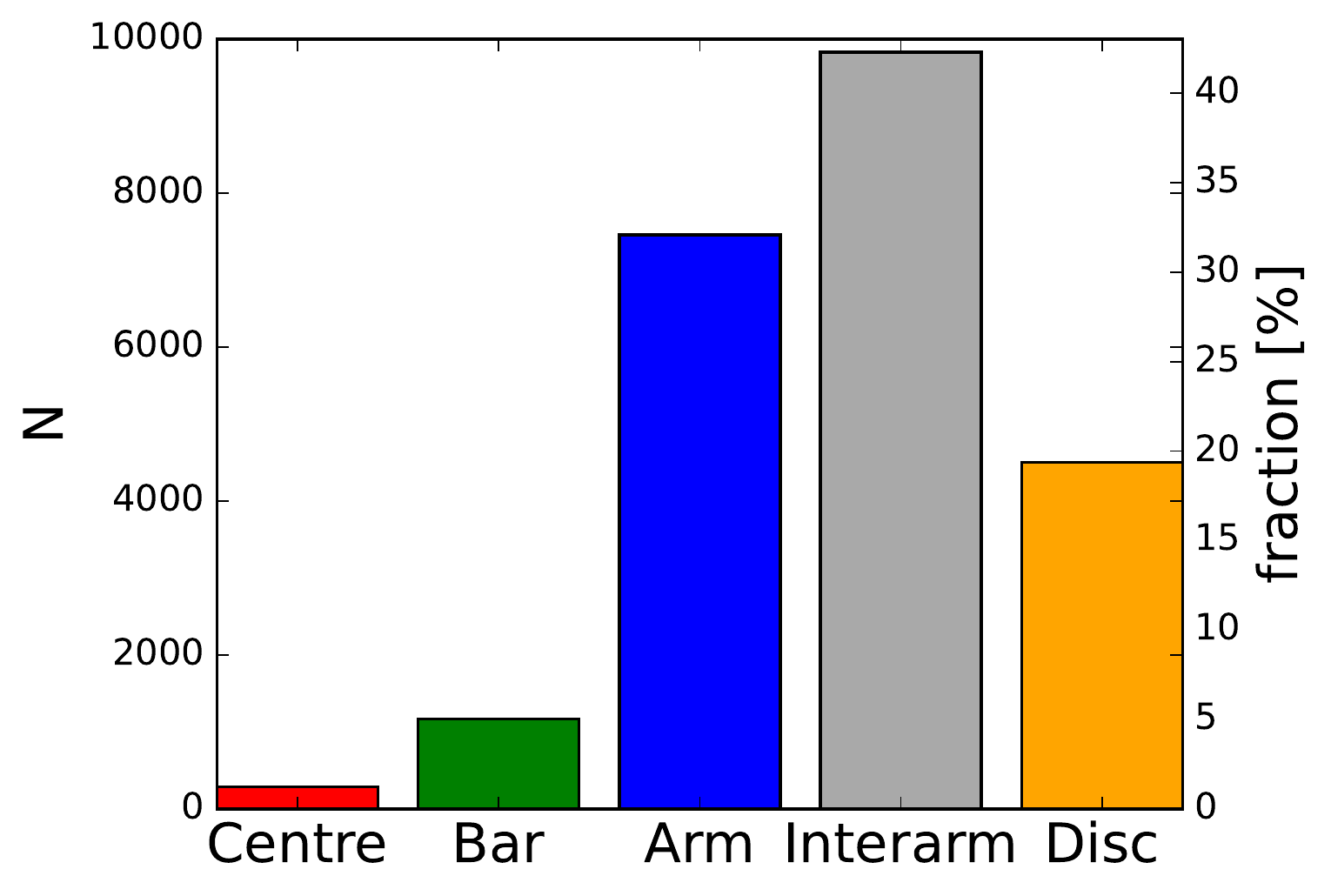}
    \includegraphics[width=3in]{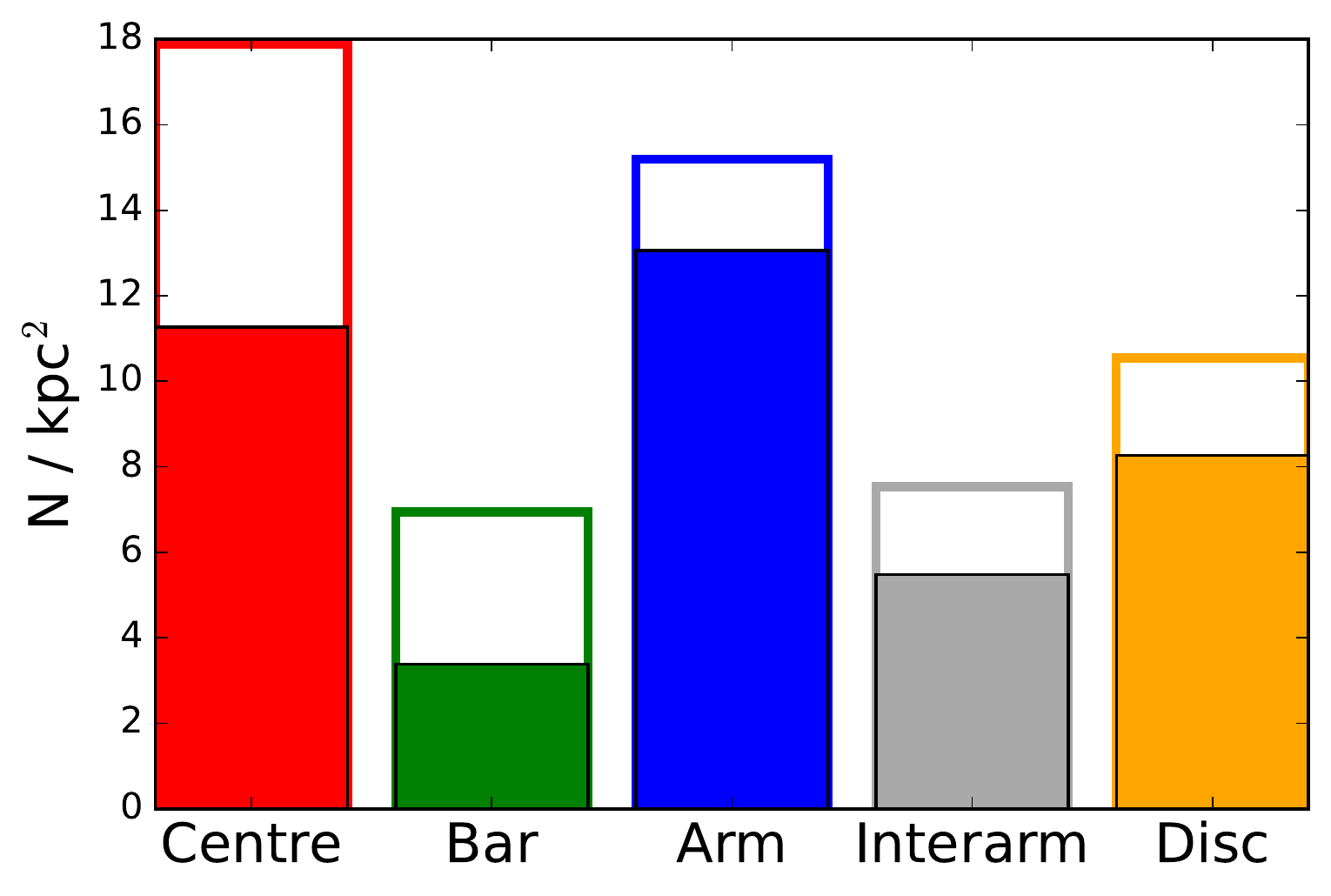}
    \caption{Distribution of \hii\ region detections across different galactic environments, both in absolute numbers (left) and in surface density (right). Unfilled bars (right) indicate the environmental distribution of all nebulae. Interarm regions contain the largest number of \hii\ regions ($\sim$40\% of the sample) but cover a large area of our fields and as such reflect a relatively low surface density.  In contrast, we identify only a small number ($\sim$ 300) of \hii\ regions in the galaxy centres, but this corresponds to a high surface density.  Nebulae not classified as \hii\ regions are over-represented in centre and bar environments, where shock excitation is more likely to contribute to the gas ionization.  }
    \label{fig:envs_bar}
\end{figure*}

The galaxies in our sample show visible structures (centre, bar, spiral arm, interarm, disc) that may reflect differences in dynamical conditions and star formation histories. To define the nebular environments, we use the stellar morphological masks that have recently been identified systematically in \cite{Querejeta2021} based on \textit{Spitzer} 3.6 $\mu$m images. 
We locate each of our nebulae with respect to the simplified environments defined in that paper, as summarised in Table \ref{tab:env_flags}. In Figure \ref{fig:envs_bar} we show both the absolute number of \hii\ regions within each environment (left) as well as the surface density of all objects in each environment (right). 

In absolute numbers, most of our \hii\ regions ($\sim$40\%) are located in interarm regions, however this environment also makes up the largest area in our fields and so correspondingly there is a relatively low number density of nebulae in this environment. In contrast, we identify only a small number ($\sim$ 300, $\sim$2\%) of \hii\ regions in the galaxy centres, but this corresponds to a high number density. 
We find the highest \hii\ region number density within spiral arm environments, and the lowest \hii\ region number density within bar environments. This reflects the fact that star formation is typically concentrated into spiral arms, and bar environments (excluding bar ends) potentially suppress star formation via bar-driven dynamics \citep{James2009}. 
Looking at the full object catalogue, including objects that do not meet our \hii\ region classification criteria, the number density of objects in the bar approximately doubles. The centre environments also contain a high number density of objects that are not classified as \hii\ regions. This is expected if AGN and dynamical shocks are significantly contributing to the gas ionization in these environments.

\begin{table*}
\begin{center}
\caption[h!]{Our Environmental flags, based on simplified assignments from \cite{Querejeta2021}.\label{tab:env_flags}}
\begin{tabular}{cccc}
\hline\hline
Label & Querejeta2021 Environment &  & Environment \\
   \hline
1 & Centre  & $\longrightarrow$  & Centre \\
2 & Bar (excluding bar ends) & \multirow{2}{*}{$\Big\}$} &  \multirow{2}{*}{Bar}  \\
3 & Bar ends  & &  \\
5 & Spiral arms inside interbar ($R_{\rm gal} < R_{\rm bar}$) & \multirow{2}{*}{$\Big\}$} &  \multirow{2}{*}{Arm}  \\
6 & Spiral arms ($R_{\rm gal} > R_{\rm bar}$) & & \\ 
4 & Interbar ($R_{\rm gal} < R_{\rm bar}$) & \multirow{3}{*}{$\Bigg\}$} &  \multirow{3}{*}{Interarm}  \\
7 & Interarm ($R_{\rm gal} > R_{\rm bar}$)  & & \\ 
8 & Outer disc ($R_{\rm gal} >$ spiral arm ends)   & & \\ 
9 & Disc ($R_{\rm gal} > R_{\rm bar}$) in galaxies without spiral masks & $\longrightarrow$ & Disc \\
  \hline
\end{tabular}
\label{table:masknotation}
\end{center}
\end{table*}

\section{Results}
\label{sec:results}
While the total stellar mass and integrated star formation rate are global properties that correlate with the integrated properties of the ISM \citep{Tremonti2004, Sanchez2020, Pessa2021}, significant secondary correlations are identified within galaxies once their nebular emission is spatially resolved. The most well known of these is the metallicity gradient \citep[for recent reviews see][]{Maiolino2019, Kewley2019}. However, gas-phase metallicity has also been shown to spatially vary with both stellar mass surface density \citep{BarreraBallesteros2016} and gas-mass surface density \citep{BarreraBallesteros2018}. Once this radial gradient is removed, higher-order variations of the metallicity are seen, along spiral arms \citep{Sanchez-Menguiano2016, Ho2017} and across discs \citep{Kreckel2019}. We revisit here that work of \citet{Kreckel2019}, an earlier analysis of spatial metallicity variations in a subset (8 out of 19) of the PHANGS--MUSE galaxies using data from an earlier version of our reduction and analysis pipeline (see also \citealt{Emsellem2022}). 

For this analysis, we consider our \hii\ region sample to be those objects that are fully contained in the field of view (`flag\_edge' = 0), are consistent with photoionization (`BPT\_NII' = `BPT\_SII' = 0 and `BPT\_OI' $<=$ 0), and where we have high confidence in our metallicity measurement (`met\_scal\_err' $<$ 0.04 dex; see Figure \ref{fig:scal_errs} and Section \ref{sec:metallicities}). By including a cut on metallicities with large uncertainties, we exclude only $\sim$1000 regions that have an average uncertainty of 0.06 dex.
We also exclude six regions with metallicity values 12+log(O/H) $<$ 8.0 as we believe these are spurious and they significantly bias our statistics (see Section \ref{sec:outliers}). For the following sections, our catalogue consists of 22,318 \hii\ regions, with between 477--2556 \hii\ regions per galaxy. 

\subsection{Radial Gradients}
\label{sec:radial_gradients}

As described above, it has been clearly established that galaxies in the local universe systematically have a lower metallicity with increasing radius (e.g. \citealt{Moustakas2010, Pilyugin2014, Sanchez2014}). These radial trends neglect prominent morphological features (spiral arms, stellar bars), though do appear to show variations in the inner and outer parts of galaxies \citep{Sanchez-Menguiano2018}. The metallicity gradients vary with stellar mass \citep{Sanchez2014} and with radius at a given mass \citep[e.g.][]{Boardman2021}, and are thought to chart the typical inside-out growth of most disc galaxies. 

\begin{figure*}
    \centering
    \includegraphics[width=7.5in]{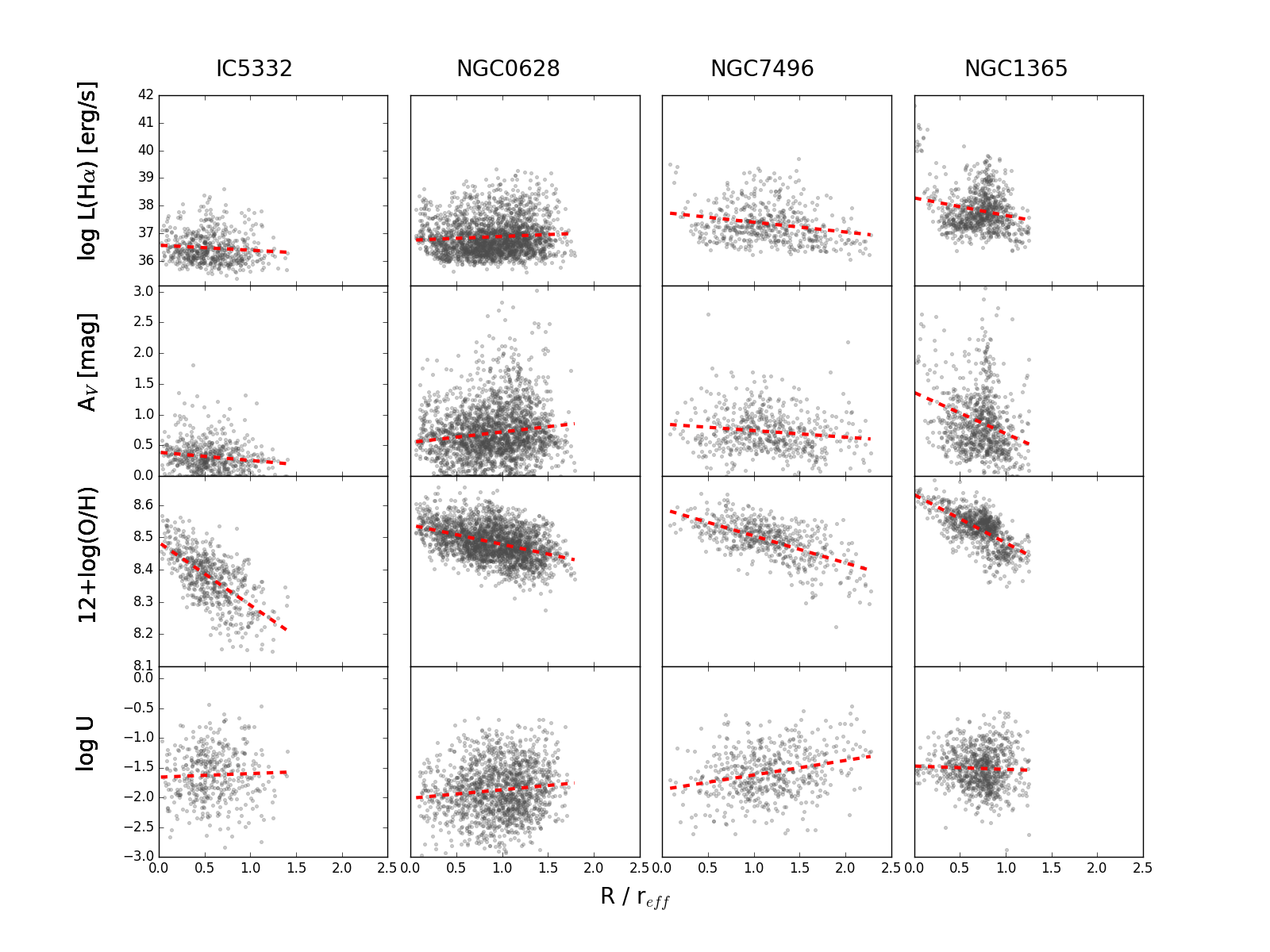}
    \caption{Radial gradients for four representative galaxies in key ISM diagnostics (from top to bottom): H$\alpha$ luminosity (L(H$\alpha$)), extinction derived from the Balmer decrement (A$_V$), metallicity (12+log(O/H)), and ionization parameter (U). For each galaxy we show the radial trends normalized to \reff. These four galaxies represent low stellar mass (IC 5332) and high stellar mass (NGC 1365) galaxies, and systems with a regular spiral pattern and no bar (NGC 0628) or strong bar and widely separated arms (NGC 7496). Most trends are flat or mildly negative, except for the metallicity gradients which show the strongest negative trends. }
    \label{fig:radial_gradients}
\end{figure*}

In Figure \ref{fig:radial_gradients} we show radial trends for a few of the key ISM diagnostics available in our \hii\ region catalogue, for a representative sample of four galaxies. These are trends in the \ha\ luminosity (L(H$\alpha$)) of individual \hii\ regions, attenuation derived from the Balmer decrement (A$_V$), metallicity (12+log(O/H)), and ionization parameter (U). Radial trends for all 19 galaxies are shown in Appendix \ref{appendix:radial}. These four galaxies in Figure \ref{fig:radial_gradients} represent a low stellar mass (IC 5332) and high stellar mass (NGC 1365) galaxy, and systems with a regular spiral pattern and no bar (NGC 0628) or strong bar and widely separated arms (NGC 7496). For each galaxy we show the radial trends scaled to \reff, to normalise the sample, and note that in most cases our coverage is limited to the inner parts of each galaxy ($<$2 \reff). We fit each gradient with a linear relation that neglects the uncertainties, as variations in these properties are expected to reflect local variations in the physical conditions and not uncertainties in the measurements. Representative values at 1 \reff\ are also given in Appendix \ref{appendix:radial}. The individual radial gradients we determine for our PHANGS--MUSE sample fall largely within the range found from the much larger IFS surveys such as CALIFA \citep{EspinosaPonce2022} and MaNGA \citep{BarreraBallesteros2022}. Beyond these radial trends, however, we see a large scatter of these properties within each galaxy due to the higher spatial resolution of our sample.

In general, galaxies show relatively flat or slightly negative slopes in L(\ha).  In many cases central starbursts are also apparent, covering scales of a few 100 pc and exhibiting high H$\alpha$ luminosities (e.g.~in NGC 1365), though we note that these measurements  could be biased by our ability to deblend neighbouring \hii\ regions given our angular resolution ($\sim$70 pc). 
There is also a floor imposed by our region identification methods and sensitivity limits, visible as a relatively sharp lower bound to our region luminosities, and tracking the variations in galaxy distances. 
Most (80\%) of the \hii\ regions we detect have L(H$\alpha$) $<$ 10$^{38}$ erg s$^{-1}$. 
About $\sim$800 ($\sim$3\%) of our \hii\ regions might be categorized as `giant \hii\ regions' (L(H$\alpha$) $>$ 10$^{39}$ erg s$^{-1}$). \hii\ region luminosity functions for each galaxy are presented in \cite{Santoro2022}. 

We find flat or slightly negative slopes in A$_V$, with typical values ranging from 0--2 mag of extinction, and a median value of 0.75 mag. We emphasise that these measurements do not represent an unbiased view of dust in the disc, as most of our nebulae are associated with star-forming regions that are expected to be dustier \citep{Calzetti1994, Kreckel2013}. We also potentially miss high attenuation, heavily embedded regions where \hb\ or even \ha\ may not be visible, though the incidence of such obscured nebulae in the local Universe is small \citep{Prescott2007}, and likely even rarer in our low-inclination galaxies. Any such obscured population will be constrained with our upcoming PHANGS--JWST observations.   

As in \cite{Kreckel2019}, we do not see significant radial gradients in the ionization parameter for any of our galaxies. This reflects that the localized ionization state of the gas is sensitive mainly to changes in the ionizing source and local gas density at the cloud interface. The flat radial gradient also increases our confidence in radial trends we uncover in metallicity. Ionization parameter variations can influence diagnostic line ratios and introduce biases depending on the metallicity prescription used. Our preferred metallicitiy prescription, the \cite{Pilyugin2016} S-calibration, is designed to minimize this bias but it is reassuring that we also do not observe any radial trends in ionization parameter. 

\begin{figure*}
    \centering
    \includegraphics[width=7in]{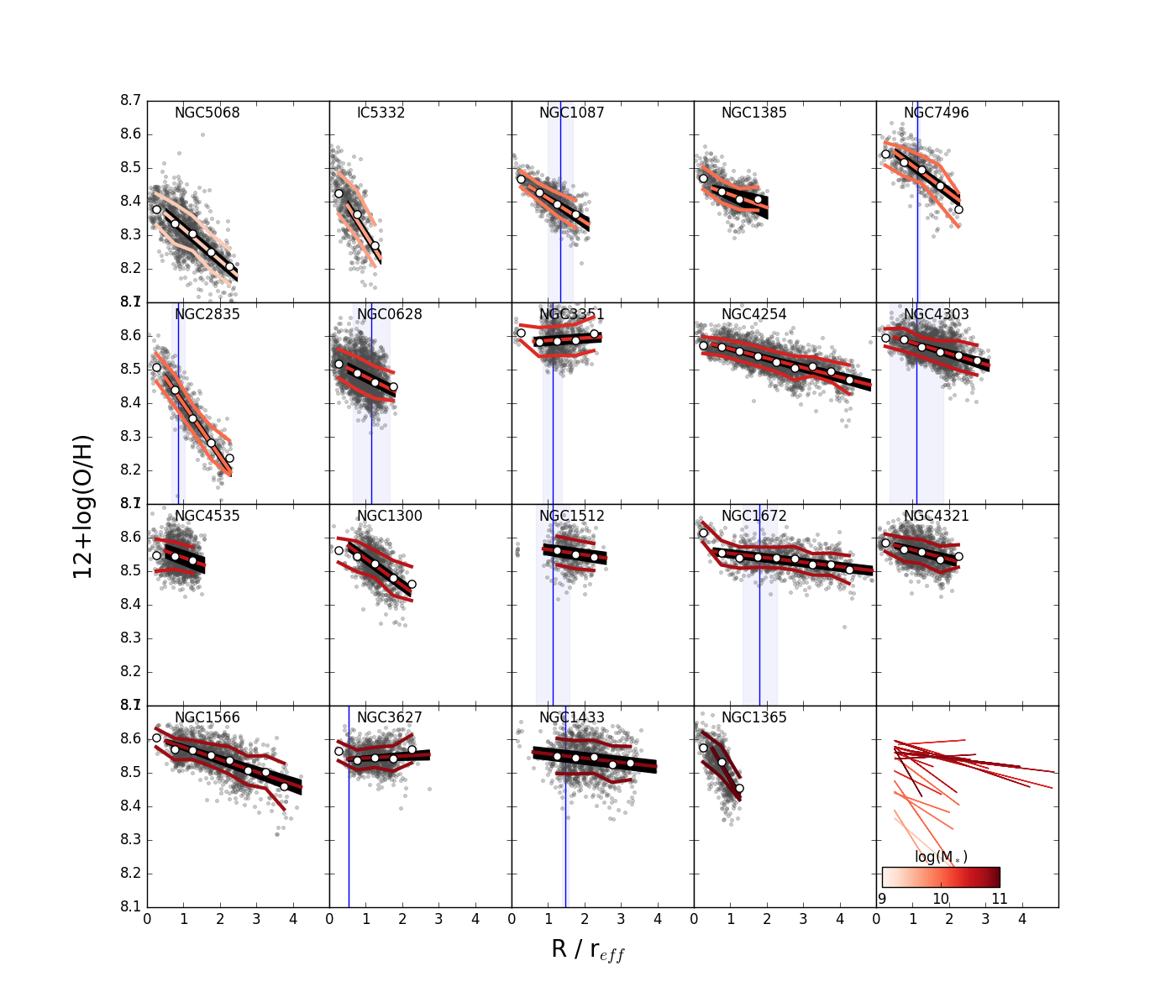}
    \caption{Metallicity 12+log(O/H) radial gradients based on the \hii\ regions within each of the 19 galaxies. Galaxies are sorted by the stellar mass, from low (top left) to high (bottom right). For each galaxy we show the radial trends normalized to \reff. A simple linear fit (central line) shows very good agreement with the median value in 0.5 \reff\  wide bins. For each bin we also track the 1$\sigma$ scatter (outer lines). These radial trends are compared to the locations of co-rotation (blue vertical lines; \citealt{Williams2021}), as measured from the stellar kinematics. The bottom right panel shows the linear trends for all galaxies overplotted, with the colour-scale indicating the total stellar mass.  }
    \label{fig:metal_gradients}
\end{figure*}

Our most pronounced radial trends are apparent in the \hii\ region metallicities (12+log(O/H)), and we show the radial metallicity gradients for all 19 galaxies in Figure \ref{fig:metal_gradients}.  A simple linear fit (solid line) shows very good agreement with the median value in 0.5 \reff\ wide bins. For the radial fit, we exclude the central 0.5 \reff\ as suggested by \citet{Sanchez-Menguiano2018}. We have also not included the uncertainties when performing our linear fit, as we allow for an intrinsic scatter due to physical variations in the gas conditions in excess of our estimated uncertainties ($\sim$0.01 dex; Figure \ref{fig:scal_errs}).  For each bin we also track the 1$\sigma$ scatter (outer lines), and find very good agreement between these binned radial trends and the linear fits, suggesting that to first order a linear fit describes the data well. This finding has been more thoroughly quantified in \cite{Williams2022}. Our galaxies clearly reflect the well-established mass-metallicity relation \citep{Tremonti2004}, with less massive galaxies exhibiting systematically lower metallicities (bottom right panel).  

For 10 of the 19 galaxies, measurements of the corotation radius are available \citep{Williams2021}. This is the location where the gas rotational dynamics and spiral pattern speed are matched, and can only be robustly measured through analysis of the stellar kinematics. Theoretical work has suggested that at this dynamical location, metallicity variations are predicted to be amplified due to the lower relative velocity between the gas and spiral pattern overdensity \citep{Spitoni2019}. We see no obvious change in the metallicity scatter at the locations of corotation, or radially with respect to corotation.  
Radial gradients for all galaxies are reported in Table \ref{tab:metal_grad_scal}. Gradients derived using alternative metallicity prescriptions are also provided in Appendix \ref{appendix:altgrads}. 

\begin{table}
\caption{Linear fit parameters for radial gradients in 12+log(O/H) using the Scal prescription, see Figure \ref{fig:metal_gradients}    }
\label{tab:metal_grad_scal}
\centering
\begin{tabular}{lrrrr}
\hline \hline
Galaxy & intercept & slope [dex/r$_{eff}$] & value at r$_{eff}$ & $\sigma$(O/H)\\
\hline
\hline
IC5332 &  8.475 $\pm$ 0.012 & -0.173 $\pm$ 0.002 & 8.302 & 0.066 \\
NGC0628 &  8.533 $\pm$ 0.014 & -0.054 $\pm$ 0.001 & 8.478 & 0.048 \\
NGC1087 &  8.479 $\pm$ 0.000 & -0.070 $\pm$ 0.010 & 8.409 & 0.032 \\
NGC1300 &  8.617 $\pm$ 0.015 & -0.079 $\pm$ 0.001 & 8.537 & 0.042 \\
NGC1365 &  8.666 $\pm$ 0.004 & -0.188 $\pm$ 0.005 & 8.477 & 0.040 \\
NGC1385 &  8.459 $\pm$ 0.002 & -0.038 $\pm$ 0.014 & 8.421 & 0.033 \\
NGC1433 &  8.569 $\pm$ 0.016 & -0.013 $\pm$ 0.001 & 8.556 & 0.051 \\
NGC1512 &  8.581 $\pm$ 0.014 & -0.016 $\pm$ 0.001 & 8.565 & 0.042 \\
NGC1566 &  8.613 $\pm$ 0.004 & -0.037 $\pm$ 0.004 & 8.576 & 0.037 \\
NGC1672 &  8.566 $\pm$ 0.009 & -0.013 $\pm$ 0.000 & 8.553 & 0.033 \\
NGC2835 &  8.555 $\pm$ 0.006 & -0.157 $\pm$ 0.002 & 8.398 & 0.040 \\
NGC3351 &  8.579 $\pm$ 0.013 & 0.007 $\pm$ 0.001 & 8.587 & 0.044 \\
NGC3627 &  8.538 $\pm$ 0.003 & 0.006 $\pm$ 0.004 & 8.544 & 0.033 \\
NGC4254 &  8.590 $\pm$ 0.004 & -0.028 $\pm$ 0.003 & 8.562 & 0.030 \\
NGC4303 &  8.613 $\pm$ 0.000 & -0.032 $\pm$ 0.006 & 8.580 & 0.034 \\
NGC4321 &  8.592 $\pm$ 0.006 & -0.028 $\pm$ 0.004 & 8.564 & 0.036 \\
NGC4535 &  8.580 $\pm$ 0.014 & -0.040 $\pm$ 0.003 & 8.541 & 0.039 \\
NGC5068 &  8.412 $\pm$ 0.013 & -0.094 $\pm$ 0.001 & 8.318 & 0.054 \\
NGC7496 &  8.588 $\pm$ 0.010 & -0.081 $\pm$ 0.003 & 8.507 & 0.045 \\
\hline
\end{tabular}
\end{table}

\subsection{Correlations with Global Properties}
\label{sec:global_properties_general}
Based on the radial trends, we explore correlations of representative derived properties with global galaxy properties. In particular, we explore trends with total stellar mass, star formation rate (SFR) and gas fraction (calculated as the sum of the H\textsc{i} and H$_2$ gas mass relative to the total gas and stellar mass). These are all global properties that are typically associated with the regulation of galaxy evolution \citep{Genzel2015}. Our galaxies span just over an order of magnitude dynamic range in these key properties. While our sample size is small compared to integral field spectral galaxy surveys like CALIFA \citep{Sanchez2012}, MaNGA \citep{Bundy2015}, or SAMI \citep{Croom2021}, our ability to robustly isolate individual \hii\ regions provides a novel opportunity to cleanly consider trends relating small scale properties to global differences. 

\begin{figure*}
    \centering
    \includegraphics[width=5in]{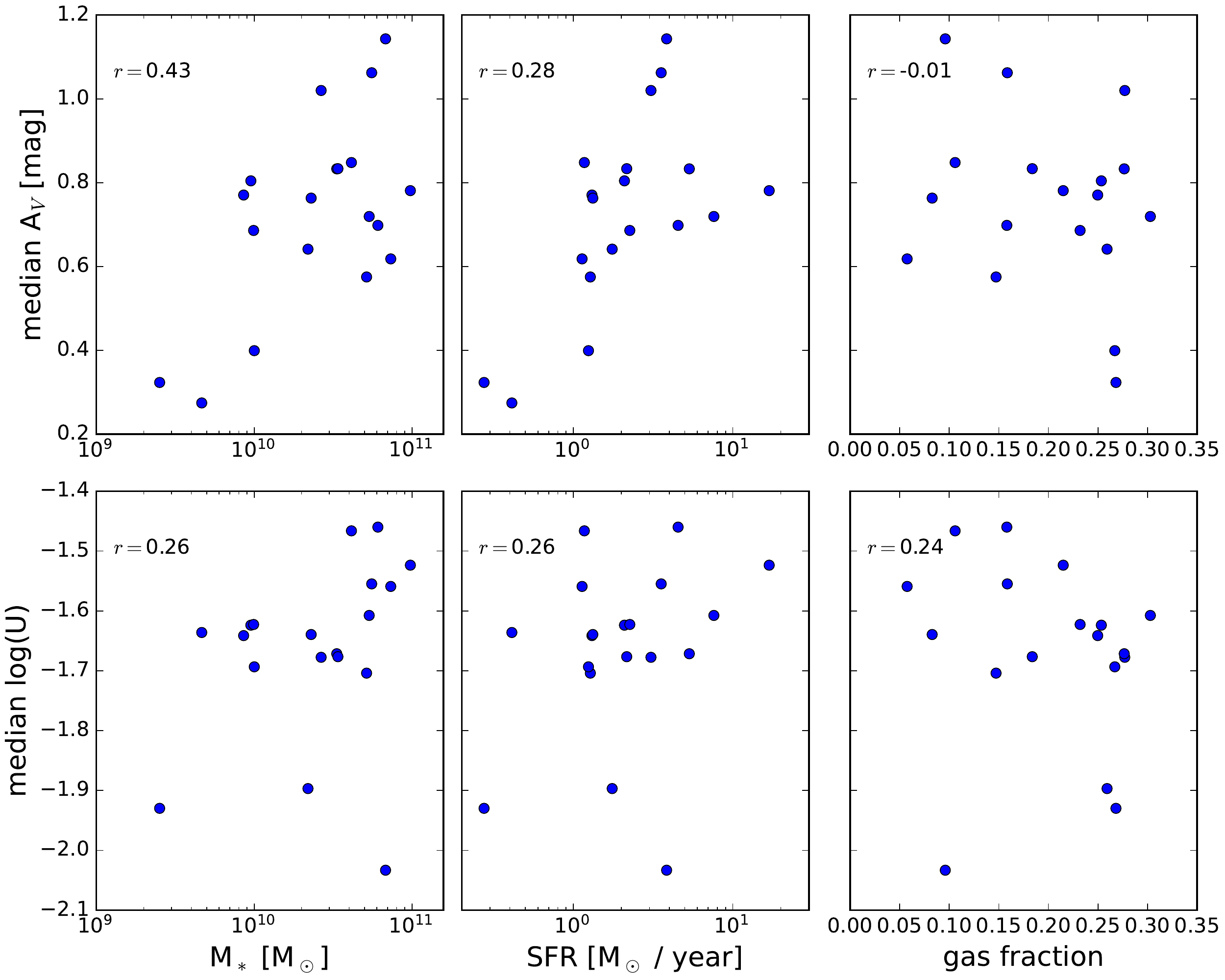}
    \caption{Comparison of median extinction (A$_V$; top) and median ionization parameter ($U$; bottom) as a function of global galaxy properties (stellar mass, star formation rate, gas fraction). In the top left of each plot we show the Pearson correlation coefficient of those quantities, $r$. A$_V$ shows modest correlations with stellar mass and SFR, reflecting an increased gas content (and hence dust content) in these systems. $U$ shows no correlations, reflecting that it is set primarily by local ISM conditions.
    \label{fig:global2}}
\end{figure*}

As both extinction (A$_V$) and ionization parameter (U) show no clear radial trends, we consider the median value measured across the galaxy disk (Figure \ref{fig:global2}). A$_V$ shows modest trends for higher values at higher stellar mass and SFR, consistent with an increased amount of gas (and hence dust) associated with these systems. We see no trend with gas fraction. Ionization parameter shows no trends with global properties, indicating that it is regulated by local physical conditions in the disk. 

\begin{figure*}
    \centering
    \includegraphics[width=5in]{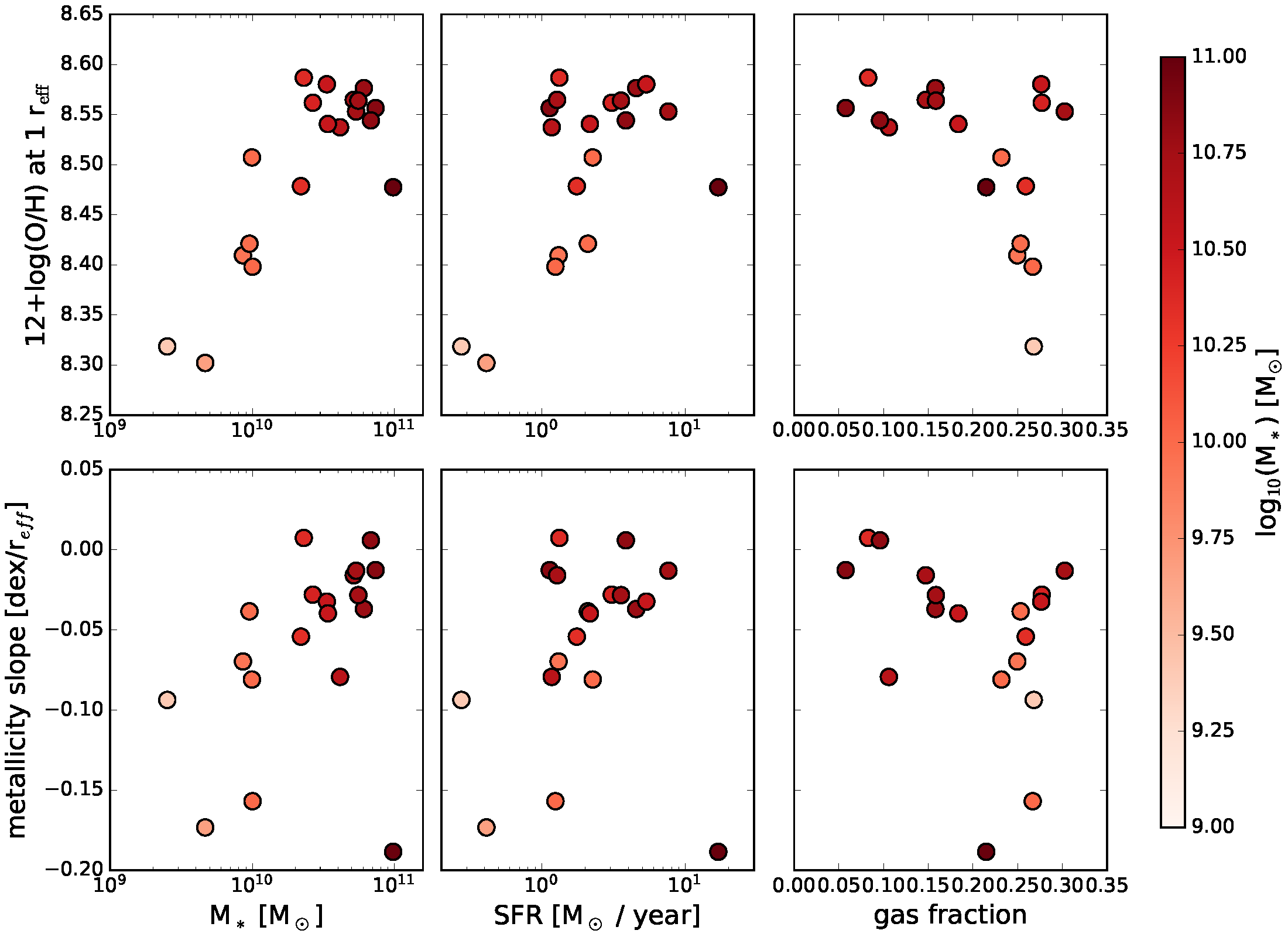}
    \caption{Comparison of parameters derived from the radial metallicity gradients with global galaxy properties (stellar mass, star formation rate, gas fraction).  These include the metallicity measured at 1 \reff (top), and the metallicity slope (bottom). Points are colour coded by the total stellar mass.  }
    \label{fig:global3}
\end{figure*}

Using our radial metallicity gradient fits, we calculate a representative metallicity at 1 \reff\ for each galaxy, and consider global trends in metallicity and metallicity slope (Figure \ref{fig:global3}). As expected, we recover the mass-metallicity relation \citep{Tremonti2004}, with more massive galaxies systematically exhibiting higher metallicities. Secondary dependencies have been reported with SFR \citep{Mannucci2010, Lara-Lopez2010, delosReyes2015}, with galaxies at lower stellar mass exhibiting lower metallicities at fixed SFR. This effect is broadly seen in our small galaxy sample.  From gas-equilibrium models, this trend has been proposed to derive primarily from a decreasing gas fraction corresponding to high SFRs \citep{Peeples2008, Peeples2009, Bothwell2013}, and to some degree this is also reflected in our sample. However, we are generally in agreement with the larger CALIFA sample of \citet{AvarezHurtado2022} in that, once the stellar mass metallcity correlation is removed, the other global properties show no obvious trends.

Trends for steeper metallicity gradients in more massive galaxies are reported in large galaxy surveys \citep{Belfiore2017, Poetrodjojo2018}, but in fact we observe the opposite trend. This could be due to the radial coverage of our sample being limited, or the predominance of bar-dominated systems (these have been observed to exhibit flatter metallicity gradients; \citealt{Zurita2021}). With those caveats, we also see trends for flatter slopes at high SFR and low gas fraction.  

\subsection{Global metallicity variations}
\label{sec:global_metal_variation}

As the linear gradient in metallicity represents the dominant first order trend, we follow the approach developed in \cite{Kreckel2019} to fit and subtract this radial gradient and examine the second order variations in metallicity, $\Delta$(O/H). We further quantify the 1$\sigma$ scatter in $\Delta$(O/H) over the entire galaxy as $\sigma$(O/H) for each galaxy (Table \ref{tab:metal_grad_scal}), to understand whether the second order variations in metallicity are driven by global galaxy properties. 

\begin{figure*}
    \centering
\includegraphics[width=5in]{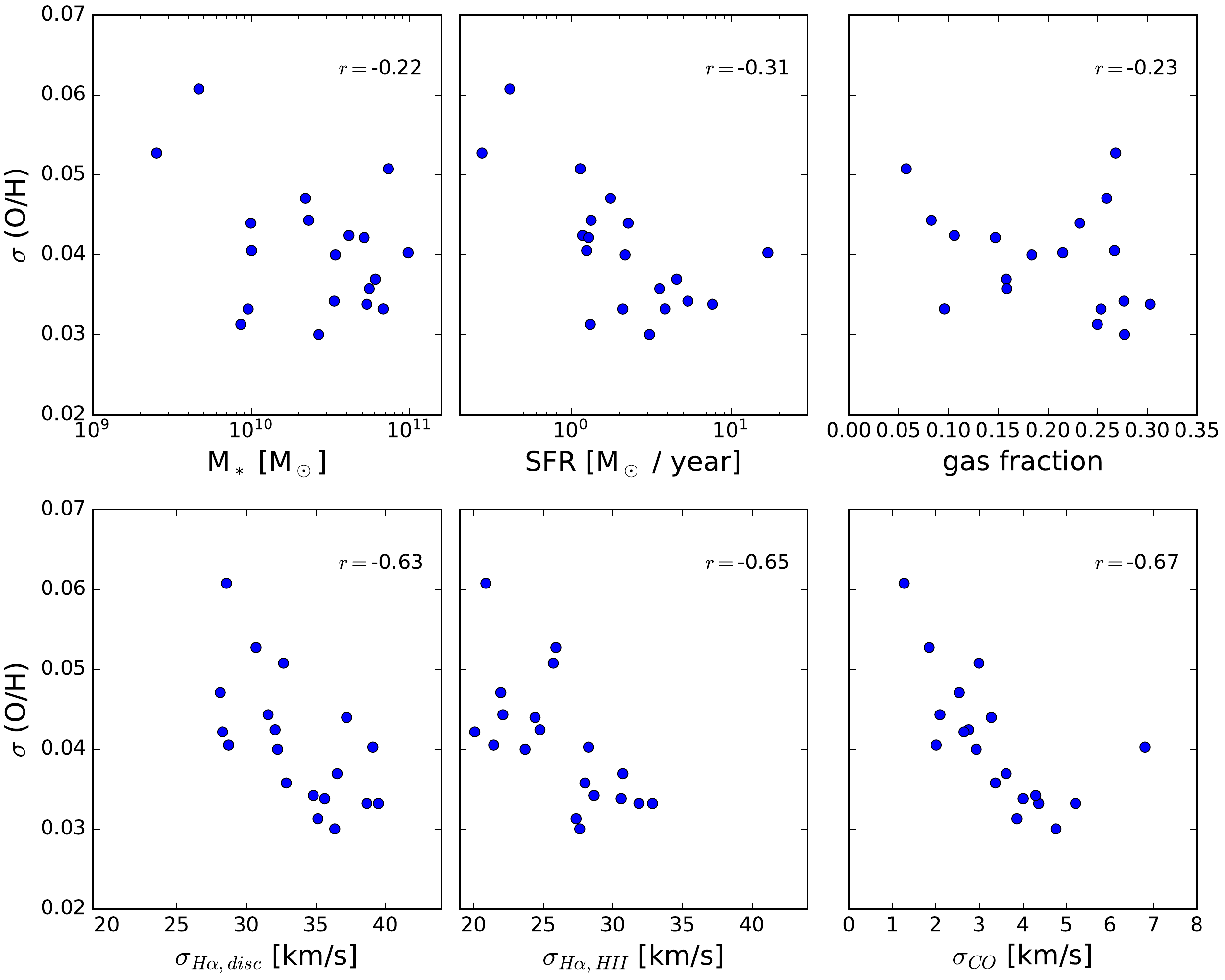}
    \caption{Correlations between the global scatter in metallicity ($\sigma$(O/H)), after removing the first order radial gradient, as a function of galaxy integrated properties. In the top right of each plot we show the Pearson correlation coefficient of those quantities, $r$. The total stellar mass (M$_*$), total star formation rate (SFR), and gas fraction show weak or absent trends. More pronounced  correlations are seen with the ionized gas velocity dispersion, as measured across the full disc ($\sigma_{\rm H\alpha, disc}$) or just the \hii\ regions ($\sigma_{\rm H\alpha, HII}$), as well as with the molecular gas velocity dispersion ($\sigma_{\rm CO}$).  }
    \label{fig:global_scatter}
\end{figure*}

We find $\sigma$(O/H) varies across the galaxy sample, with values ranging from 0.03 -- 0.06 dex. These values do not change significantly ($<$0.005 dex) if we impose a stricter cut on our metallicity uncertainties. In the top panels of Figure \ref{fig:global_scatter} we show how $\sigma$(O/H) correlates with different global galaxy properties, including total stellar mass (M$_*$), total star formation rate (SFR), and gas fraction (calculated as the sum of the H\textsc{i} and H$_2$ gas mass relative to the gas plus stellar mass), all properties which might be expected to regulate mixing in the disc \citep{Krumholz2018}. We find a weak correlation with M$_*$ and SFR, and no correlation with gas fraction.  

In \cite{Kreckel2020}, the mixing scale for metals, as quantified via the two point correlation function, was found to display similar weak correlations with SFR. However, those authors found the most pronounced correlation with the gas velocity dispersion, indicating that the homogeneity of the metal distribution in the gas (and corresponding mixing scale length) was regulated by gas turbulence. To test this, we consider $\sigma$(O/H) as a global measure of this metal distribution homogeneity and compare it with three tracers of the multi-phase gas velocity dispersion. In the bottom panels of Figure \ref{fig:global_scatter}, we show two constraints from the ionized gas: the median ionized gas velocity dispersion, measured across the entire MUSE map ($\sigma_{\rm H\alpha, disc}$) and the median ionized gas velocity dispersion measured only within the \hii\ regions ($\sigma_{\rm H\alpha, HII}$). For both of these we consider only pixels or regions where the H$\alpha$ emission achieves a S/N $>$ 20, to minimize uncertainties introduced by the low spectral resolution of MUSE. We also correct for the instrumental dispersion ($\sim$49 \kms\ at \ha, as reported in \citealt{Bacon2017}).  The disc as a whole shows typically higher dispersions ($\sim$30--35 \kms) compared to the \hii\ regions ($\sim$20--30 \kms), reflecting elevated gas dispersion in the diffuse ionized gas \citep{Moiseev2012, Moiseev2015, DellaBruna2020}. Both show positive correlations with $\sigma$(O/H).  The tightest correlation is seen when considering the molecular gas velocity dispersion ($\sigma_{\rm CO}$), measured from the median value within the `strict' second moment maps \citep{Leroy2021}. These values are significantly smaller ($\sim$2--4 \kms), reflecting the thin mid-plane distribution of this colder and dense ISM component.  
These correlations between the global scatter in metallicities (relative to the radial gradients) and the turbulent state of the ISM (in both the ionized and molecular material) demonstrate convincingly that the ISM dynamics play a critical role in regulating the mixing of metals across galaxy discs. 

\subsection{Local metallicity variations}
\label{sec:local_metal_variation}

\begin{figure*}
    \centering
    \includegraphics[width=7in]{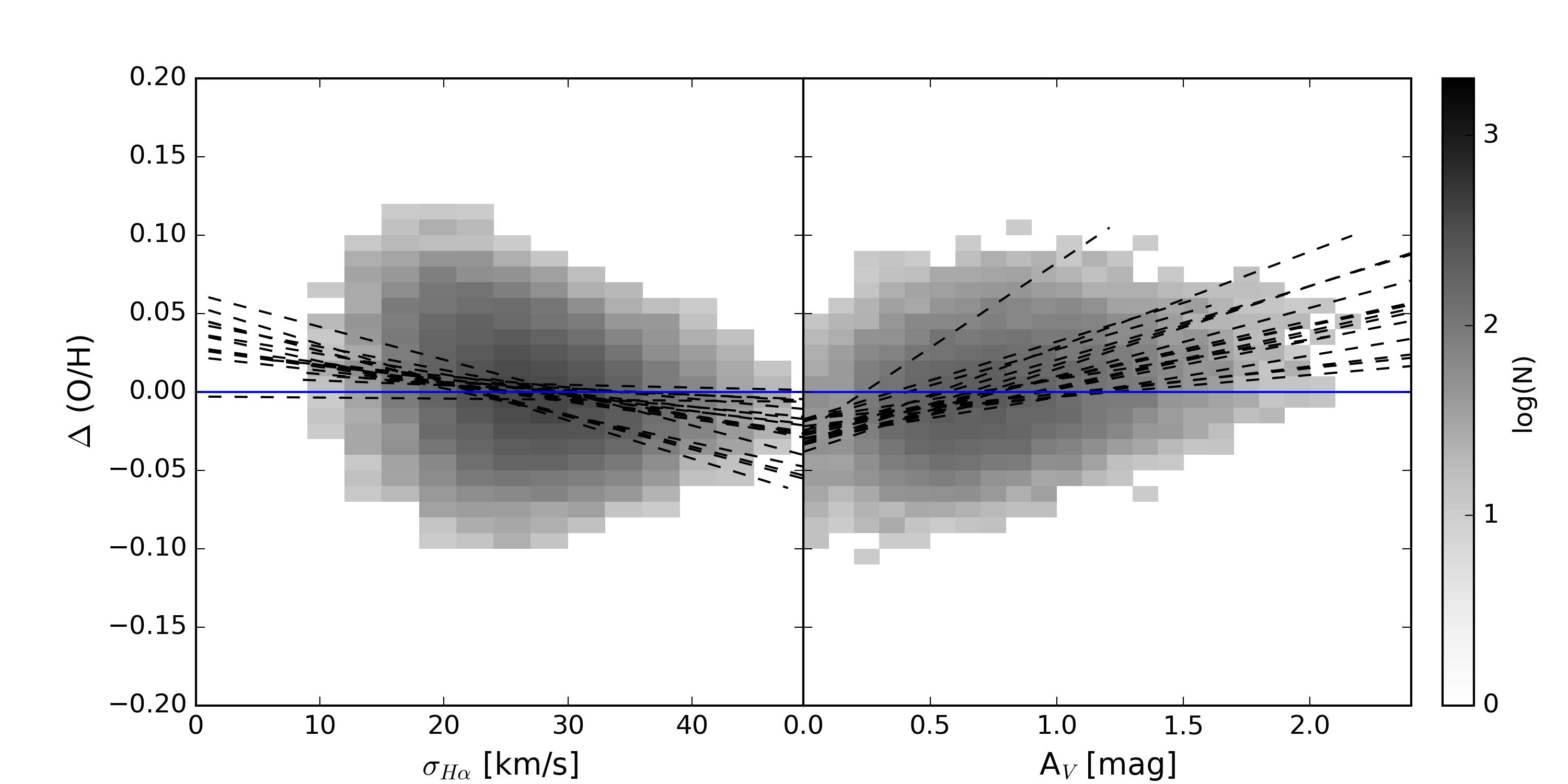}
    \caption{Metallicity variations measured after removing the radial gradient ($\Delta$(O/H)) show a negative correlation with the \ha\ velocity dispersion ($\sigma_{H\alpha}$) measured across the \hii\ region (left), and a positive correlation with the A$_V$ measured from the Balmer decrement (right).  The greyscale indicates the distribution density, and linear fits performed for individually galaxies are shown as dashed lines. Relative to a perfectly linear metallicity gradient (blue line), enriched regions are found at low velocity dispersions and dusty environments.} 
    \label{fig:DOH_sigma}
\end{figure*}

In \citet{Kreckel2019} we identified a strong correlation between $\Delta$(O/H) and ionization parameter (as traced by \siii/\sii). 
Here, we revisit secondary correlations between $\Delta$(O/H) and other local ISM physical conditions. 
Figure \ref{fig:DOH_sigma} compares $\Delta$(O/H) with the H$\alpha$ velocity dispersion measured across the integrated \hii\ region spectra and with the A$_V$ measured via the Balmer decrement. There is a weak negative correlation with velocity dispersion echos the result found globally in Figure \ref{fig:global_scatter}, and is reflected systematically within individual galaxies (dashed lines).  Given the low instrumental velocity resolution ($\sim$49 \kms\ at \ha) and moderately large integrated scales ($\sim$70 pc), we expect that our determined ionized gas velocity dispersion traces predominantly the larger scale ISM turbulence rather than local cloud turbulence \citep[likely contributing only $\sim$10 \kms;][]{Relano2005, Medina2014}, though we cannot exclude the possibility that some of these systems experience strong stellar winds \citep[which can contribute to expansion velocities by as much as $\sim$60 \kms;][]{Egorov2014, Egorov2017}.  
We also identify a positive correlation between $\Delta$(O/H) and A$_V$, which is again present within individual galaxies (dashed lines) though with more variations between galaxies.  However, both of these correlations could also arise from the correlation of $\Delta$(O/H) with \ha\ luminosity \citep[see Figure 5 in][]{Kreckel2019}.  

\begin{figure*}
    \centering
    \includegraphics[width=\textwidth]{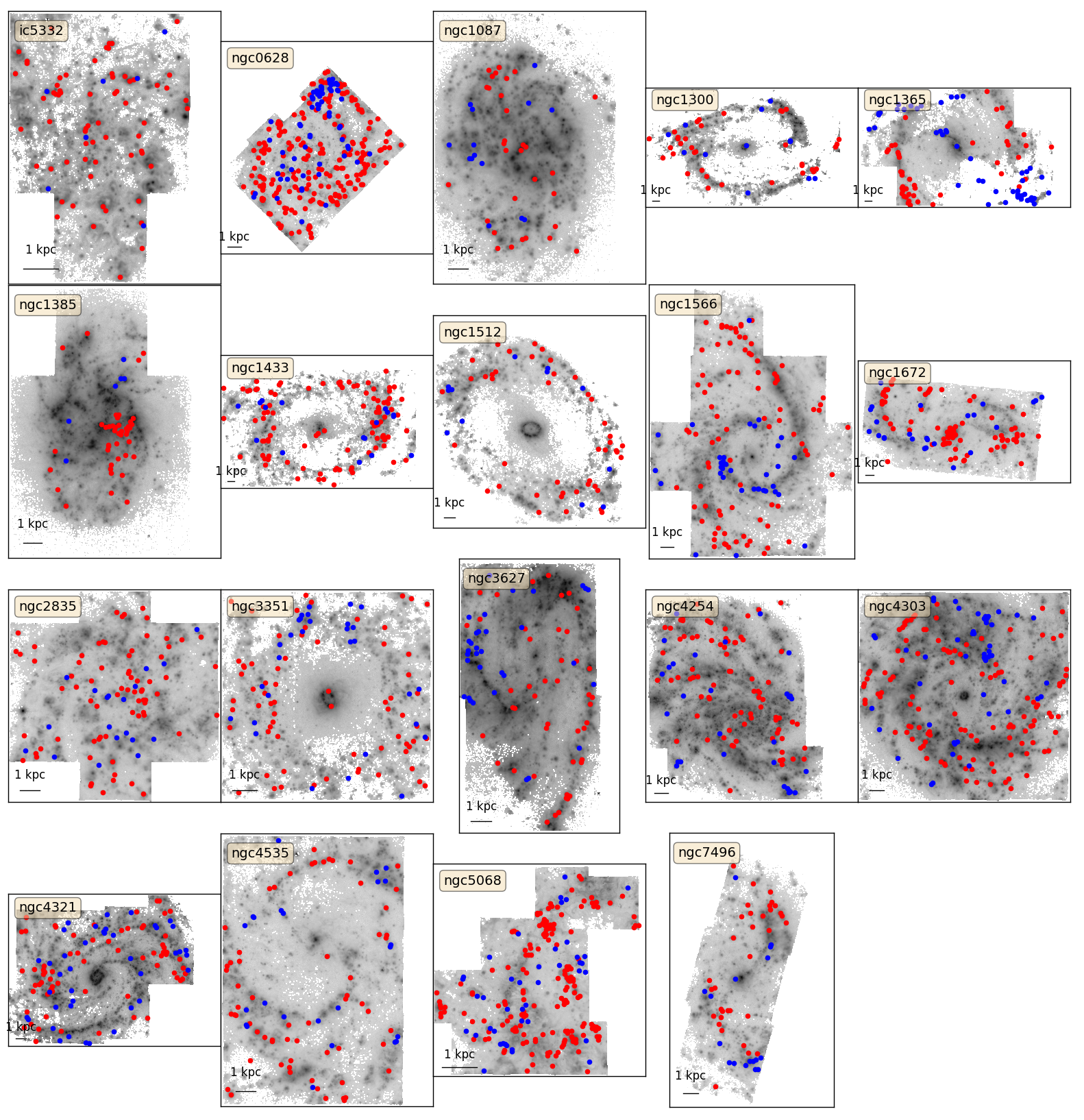}
    \caption{Maps of the \ha\ emission overlaid with locations of \hii\ regions with particularly high (red) or low (blue) metallicities compared to the radial gradient. Here we have required low statistical uncertainties ($<$0.01 dex) and be offset from the radial trend by more than 0.05 dex. While patterns appear in some galaxies relative to the spiral pattern (e.g. NGC 1672, NGC 1365), and spatial clustering is apparent in many systems (e.g. NGC 2835, NGC 3627), variations in the metallicity patterns are not systematically observed in all galaxies.   
    \label{fig:metal_outlier_maps}}
\end{figure*}

In Figure \ref{fig:metal_outlier_maps} we qualitatively examine the locations of regions with particularly high and low metallicity (relative to the radial gradient) within the galaxy discs. Here, we make a simple cut and highlight \hii\ regions with $\Delta$(O/H) $>$ 0.05 in red, and \hii\ regions with $\Delta$(O/H) $< -0.05$ in blue. While in some cases the enriched regions appear concentrated along spiral arms or at bar-ends, they can also be found distributed across the galaxy discs. Similarly, the regions with decreased abundances show some clustering (reflecting the homogeneity on kpc scales quantified in \citealt{Kreckel2020}), but no obvious patterns with galaxy environments. There is no clear difference in  $\Delta$(O/H) between the different environmental masks of \citet{Querejeta2021}, as was also shown by \cite{Williams2022}. This reflects a complicated relation between enrichment patterns and individual galaxy dynamics. A more detailed analysis of a larger sample of \hii\ regions with a larger dynamic range in these quantities will be need to disentangle what drives these second-order metallicity variations in galaxies.

\section{Discussion}
\label{sec:discussion}
\subsection{Metallicity variations}

As reported in \cite{Williams2022}, our metallicity gradients are dominated by linear radial trends (Figure \ref{fig:metal_gradients}) following earlier works on larger samples \citep[e.g.]{Sanchez2014,EspinosaPonce2022, BarreraBallesteros2022}. Following \citet{Sanchez-Menguiano2018} we do not measure the gradient within 0.5\reff, however unlike that work we see no consistent  indication for a flattening within our galaxy sample at larger radii. Our smaller sample as compared to the 109 galaxies in \citet{Sanchez-Menguiano2018} mean its harder to draw conclusions why this is, but the higher physical resolution of the PHANGS--MUSE sample ($\sim70$\, pc) as compared to theirs ($\sim460$\,pc) could be one possible reason.  Second-order variations are small, typically below $\sim$0.05 dex, and reflect a remarkable level of homogeneity in the metallicity distribution across galaxies. They do not show any correlation with co-rotation radius, which has been predicted by simulations to influence the efficiency of mixing within the disc \citep{Spitoni2019}. 

We establish that there is a weak global correlation between the magnitude of metallicity variations ($\sigma$(O/H)) and star formation rate (Figure \ref{fig:global_scatter}, top), but tighter correlations are observed with measures of the global median gas velocity dispersion (Figure \ref{fig:global_scatter}, bottom). This holds when considering both ionized and molecular gas phases. This is consistent with more turbulent ISM conditions leading to mixing on larger scales, resulting in overall more homogeneity in the metal distribution. We do not aim to investigate the source of turbulence in this paper, and careful work will be needed to disentangle the effects of star formation, secular dynamical processes and external gas accretion processes.  

We identify \hii\ regions with metallicities significantly different from the linear radial gradients, and observe that on local scales this correlation holds, with enriched \hii\ regions correlating with lower local ionized gas velocity dispersions (Figure \ref{fig:DOH_sigma}). These also correlate with dustier local environments, as traced by the Balmer decrement. Assuming gas and dust are well mixed, this would suggest that more enriched regions are associated with higher density gas. Together with the trend established with \ha\ velocity dispersion, this leads to a picture where a denser, calmer ISM facilitates localized pockets of enrichment.  At the other end of the spectrum, relatively more metal poor gas is associated with lower gas densities and increased turbulence. 
The local metallicity variations we report are consistent with the picture developed in \cite{Kreckel2020}, which quantified the mixing scales within eight of these 19 galaxies and determined that metallicity variations are likely reflecting dilution rather than pollution of the ISM.  

Another key result from \cite{Kreckel2019} in their initial study of eight of the PHANGS--MUSE targets is the identification of systematic azimuthal variations in the metallicity distribution. This has been confirmed for the full sample of 19 galaxies in \cite{Williams2022}, who analyzed interpolated metallicity maps. What remains less clear is how these variations may or may not correlate with galaxy environments (centre, bar, spiral arm, interarm).  Looking at integrated environments across the sample, \cite{Williams2022} were unable to recover any systematic trends aside from systematic enrichment of galaxy centres. This is somewhat in conflict with previous results on individual galaxies \citep{Ho2017, Vogt2017, Ho2018}, and claims based on growing samples \citep{Sanchez-Menguiano2019} for correlations in metallicity variations with spiral arms. In the initial sample of eight galaxies of \cite{Kreckel2019}, half were found to have variations correlating with spiral structure but often in only a single spiral arm. 

We do not revisit this interesting topic as we believe it requires careful dynamical considerations, tailored to each galaxy, but it demonstrates the complicated abundance patterns in relation to the galaxy environments (Figure \ref{fig:metal_outlier_maps}).  While in some galaxies the enriched regions (in red) appear to strongly trace the spiral pattern (e.g. NGC~1365, NGC~1566, NGC~1672), they can also generally be found throughout the entire disk. Qualitatively, the regions with reduced metallicity (in blue) often appear somewhat clustered and located at bar-ends. These maps highlight the challenges in establishing the role of galaxy environment and the role of gas flows in regulating the enrichment patterns in galaxy discs. 

\subsection{Missing nebulae and objects failing the BPT cuts}

Our catalogue consists of objects that are selected to be bright in H$\alpha$, but as is apparent in the BPT diagrams (Figure \ref{fig:bpt}) these are not all nebulae where the ionization is dominated by photoionization from young massive stars. Based on our consideration of diagnostic line ratios (Section \ref{sec:bpt}), this results in a sample of 23,244 
nebulae that we classify as \hii\ regions. However, it leaves 7,546 
objects in our catalogue for which we provide no definitive classification. These could be \hii\ regions blended with strong DIG or AGN emission, SNRs or PNe. Note that 609 objects are already excluded entirely from this analysis as they fall at the field edge.

Of the unclassified objects, we find that 4,688
are labelled as `composite' based on the \oiii/\hb\ vs. \nii/\ha\ BPT diagnostic (BPT\_NII = 1), and are quite likely \hii\ regions.
The commonly used BPT demarcation empirically established by \cite{Kauffmann2003} was developed for classification of central kpc-scale and integrated galaxy spectra. Recent work has begun to explore whether this parameter space is sufficiently represented once outer disc environments are considered, and wider parameter space including kinematic diagnostics are included \citep{Law2021}. With our work, we consider even smaller physical scales ($<$100 pc), and indeed recent modelling has shown that individual \hii\ regions throughout their evolution may populate the `composite' regions of the BPT diagram (falling between the \citealt{Kauffmann2003} and \citealt{Kewley2001} demarcations) for short periods during their earliest phase of evolution \citep{Pellegrini2020}. One of the main long-term science goals for producing this catalogue is to provide the necessary database of high-quality emission-line fluxes necessary to continue such detailed comparisons with cutting-edge models.  Within PHANGS, ongoing work applies a bayesian framework to match line ratios in emission line objects with different model grids, with the goal of establishing new classification methods (Congiu et al. in prep). 

Active Galactic Nuclei (AGN) provide another potential source of gas excitation, and are present in 7 (37\%) of our galaxies (as labelled in Table \ref{tab:galaxies}), with four of these AGN hosting molecular gas outflows (three galaxies without AGN also host molecular gas outflows; \citealt{Stuber2021}).   In some cases (e.g.~NGC 1365; \citealt{Venturi2018}) they represent a remarkable dominant source of ionization, with \oiii\ bright ionization cones visible across the central kpc of the galaxy and extending over nearly the full MUSE field of view. In cases of lower luminosity AGN (NGC 1433, NGC 4303, NGC 7496), it can be difficult to spatially isolate any AGN contributions, as they appear as extended ionized structures associated with (presumably) outflowing material, and seen in projection with  \hii\ regions will bias the emission line diagnostics.  Visualization of the emission line maps for all galaxies is available in \cite{Emsellem2022}. 

Supernova remnants (SNRs) are the network of shocks caused by supernova explosions as they expand into the surrounding ISM. They are often identified via their strong \sii\ and \ha\ emission \citep[e.g.][]{Long2022}.  As predicted by shock models \citep{Allen2008}, their diagnostic line ratios populate regions of the BPT diagrams that partially overlap with  photoionization. With typical sizes of less than 100\, pc, most of these objects are expected to be unresolved in our data. Ongoing projects within the PHANGS collaboration are building catalogues of SNRs, focusing on detection of these objects via their distinctive line ratios and broadened line kinematics (Li et al. in prep, Congiu et al. in prep). Preliminary results suggest  between 1000--5000 SNRs are present in our data, and may make up a significant fraction of the unclassified objects. 

Planetary nebulae (PNe) are shells of gas expelled by intermediate mass stars \citep[1-8 M$_\odot$;][]{Parker2022} and ionized by the central source in the end phase of their life.  They are particular bright in \oiii\ line emission \citep{Parker2022}, but can also emit strongly in \ha, and are unresolved at the distances of our target galaxies. Given that the central stars exhibit a harder ionizing spectrum than typical \hii\ regions, they could be expected to fall in regions of the BPT diagrams traditionally populated by AGN. 
Recently, \cite{Scheuermann2022} used the PHANGS--MUSE observations to identify PNe, with selection based on \oiii\ emission and source classification confirmed based on diagnostic line ratios. They find a total of 899 PNe across these 19 galaxies,  193 of which are  within 1$^{\prime\prime}$ of sources in our nebular catalogue. 158 of these (82\%) are unclassified, failing our BPT cuts, and making up a very small fraction of the unclassified objects. Overall, these make up a very small fraction of our nebular catalogue, which is unsurprising as many of the PNe are faint or undetected in \ha.

\subsection{Odd/misclassfied nebulae}
\label{sec:outliers}

Because of our automated approach to object identification, line fitting, and object classification, our catalogue contains objects that appear as unusual outliers in critical diagnostics. We examine the spectra of a subset of these objects to understand if there are any systematic problems and provide some explanation for these outliers.

One category of unusual object are \hii\ regions identified as having particular large \ha\ velocity dispersions, and we look at the 12 objects $\sigma_{H\alpha}$ > 200 km s$^{-1}$. 
One object (NGC1385, \#12) clearly corresponds to a foreground star missed by our flagging. 
We find that the central AGN in both NGC 1365 and NGC 1566 end up classified as \hii\ regions because we perform only single Gaussian fits to our emission lines, and the underlying broad line component results in a biased fit.    
Two other objects (\#589 in NGC 1300 and \#35 NGC 3627) appear to be a result of poorly constrained continuum fits and the misclassification of noise as spectral lines within the spectra. 
The remaining seven objects clearly contain a broader secondary line component that is coincident with the \ha\ line emission. Whether this is a signature of massive stars driving strong winds or due to a background galaxy seen in projection is difficult to distinguish with these spectra alone. PHANGS--HST imaging reveals unresolved bright sources coincident with two of the nebulae, but no obvious background galaxy or bright stellar source at the other positions. 

Six objects that we identify as \hii\ regions result in inferred metallicity measurements that are unusually low, 12+log(O/H) $<$ 8.0, and in fact all appear to be spurious sources. On closer inspection, two are faint foreground stars that had not been identified, one corresponds to an object where the stellar continuum is poorly constrained and noise peaks are fit as emission lines, and the remaining three show secondary broadened Gaussian profiles that bias the \ha\ fit and skew the resulting line ratio diagnostics. Many of these exhibit S/N$\sim$10 in fainter lines (\sii, \oiii), suggesting more stringent S/N cuts would be effective in excluding spurious sources. 

Finally, our observation of NGC 1672 happened to occur only shortly after a supernova event in the galaxy, AT2017gax, and as a result is poorly fit in our catalogue and has ID \#429. 

\subsection{Impact of Resolution}
\label{sec:blending}

\begin{figure}
    \centering
    \includegraphics[width=3.5in]{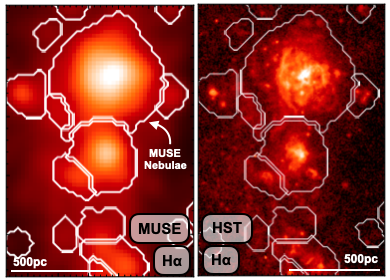}
    \caption{A comparison of MUSE and HST \ha\ imaging in a star-forming complex of NGC~1672 \citep{Barnes2022}. While it is clear that MUSE does separate distinct complexes, the high $\sim$8~pc HST resolution reveals complicated \ha\ morphologies including clustered compact regions which cannot be distinguished from neighbouring brighter regions at the 70~pc MUSE resolution. }
    \label{fig:MUSE_HST}
\end{figure}

Given that star formation is often observed to be clustered (that is, star formation occurs in high gas density regions like spiral arms), one limiting factor in constructing our nebular catalogue is the physical resolution of our observations. In Table \ref{tab:galaxies} we list the physical resolution within each galaxy, which reflects both the seeing conditions during the observation and the distance to the source. Values range from 23--104 pc, sufficient to identify star forming regions that are isolated but likely insufficient to separate clustered star-forming complexes. In comparison with HST \ha\ narrowband imaging (Figure \ref{fig:MUSE_HST}), it is apparent that objects identified as a single source in MUSE can be comprised of several neighbouring complexes.  We are not able to fully account for this effect without  additional HST \ha\ imaging, which is currently only available for a handful of our targets (NGC~628, NGC~1672, NGC~3351) but will be available for the full sample with an upcoming PHANGS HST narrow band survey. 

The effect of blending of objects within this catalogue is discussed in \cite{Santoro2022} in relation to its impact on the \hii\ region luminosity function.  Those authors estimate the mean separation between \hii\ regions per galaxy and find that it shows only a modest correlation with changes in the \hii\ region luminosity function slope and cannot be responsible for differences in luminosity function slope that are observed between galaxies. 

\begin{figure*}
    \centering
    \includegraphics[width=7in]{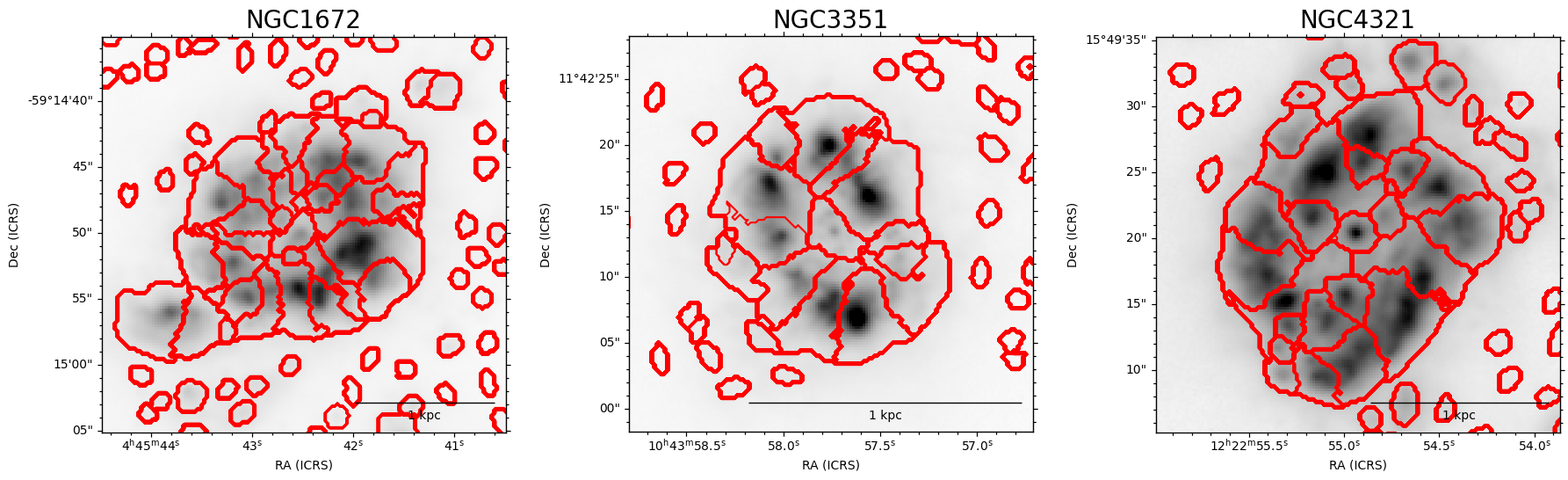}
    \caption{Galaxy centres from NGC~1672, NGC~3351 and NGC~4321, which host nuclear star-forming rings.  \ha\ emission (greyscale) shows clustered and clumpy star-forming regions, which are only moderately well separated into individual \hii\ regions (red contours). Region sizes in this particular environment are also significantly overestimated. }
    \label{fig:starburstrings}
\end{figure*}

Another location where blending becomes a clear issue is in the centres of galaxies hosting starbursting rings, where the extreme spatial concentration of bright \hii\ regions make decomposition challenging. Region identification is further complicated by the increased levels of diffuse ionized gas, associated with the high stellar density contributing an additional ionization component (see Section \ref{sec:dig};  \citealt{Belfiore2022}). Three galaxies (Figure \ref{fig:starburstrings}) host particularly bright nuclear star-forming rings (NGC 1672, NGC 3351, NGC 4321), and we caution against over-interpretation of these integrated nebular fluxes without more careful spatial decomposition or deblending of objects. 

While the HST comparisons and \hii\ region luminosity functions suggest blending of nebulae is not a huge issue (excluding starbursting rings), even if it is occurring, the resolutions we achieve with MUSE are generally insufficient for accurate determination of \hii\ region sizes. This has lead us to exclude any size determination from our catalogue. This issue was been highlighted in comparisons of MUSE \hii\ regions with HST narrow band \ha\ imaging \citep{Barnes2022}, where it is apparent that at $\sim$70~pc scales MUSE is unable to provide robust quantification of sizes or morphologies \citep{Hannon2022}. 
In our catalogue, we therefore provide the number of 0.2\arcsec $\times$ 0.2\arcsec pixels associated with the spatial mask of each \hii\ region, as a way of distinguishing the larger objects and derive surface brightnesses.

\subsection{Diffuse Ionized Gas}
\label{sec:dig}

A prominent feature in our deep emission line maps is the pervasive diffuse ionized gas (DIG) component. As was shown in \cite{Zurita2000} and more recently revisited for the PHANGS--MUSE galaxies \citep{Belfiore2022}, the bulk of this emission is spatially correlated with \hii\ region locations, and can be explained well by a model where ionizing photons leak from their \hii\ regions and propagate out to $\sim$kpc scales in the disc.  However, with an additional consideration of the emission line ratios in the gas, it is clear that an additional contribution from hot low-mass evolved stars (HOLMES) is required \citep{Belfiore2022}.  

Systematic differences in line ratios observed in \hii\ regions compared to the DIG were first established in Milky Way observations \citep{Haffner2009},  and have the potential to significantly bias line flux measurements particularly for \sii\ and \nii\ which are emitted strongly in the DIG. However, the irregular spatial distribution and  low-surface brightness of the DIG makes it difficult to model and subtract this component. In our construction of the nebular catalogues, we have not applied corrections for the DIG along the line of sight. As most of our objects are bright, exhibiting median \ha\ luminosities of $2 \times 10^{37}$ ergs s$^{-1}$ and \ha\ surface brightnesses of $2 \times 10^{39}$ ergs s$^{-1}$ kpc$^{-2}$,  we expect that the DIG should not strongly impact our line ratios or derived quantities (e.g.~dust attenuation, metallicity, ionization parameter). However, we still caution that physical quantities interpreted from integrated line fluxes may still be impacted, especially for \hii\ regions with low surface brightnesses or in crowded environments.

\section{Conclusions}
\label{sec:conclusion}

We present a new catalogue of 30,790 nebulae, selected morphologically in \ha\ images, that are fully contained within the PHANGS--MUSE coverage of 19 nearby star-forming spiral galaxies. For all nebulae, we construct an integrated spectrum and measure integrated line fluxes and line kinematics for strong lines across the optical (4800--9300\AA) spectrum (Table \ref{tab:lines}). We calculate derived properties, including dust attenuation via the Balmer decrement, characterise their galactic environments (centre, bar, spiral arm, interarm, disk), and classify objects based on their emission line (BPT) diagnostics. For the 23,244 nebulae that we classify as \hii\ regions (Section \ref{sec:bpt}), we calculate the gas-phase metallicity and ionization parameter. A full list of properties characterised in our catalogue is included in Table \ref{tab:catalog_data}. 

We examine the radial gradients of the nebula physical properties within galaxies, finding that the metallicity gradients demonstrate the most pronounced trends (Table \ref{tab:metal_grad_scal}). We also look in more detail at the residual variation in metallicity ($\Delta$(O/H)), after fitting and subtracting the linear radial trends in log(O/H), and quantify the global scatter in this residual metallicity within each galaxy ($\sigma$(O/H)). As was previously shown using a sub-sample of our galaxies \citep{Kreckel2019}, we recover small scatters with  $\sigma(O/H)=$0.03 -- 0.07 dex, and show that this scatter correlates weakly with the global star formation rate and correlates strongly with both ionized and molecular gas velocity dispersions.  Local metallicity variations show further support for these trends with the turbulent condition of the ISM, and we find that enriched regions preferentially show lower \ha\ velocity dispersion and higher dust attenuation, reflecting a correspondence with calmer pockets of the ISM. These correlations between the metallicity variations (relative to the radial gradients) and the turbulent state of the ISM (in both the ionized and molecular material) demonstrate convincingly that the ISM dynamics play a critical role in regulating the mixing of metals across galaxy discs. Correlations with galaxy environment are qualitatively apparent in some galaxies, but systematic trends are less clear and more detailed dynamical modelling is required.

With this catalogue, we do not yet include a full analysis of the temperature sensitive auroral emission lines (\nii5755, \siii6313, \oii7320,7330), which are contained within our wavelength range and detected in about a thousand \hii\ regions.  Careful fitting of these faint lines ($\sim$1\% of the intensity of \ha) is required to determine robust line fluxes, and will be the subject of future work.  In combination with the collisionally excited strong-lines, we aim to derive electron temperatures and simultaneously (using the \sii\ density diagnostic) determine electron densities for a sub-sample of our catalogue.

This catalogue of the young, ionized nebulae represents a key parameter when developing a model of the baryon cycle within galaxies on resolved ($<$100~pc) scales.  
Within the PHANGS collaboration, we aim to quantify key stages in this process by characterising with PHANGS--ALMA the molecular gas \citep{Sun2020} and individual giant molecular clouds (GMCs; \citealt{Rosolowsky2021}; Hughes et al. in prep), and with PHANGS--HST  the individual star clusters and stellar associations (\citealt{Turner2021}; Larson et al. in prep). Comparison of the \hii\ region and GMC distributions reveal relatively little spatial overlap \citep{Kreckel2018}, indicative of short cloud disruption timescales \citep{Kim2021, Chevance2022}. Ongoing work links individual \hii\ regions with their parent GMC, to determine if stellar feedback has a measurable impact on molecular cloud properties (Zakardjian et al. in prep), and searches for super-bubble morphologies in the molecular gas, to quantify feedback energetics (Watkins et al. in prep). Cross-matching our \hii\ region catalogue with young stellar associations provides quantitative constraints on the mass and age of the stars powering these ionized nebulae, uniquely enabling us to constrain the evolutionary sequence and link the ionizing photon budget to the ionization state of the gas (\citealt{Barnes2022}, Scheuermann et al. submitted, Egorov et al. in prep).  In the near future, these results will be further complemented by PHANGS--JWST maps of the earliest embedded phases of star formation and constraints on the dust chemistry, further completing our view of the complete star-formation cycle.

\section*{Acknowledgements}

This work was carried out as part of the PHANGS collaboration.

Based on observations collected at the European Southern Observatory under ESO programmes 1100.B-0651, 095.C-0473, and 094.C-0623 (PHANGS--MUSE; PI Schinnerer), as well as 094.B-0321 (MAGNUM; PI Marconi), 099.B-0242, 0100.B-0116, 098.B-0551 (MAD; PI Carollo) and 097.B-0640 (TIMER; PI Gadotti). 

KK, FS, and OE gratefully acknowledge funding from the German Research Foundation (DFG) in the form of an Emmy Noether Research Group (grant number KR4598/2-1, PI Kreckel). 
KK, EJW and SCOG acknowledge support from the Deutsche Forschungsgemeinschaft (DFG, German Research Foundation) – Project-ID 138713538 – SFB 881 (“The Milky Way System”, subprojects B1, B2, B8 and P1). 
SCOG also acknowledges funding from the European Research Council via the ERC Synergy Grant ``ECOGAL -- Understanding our Galactic ecosystem: From the disk of the Milky Way to the formation sites of stars and planets'' (project ID 855130) and from the Heidelberg Cluster of Excellence (EXC 2181 - 390900948) ``STRUCTURES: A unifying approach to emergent phenomena in the physical world, mathematics, and complex data'', funded by the German Excellence Strategy. ATB and FB would like to acknowledge funding from the European Research Council (ERC) under the European Union’s Horizon 2020 research and innovation programme (grant agreement No.726384/Empire).
FS, ES, and TGW acknowledge funding from the European Research Council (ERC) under the European Union’s Horizon 2020 research and innovation programme (grant agreement No. 694343).
E.C. acknowledge support from ANID Basal projects ACE210002 and FB210003.
PSB acknowledge support from the project project PID2019-107427-GB-31 funded by the MCIN/AEI/10.13039/50110001103.
This research has made use of the NASA/IPAC Extragalactic Database (NED) which is operated by the Jet Propulsion Laboratory, California Institute of Technology, under contract with NASA. It also made use of a number of python packages, namely the main \textsc{astropy} package \citep{Astropy+2013,Astropy+2018}, \textsc{numpy} \citep{Harris+2020} and \textsc{matplotlib} \citep{Hunter+2007}.

\section*{Data Availability}

The MUSE data underlying this work are presented in \citet{Emsellem2022} and are available at the ESO archive \footnote{https://archive.eso.org/scienceportal/home?data\_collection=PHANGS} and CADC archive \footnote{https://www.canfar.net/storage/vault/list/phangs/RELEASES/PHANGS-MUSE}. The catalogue of all identified nebulae, along with fits files containing masks of the individual nebulae locations, are published along with this paper and available in the online supplementary material of the journal as well as at the CADC archive.



\bibliographystyle{mnras}
\bibliography{nebular} 

\clearpage

\appendix

\section{Image Atlas}
\label{appendix:atlas}

\begin{figure*}
    \centering
    \includegraphics[width=7in]{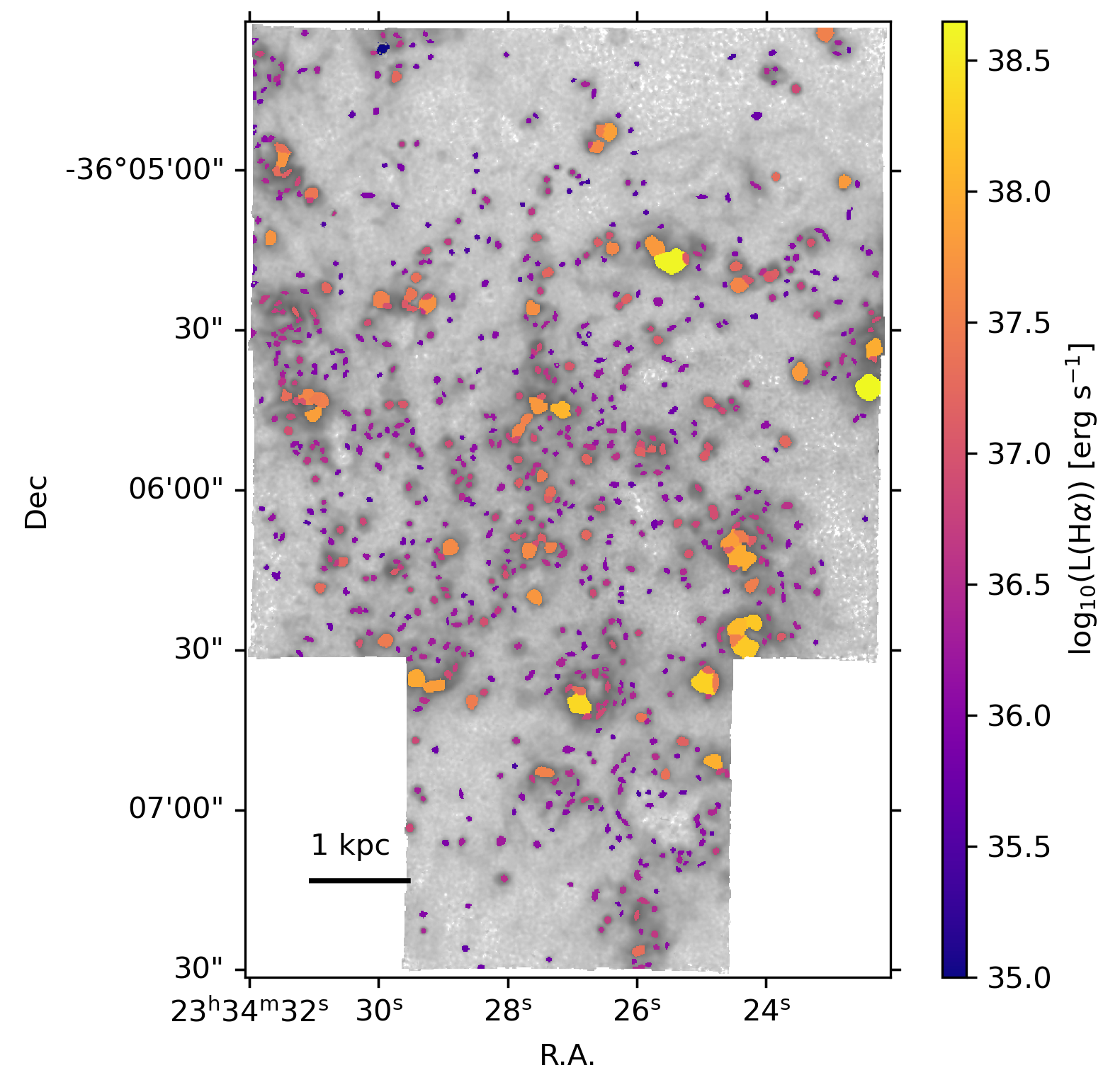}
    \caption{The spatial distribution of nebulae in the galaxy IC5332. The background greyscale image shows the \ha\ emission in log scale, and the colour of the nebulae indicates their intrinsic (dust corrected) \ha\ luminosity.}
    \label{fig:atlas_IC5332}
\end{figure*}

\begin{figure*}
    \centering
    \includegraphics[width=7in]{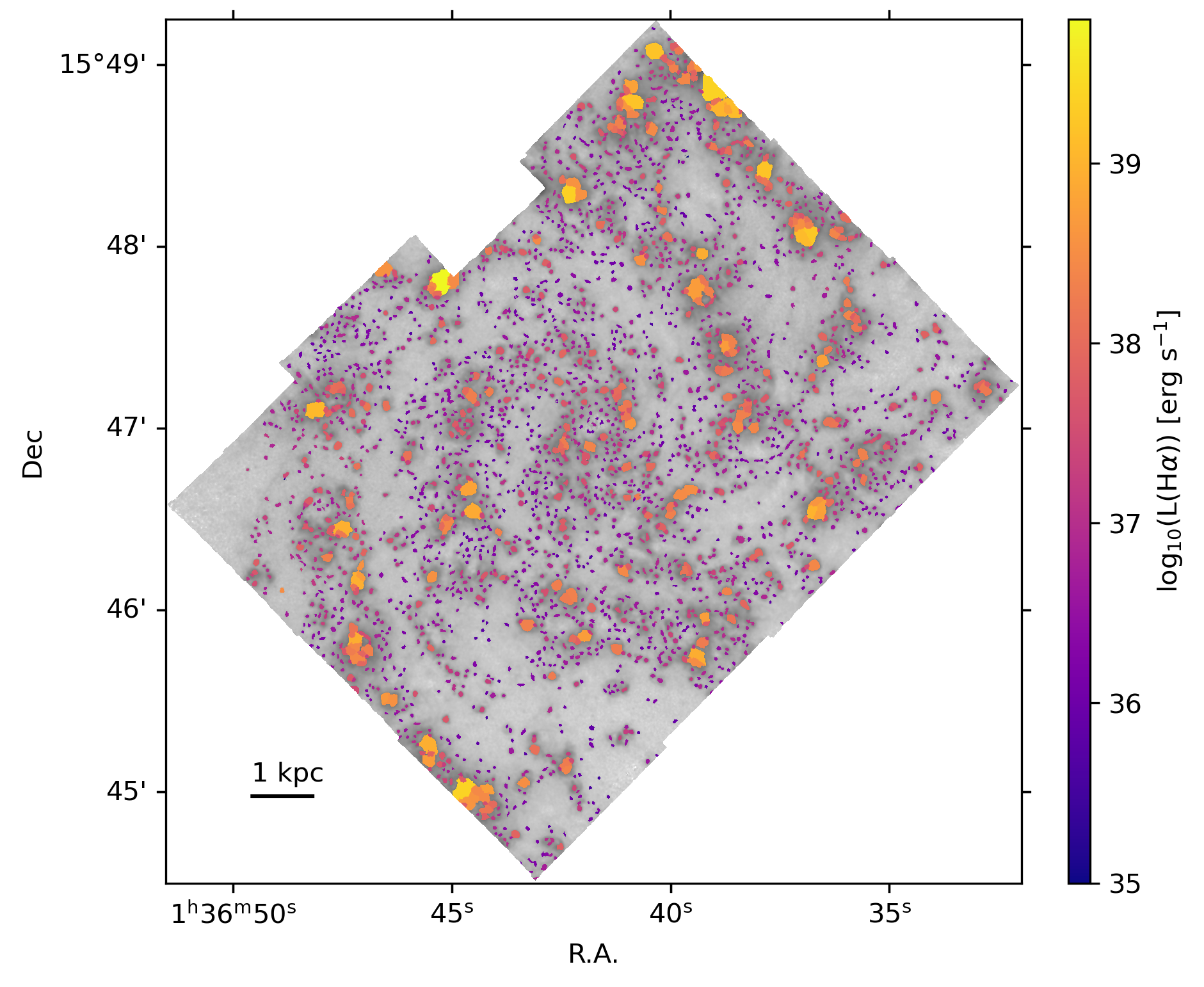}
    \caption{As in figure \ref{fig:atlas_IC5332} but for NGC 628.}
    \label{fig:atlas_NGC628}
\end{figure*}

\begin{figure*}
    \centering
    \includegraphics[width=7in]{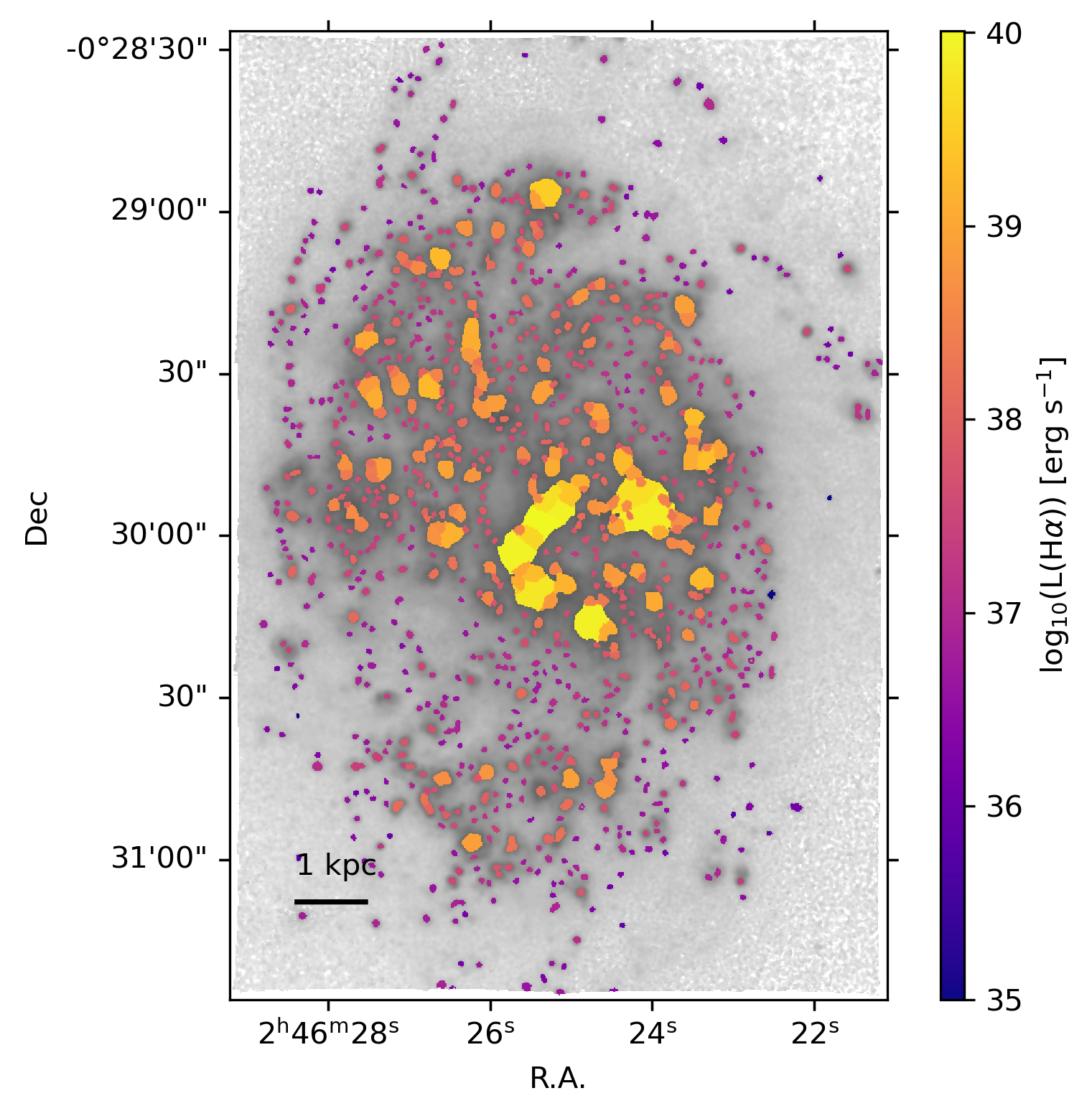}
    \caption{As in figure \ref{fig:atlas_IC5332} but for NGC 1087}
    \label{fig:atlas_NGC1087}
\end{figure*}

\begin{figure*}
    \centering
    \includegraphics[width=7in]{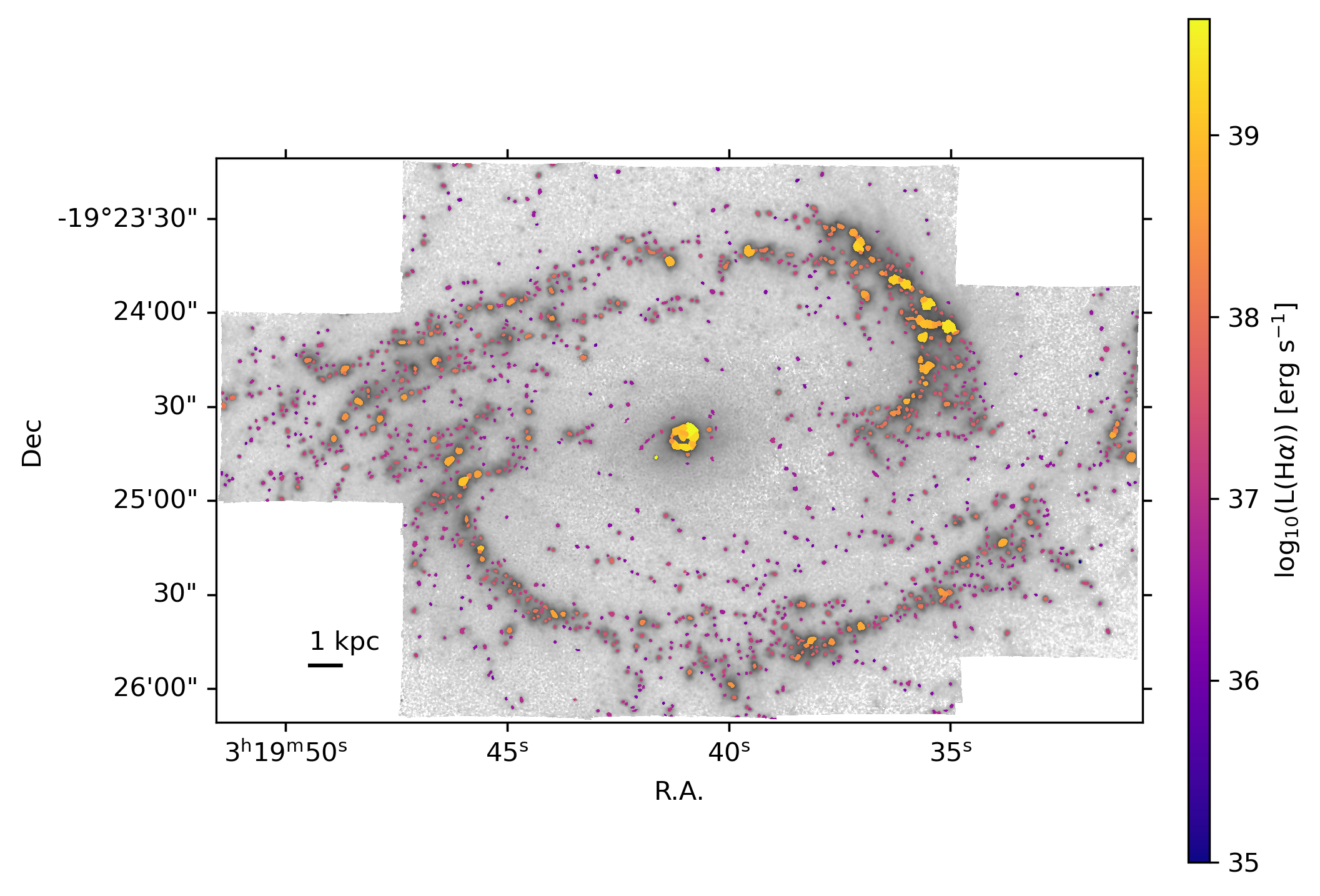}
    \caption{As in figure \ref{fig:atlas_IC5332} but for NGC 1300}
    \label{fig:atlas_NGC1300}
\end{figure*}

\begin{figure*}
    \centering
    \includegraphics[width=7in]{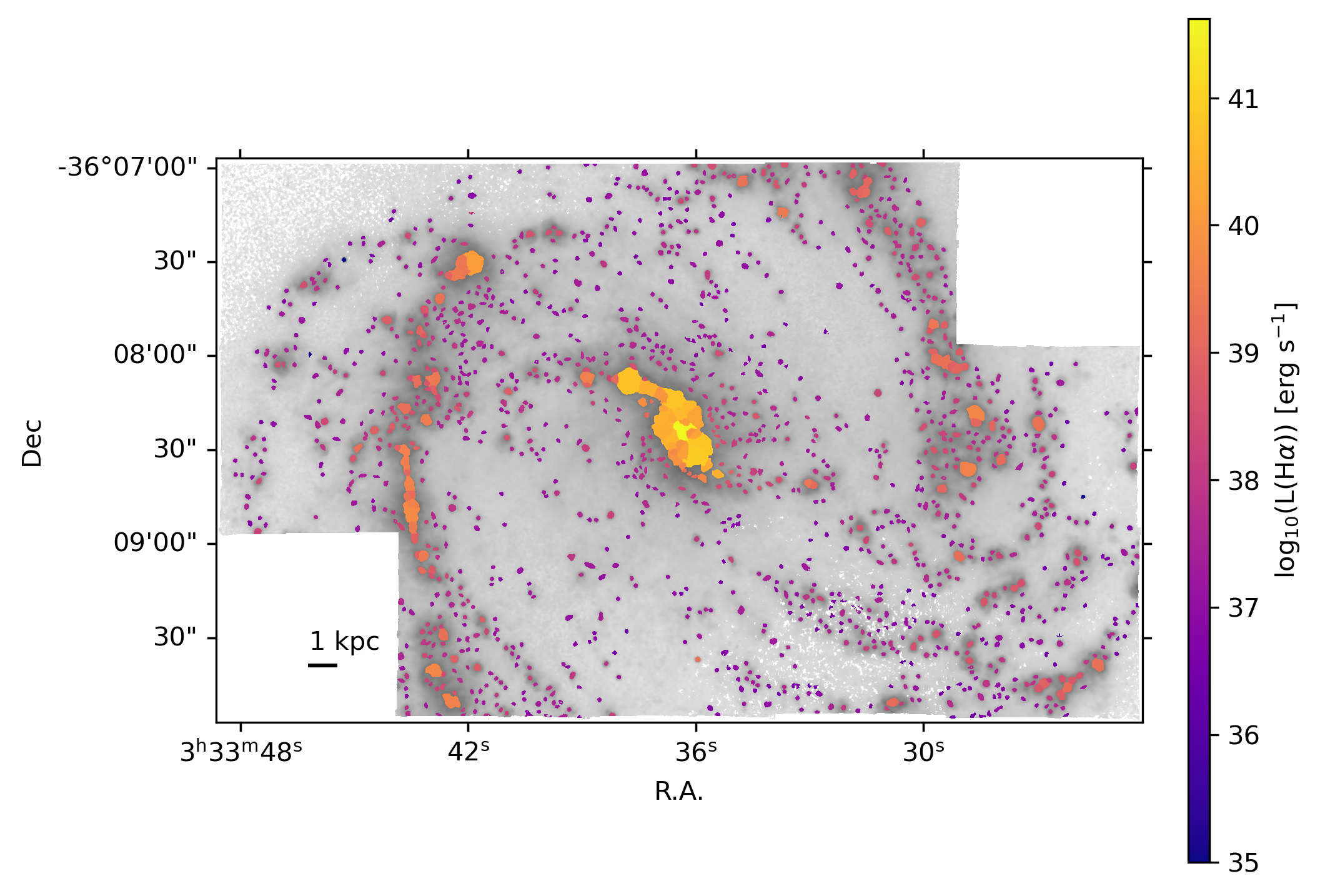}
    \caption{As in figure \ref{fig:atlas_IC5332} but for NGC 1365}
    \label{fig:atlas_NGC1365}
\end{figure*}

\begin{figure*}
    \centering
    \includegraphics[width=7in]{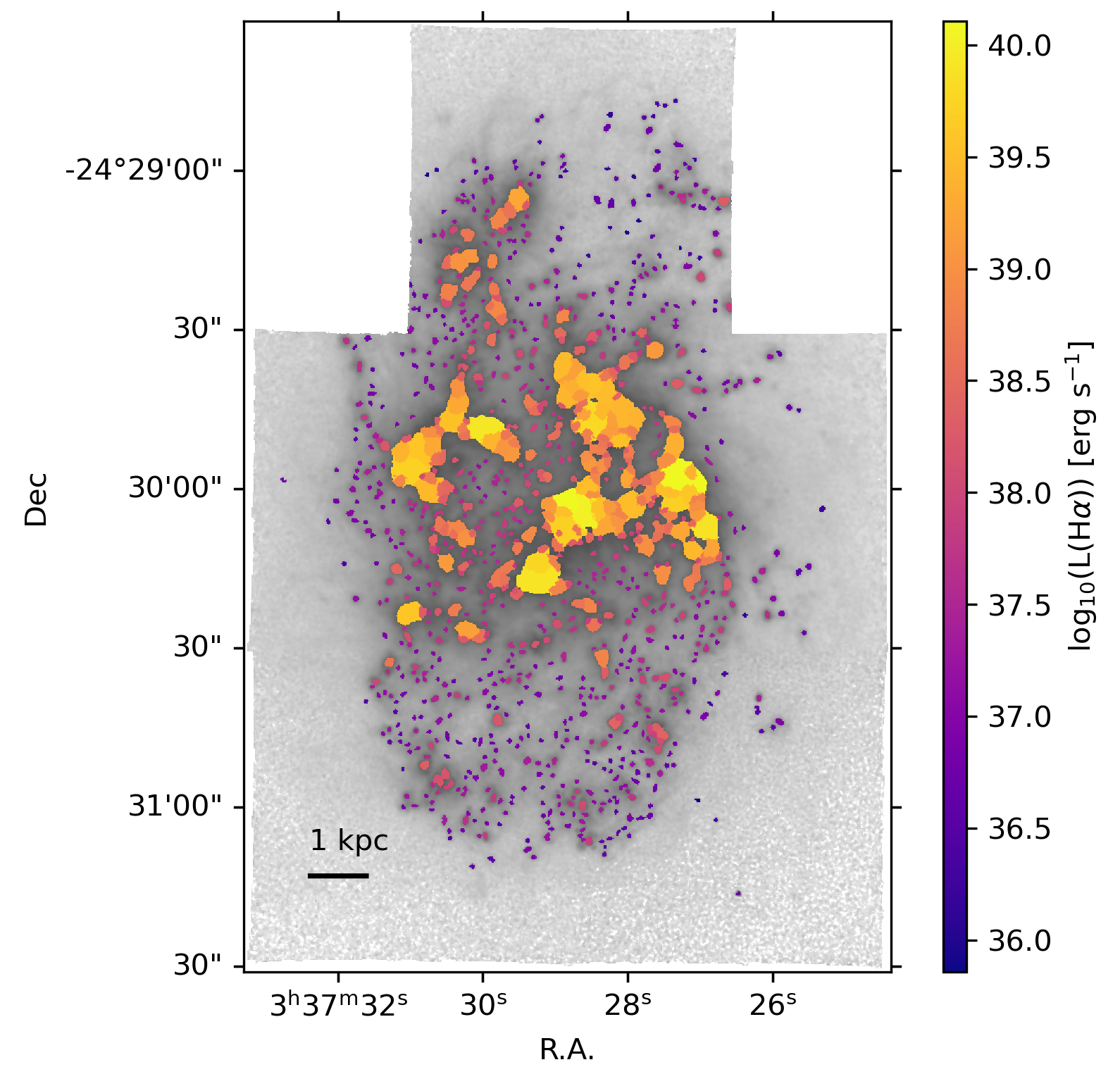}
    \caption{As in figure \ref{fig:atlas_IC5332} but for NGC 1385}
    \label{fig:atlas_NGC1385}
\end{figure*}

\begin{figure*}
    \centering
    \includegraphics[width=7in]{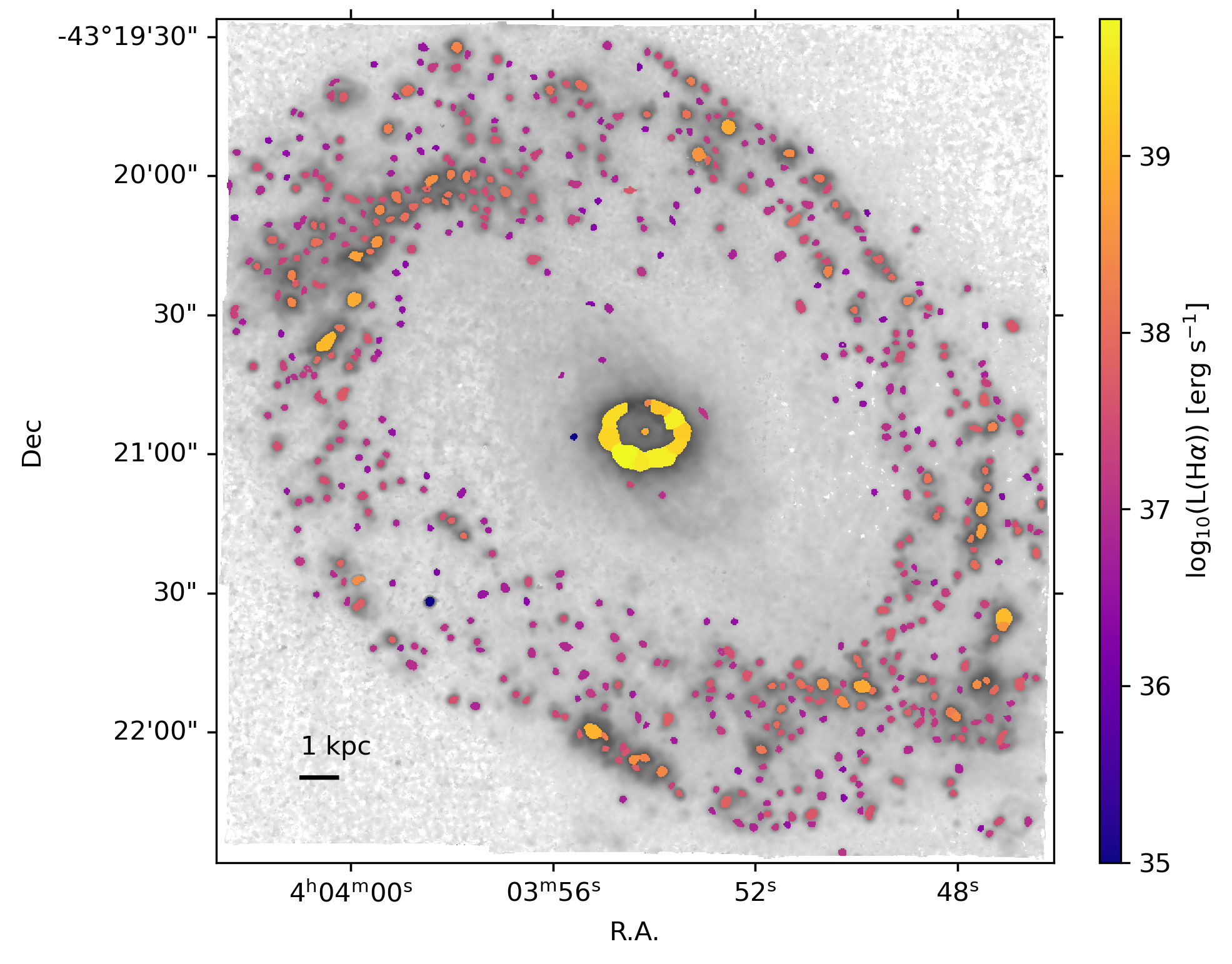}
    \caption{As in figure \ref{fig:atlas_IC5332} but for NGC 1512}
    \label{fig:atlas_NGC1512}
\end{figure*}

\begin{figure*}
    \centering
    \includegraphics[width=7in]{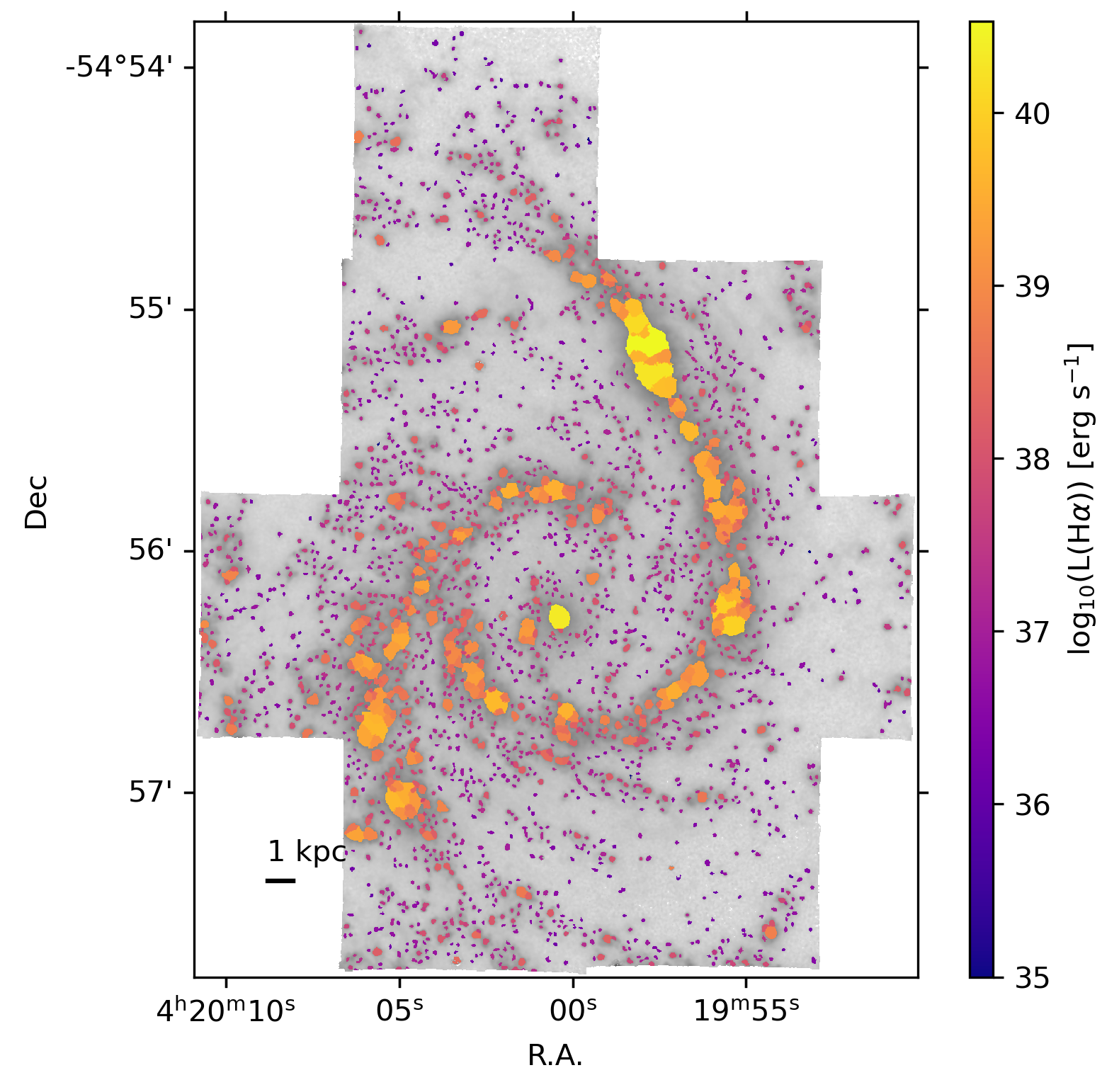}
    \caption{As in figure \ref{fig:atlas_IC5332} but for NGC 1566}
    \label{fig:atlas_NGC1566}
\end{figure*}

\begin{figure*}
    \centering
    \includegraphics[width=7in]{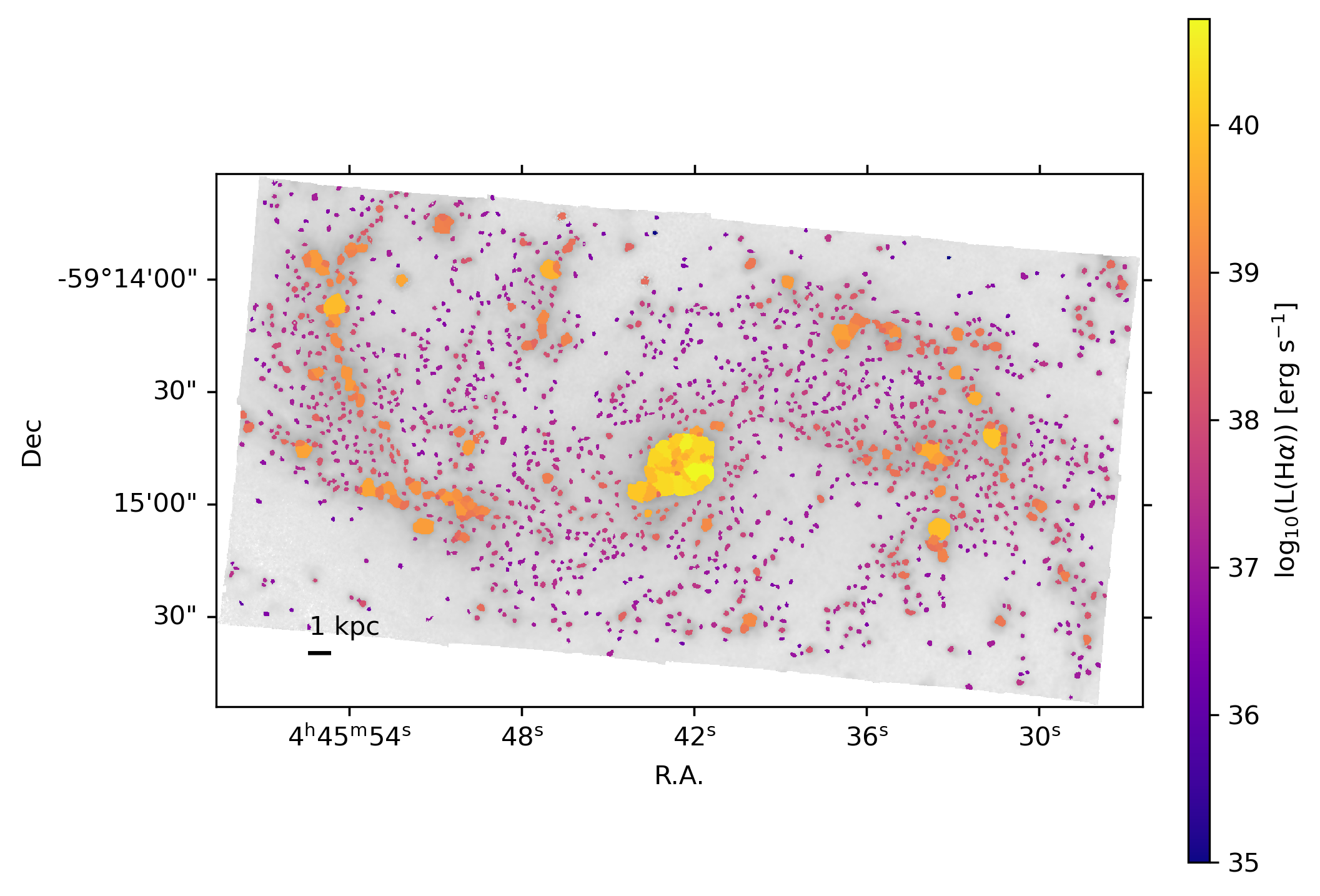}
    \caption{As in figure \ref{fig:atlas_IC5332} but for NGC 1672}
    \label{fig:atlas_NGC1672}
\end{figure*}

\begin{figure*}
    \centering
    \includegraphics[width=7in]{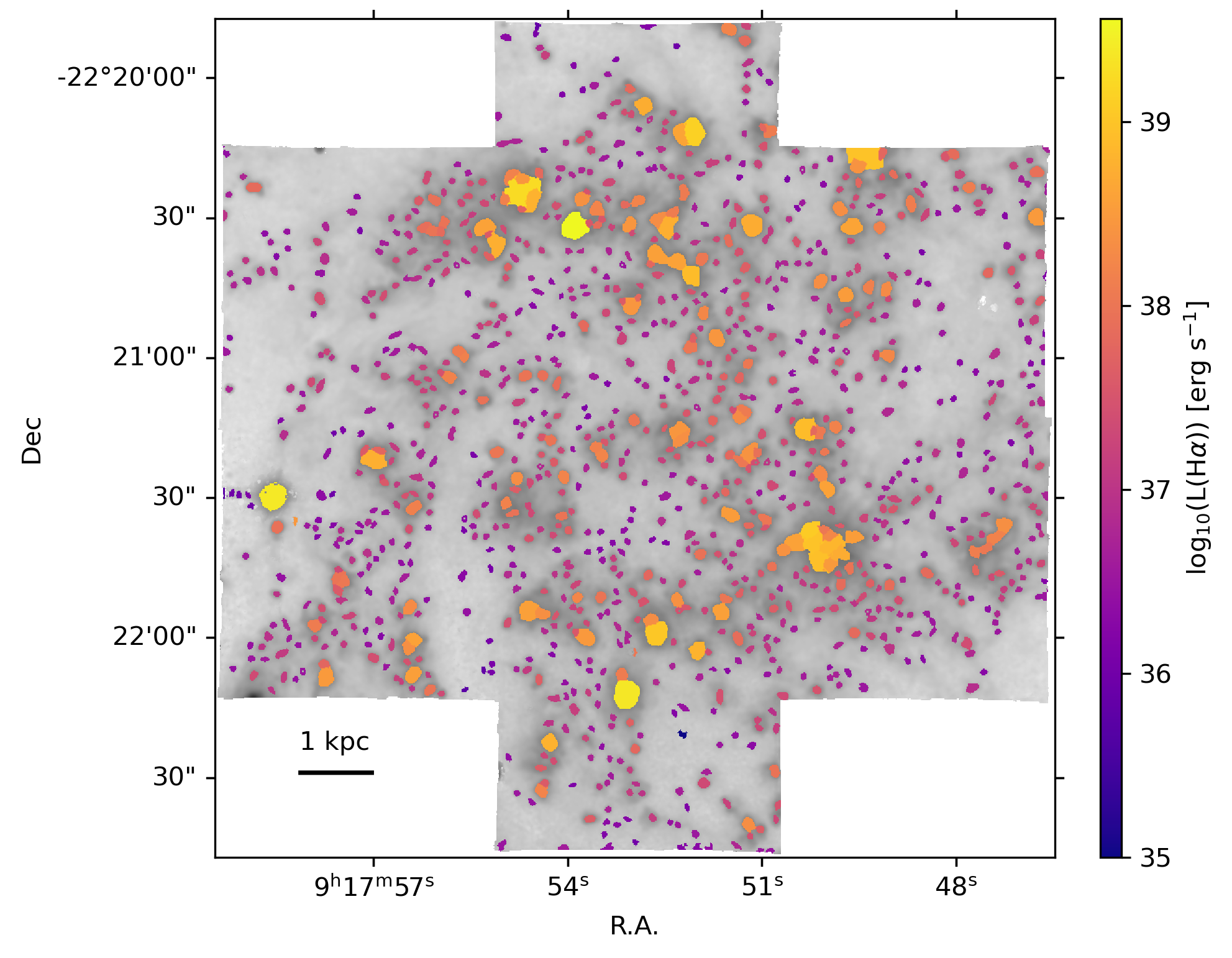}
    \caption{As in figure \ref{fig:atlas_IC5332} but for NGC 2835}
    \label{fig:atlas_NGC2835}
\end{figure*}

\begin{figure*}
    \centering
    \includegraphics[width=7in]{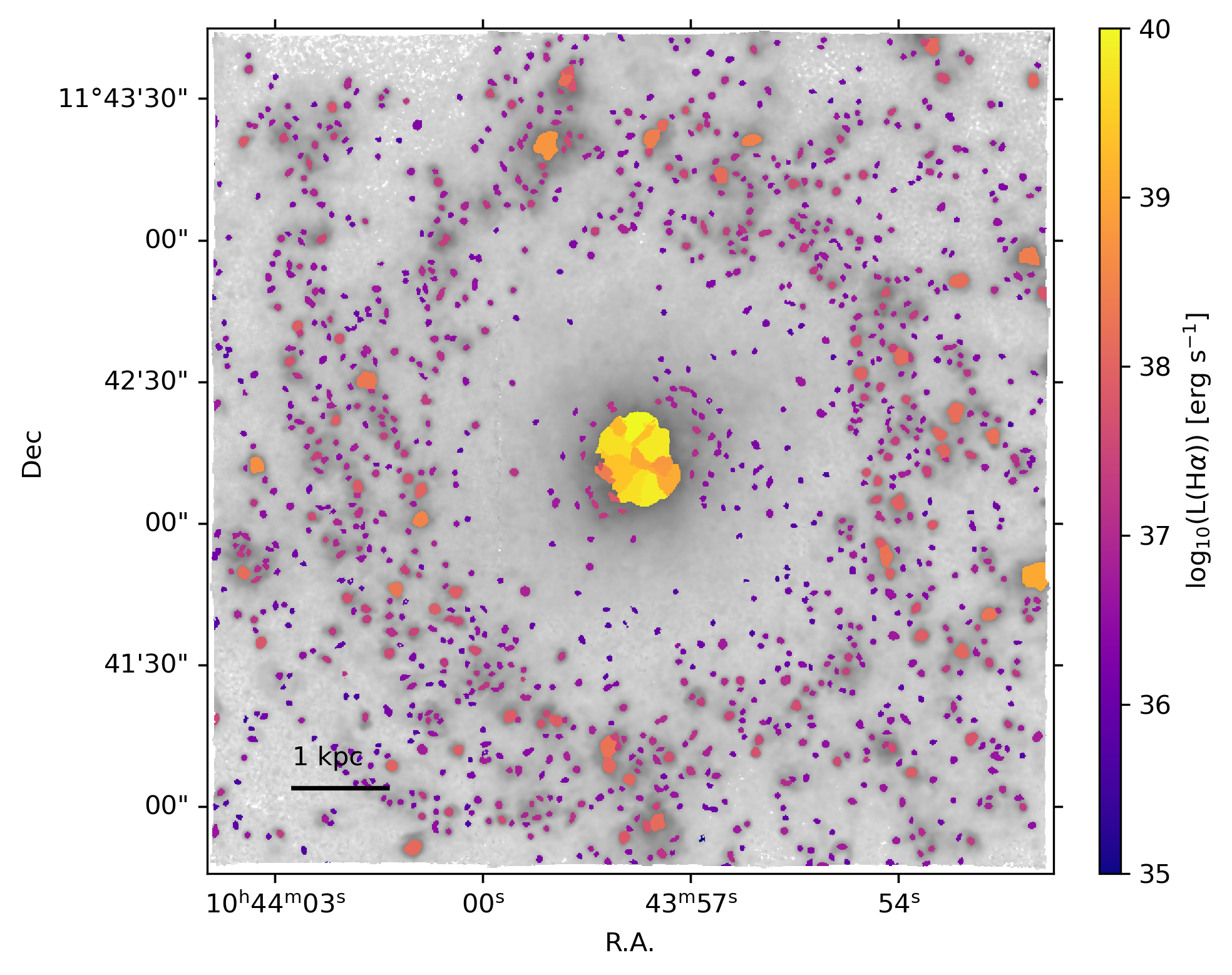}
    \caption{As in figure \ref{fig:atlas_IC5332} but for NGC 3351}
    \label{fig:atlas_NGC3351}
\end{figure*}

\begin{figure*}
    \centering
    \includegraphics[width=7in]{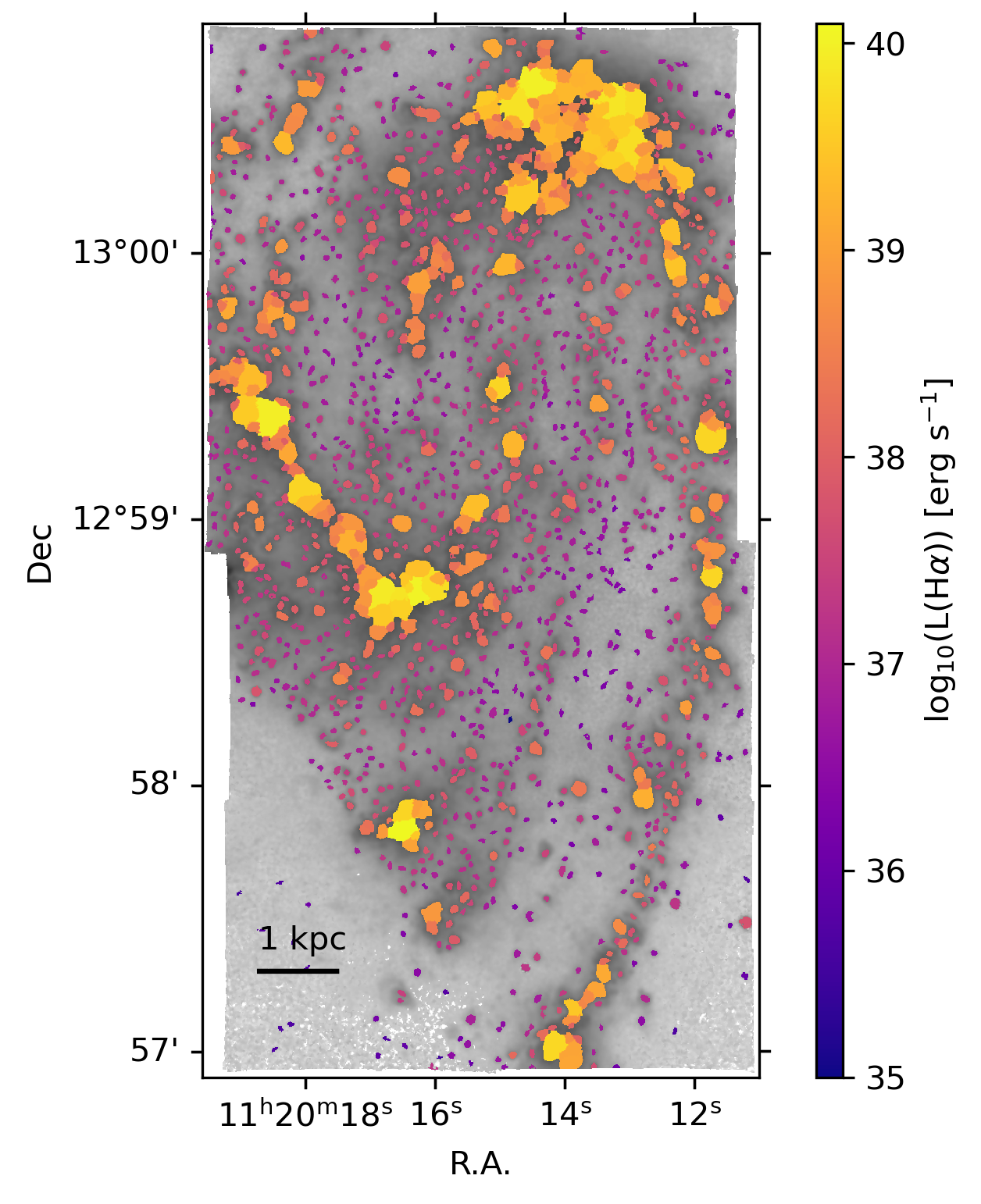}
    \caption{As in figure \ref{fig:atlas_IC5332} but for NGC 3627}
    \label{fig:atlas_NGC3627}
\end{figure*}

\begin{figure*}
    \centering
    \includegraphics[width=7in]{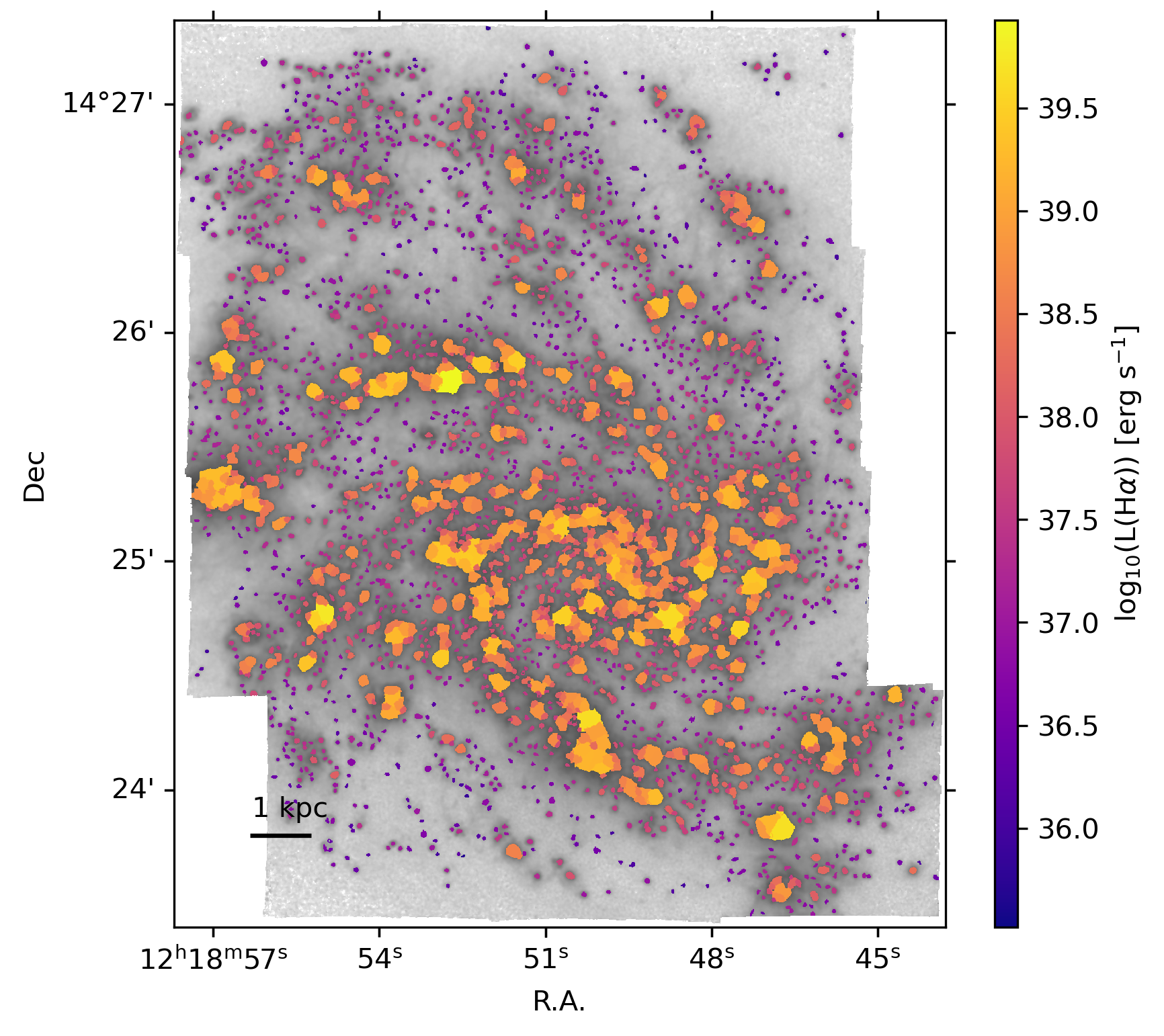}
    \caption{As in figure \ref{fig:atlas_IC5332} but for NGC 4254}
    \label{fig:atlas_NGC4254}
\end{figure*}

\begin{figure*}
    \centering
    \includegraphics[width=7in]{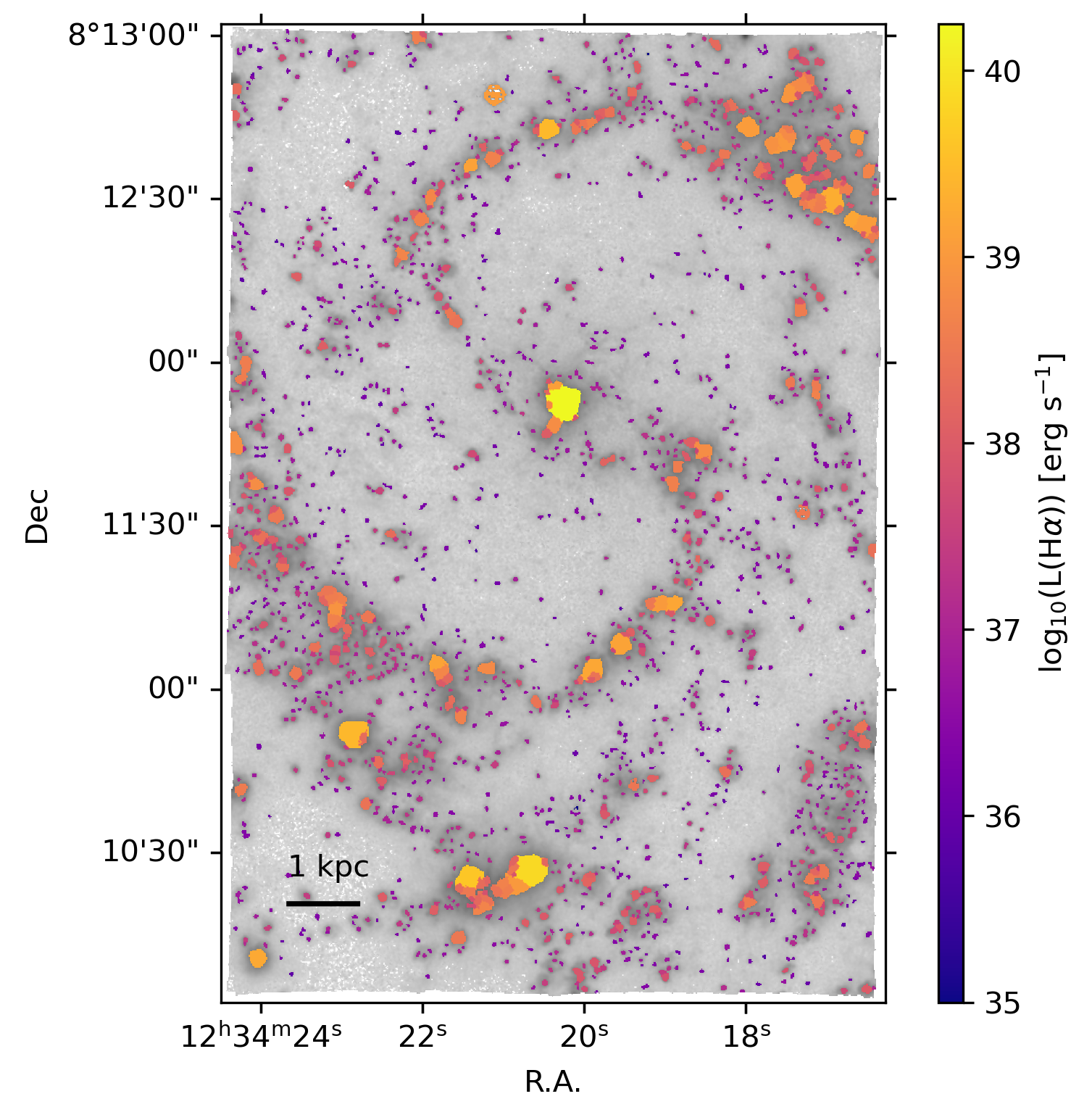}
    \caption{As in figure \ref{fig:atlas_IC5332} but for NGC 4535}
    \label{fig:atlas_NGC4535}
\end{figure*}

\begin{figure*}
    \centering
    \includegraphics[width=7in]{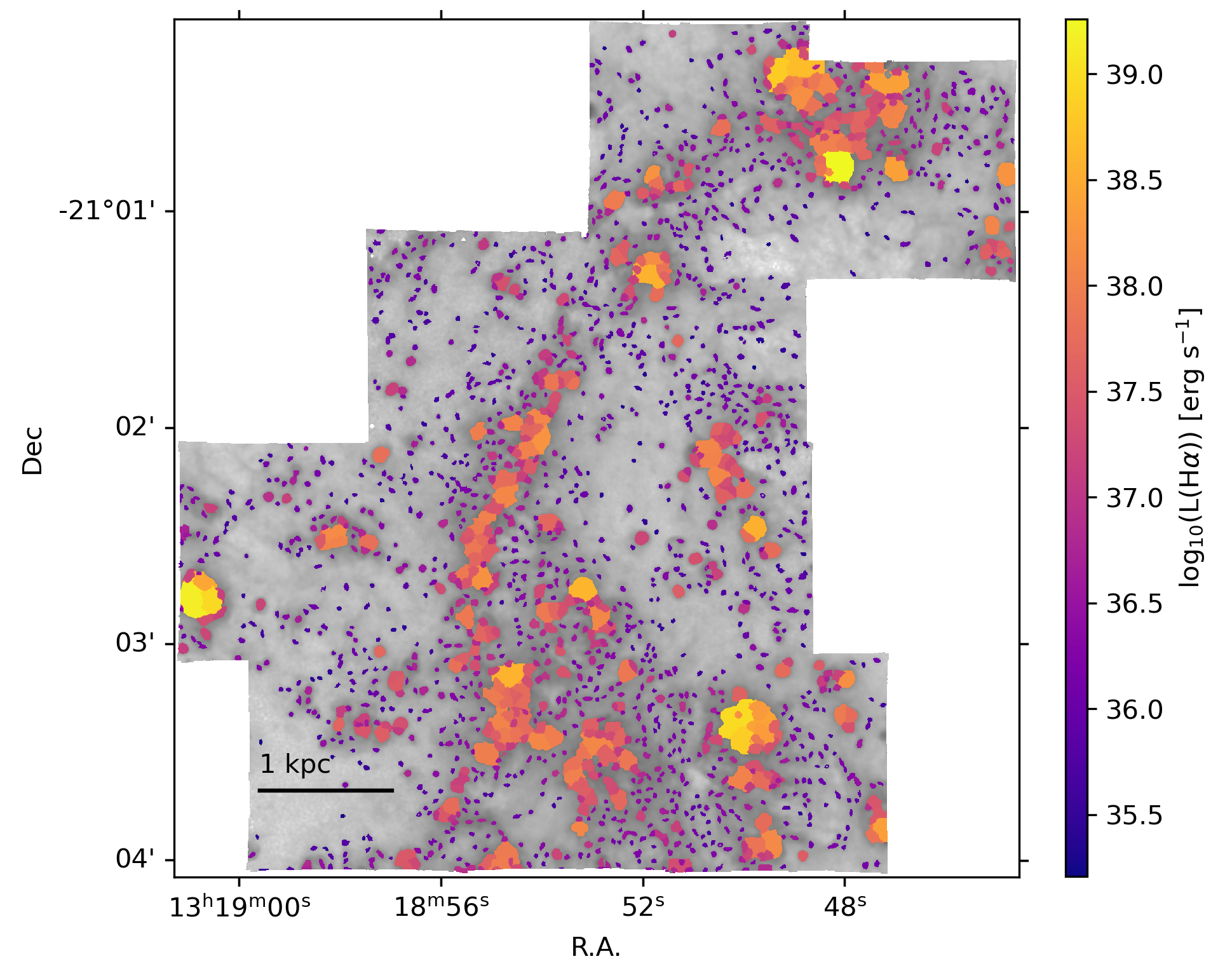}
    \caption{As in figure \ref{fig:atlas_IC5332} but for NGC 5068}
    \label{fig:atlas_NGC5068}
\end{figure*}

\begin{figure*}
    \centering
    \includegraphics[width=7in]{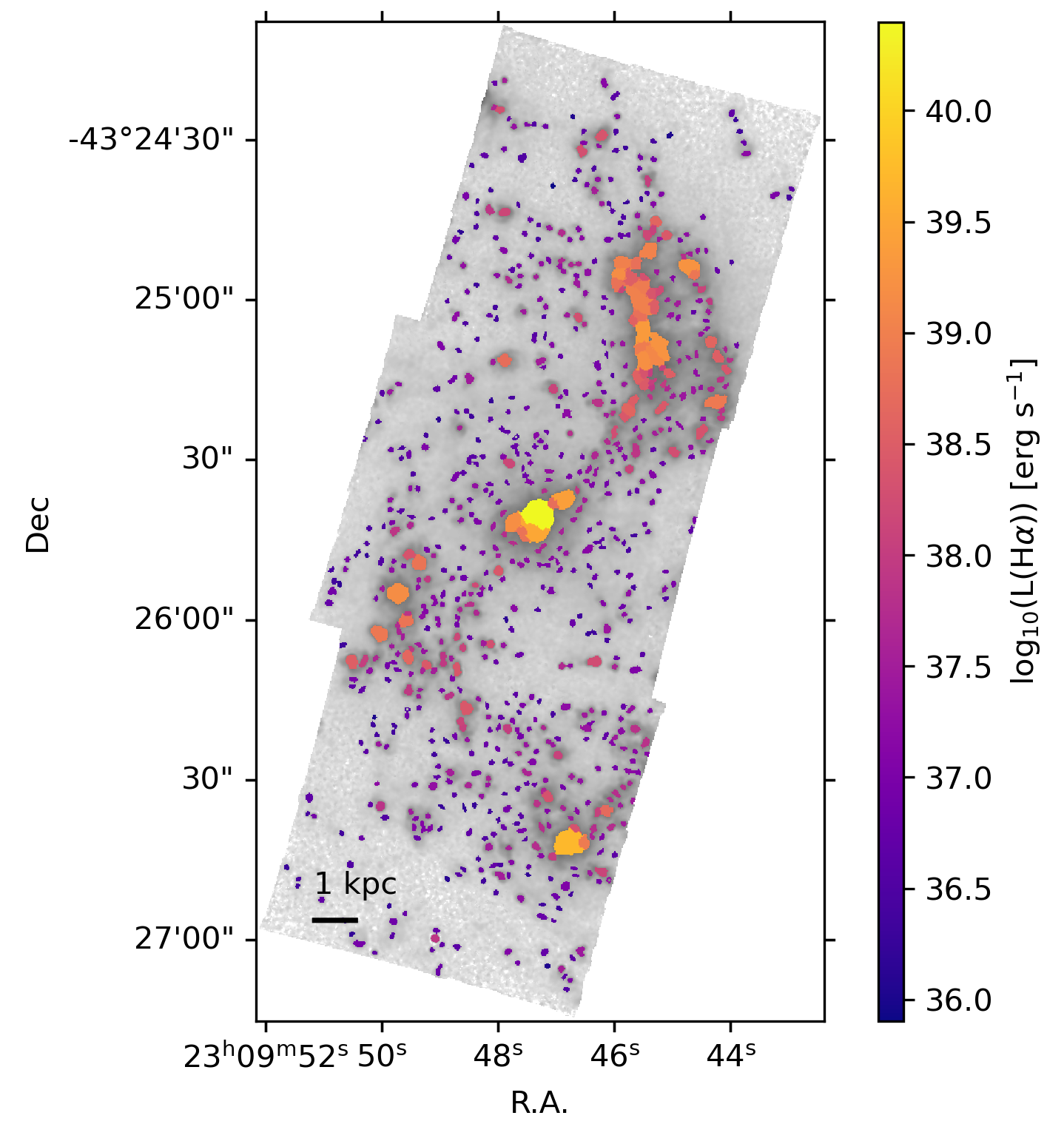}
    \caption{As in figure \ref{fig:atlas_IC5332} but for NGC 7496}
    \label{fig:atlas_NGC7496}
\end{figure*}

To provide a visual impression of the size and distribution of our identified nebulae, we provide images (Figures \ref{fig:atlas_IC5332} - \ref{fig:atlas_NGC7496}) of the \ha\ emission in all 19 galaxies with the nebular catalogue masks overplotted with the colours indicating the intrinsic (dust corrected) \ha\ luminosity of each nebulae. 

\section{BPT diagrams for all galaxies}
\label{appendix:bpt}

\begin{figure*}
    \centering
    \includegraphics[width=7in]{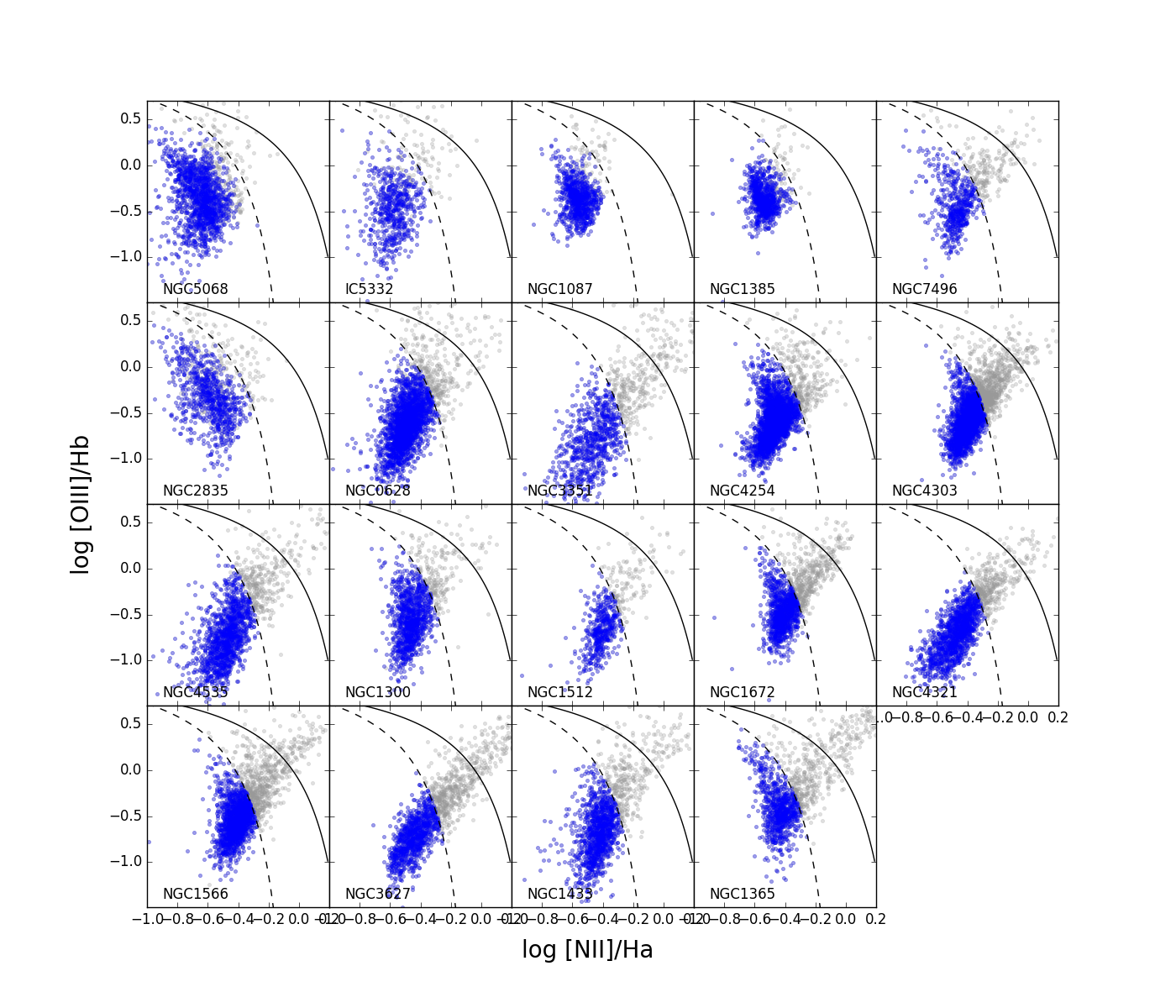}
    \caption{\oiii/\hb\ as a function of \nii/\ha\ for each galaxy individually. Galaxies are ordered by stellar mass from low (top left) to high (bottom right). The full nebular catalogue is shown in grey, \hii\ regions identified as consistent with photoionization across all three BPT diagnostics (see Section \ref{sec:bpt} and Table \ref{tab:bpt_flags}) are marked in blue. Overlaid are the \citet{Kauffmann2003} diagnostic curve (dashed line) and \citet{Kewley2006} diagnostic curves (solid line). }
    \label{fig:bpt_nii_all}
\end{figure*}

\begin{figure*}
    \centering
    \includegraphics[width=7in]{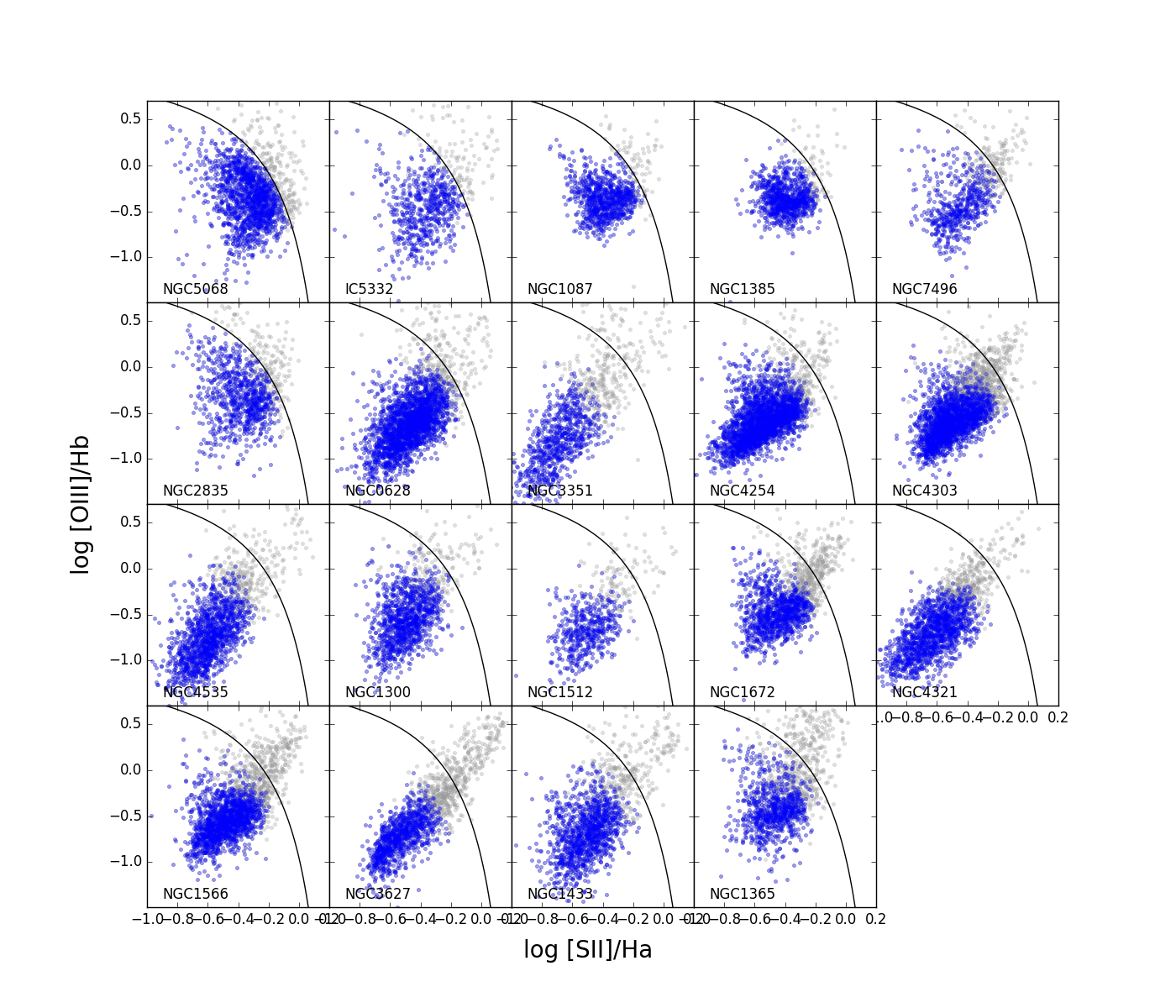}
    \caption{\oiii/\hb\ as a function of \sii/\ha\ for each galaxy individually. Galaxies are ordered by stellar mass from low (top left) to high (bottom right). The full nebular catalogue is shown in grey, \hii\ regions identified as consistent with photoionization across all three BPT diagnostics (see Section \ref{sec:bpt} and Table \ref{tab:bpt_flags}) are marked in blue. Overlaid is the \citet{Kewley2006} diagnostic curve (solid line). }
    \label{fig:bpt_sii_all}
\end{figure*}

\begin{figure*}
    \centering
    \includegraphics[width=7in]{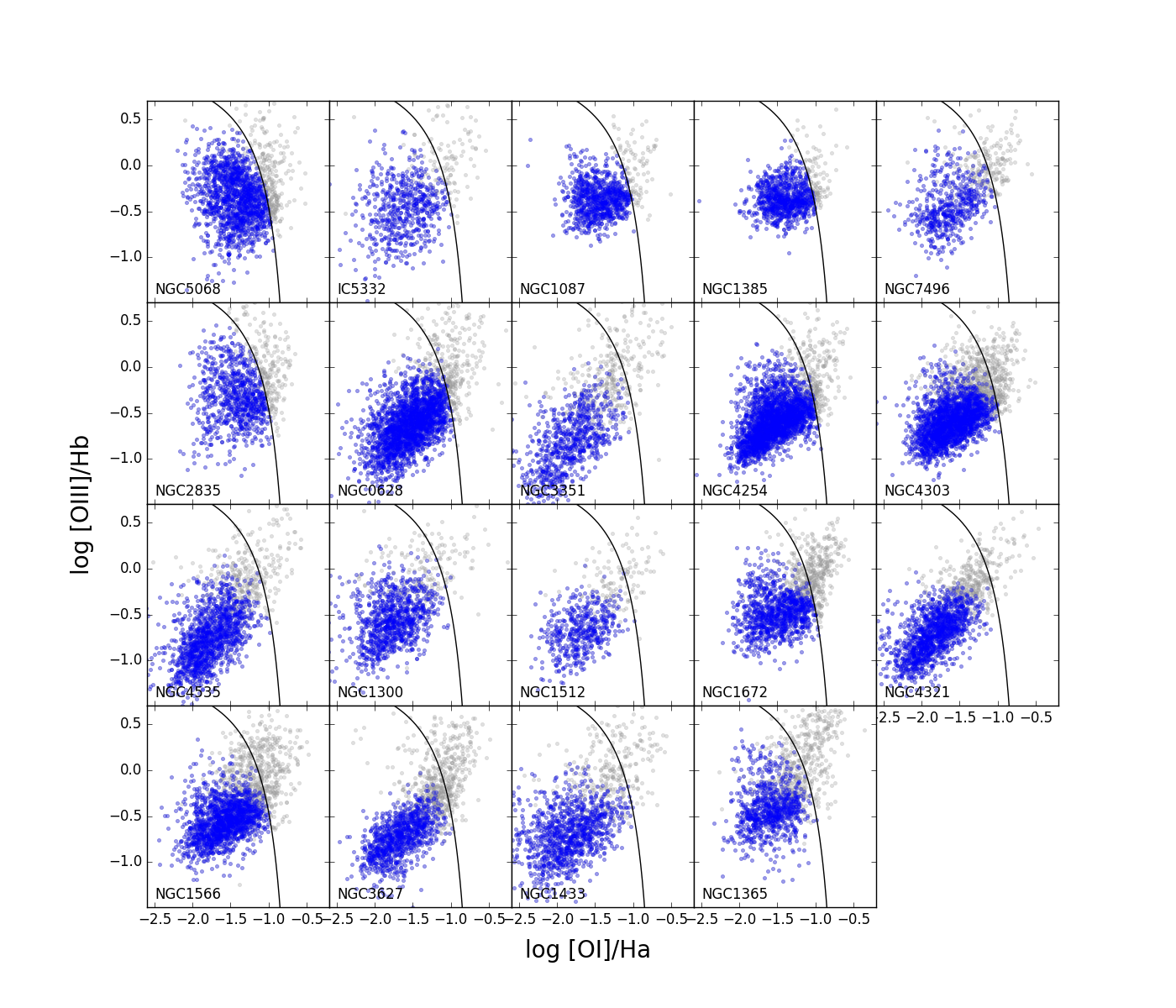}
    \caption{\oiii/\hb\ as a function of \oi/\ha\ for each galaxy individually. Galaxies are ordered by stellar mass from low (top left) to high (bottom right). The full nebular catalogue is shown in grey, \hii\ regions identified as consistent with photoionization across all three BPT diagnostics (see Section \ref{sec:bpt} and Table \ref{tab:bpt_flags}) are marked in blue. Overlaid is the \citet{Kewley2006} diagnostic curve (solid line).}
    \label{fig:bpt_oi_all}
\end{figure*}

BPT \citep{Baldwin1981} diagnostics are commonly used to distinguish the ionization source both in individual ionized nebulae and across integrated galaxy spectra. We demonstrate the range of line ratios observed within our complete nebular catalogue in Section \ref{sec:bpt} and Figure \ref{fig:bpt}, and include here the breakdown for each individual galaxy (Figures \ref{fig:bpt_nii_all}-\ref{fig:bpt_oi_all}). In all diagrams, regions consistent with photoionization across all three diagnostics are marked in blue. The complete nebular catalogue is shown in grey. Overlaid are the \citet{Kauffmann2003} diagnostic curve (dashed line) and \citet{Kewley2006} diagnostic curves (solid lines).

\section{Radial gradients for all galaxies}
\label{appendix:radial}

\begin{figure*}
    \centering
    \includegraphics[width=7in]{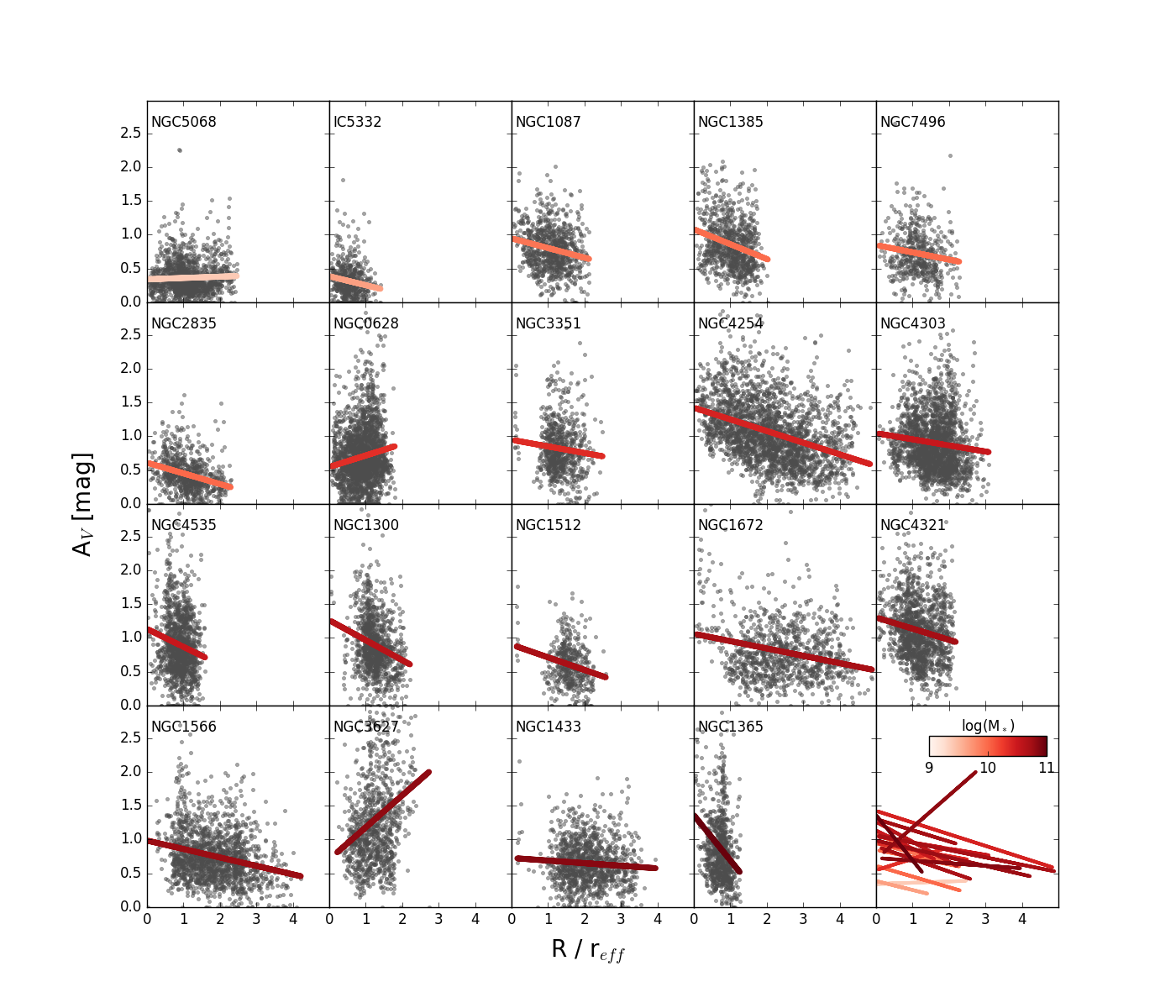}
    \caption{A$_V$ radial gradients.}
    \label{fig:ebv_gradient}
\end{figure*}

\begin{figure*}
    \centering
    \includegraphics[width=7in]{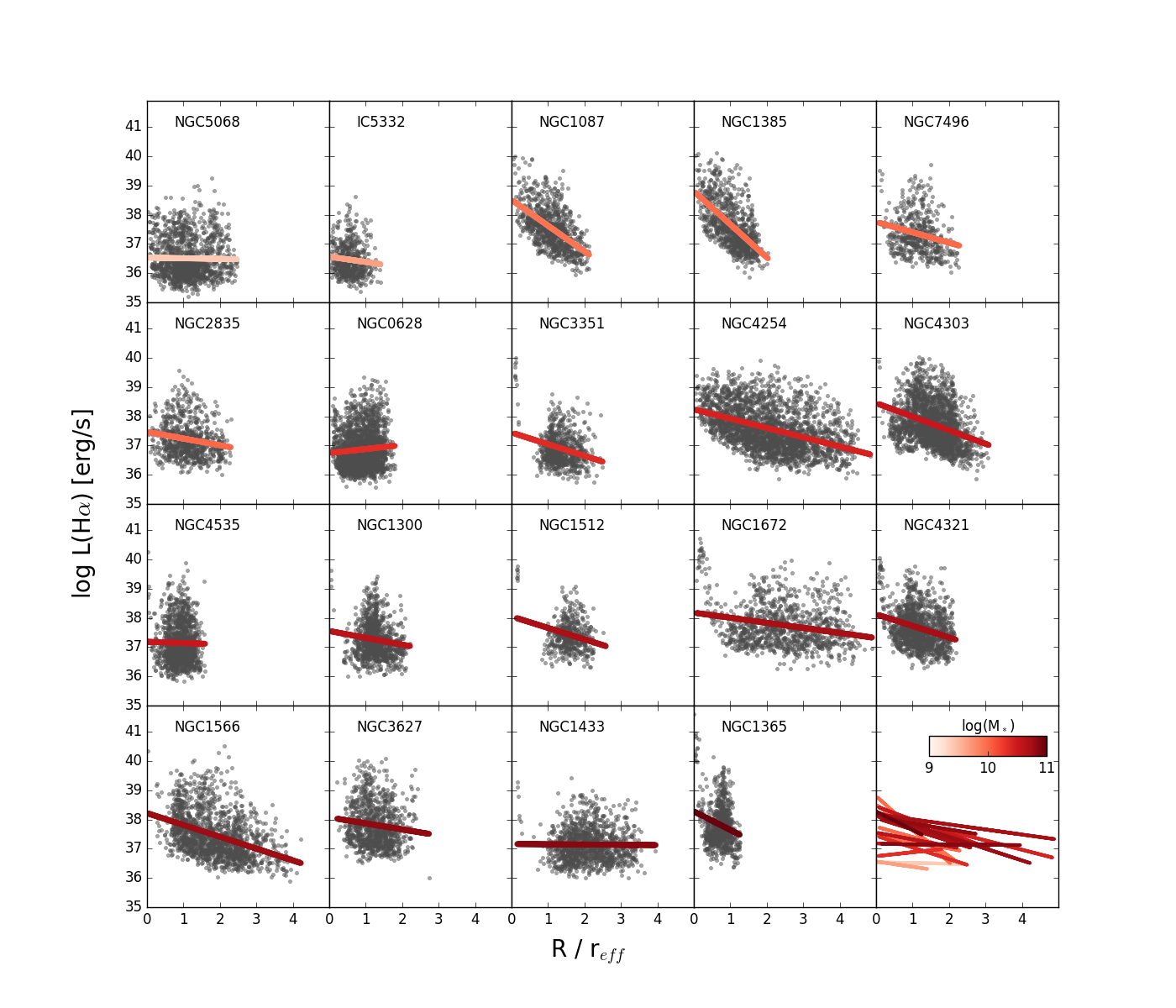}
    \caption{L(H$\alpha$) radial gradients.}
    \label{fig:LHa_gradient}
\end{figure*}

\begin{figure*}
    \centering
    \includegraphics[width=7in]{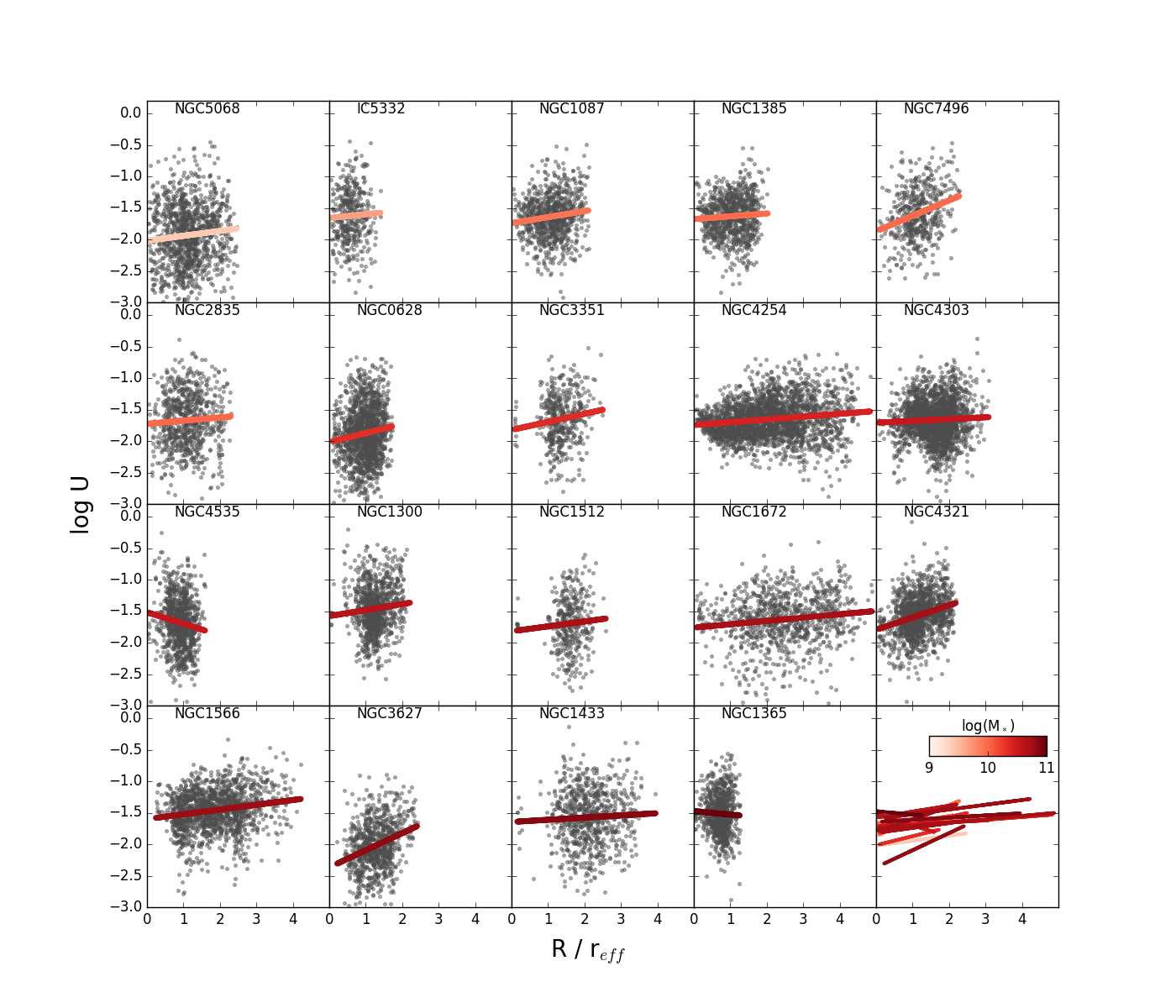}
    \caption{Ionization parameter (U) radial gradients.}
    \label{fig:U_gradients}
\end{figure*}

\begin{figure*}
    \centering
    \includegraphics[width=7in]{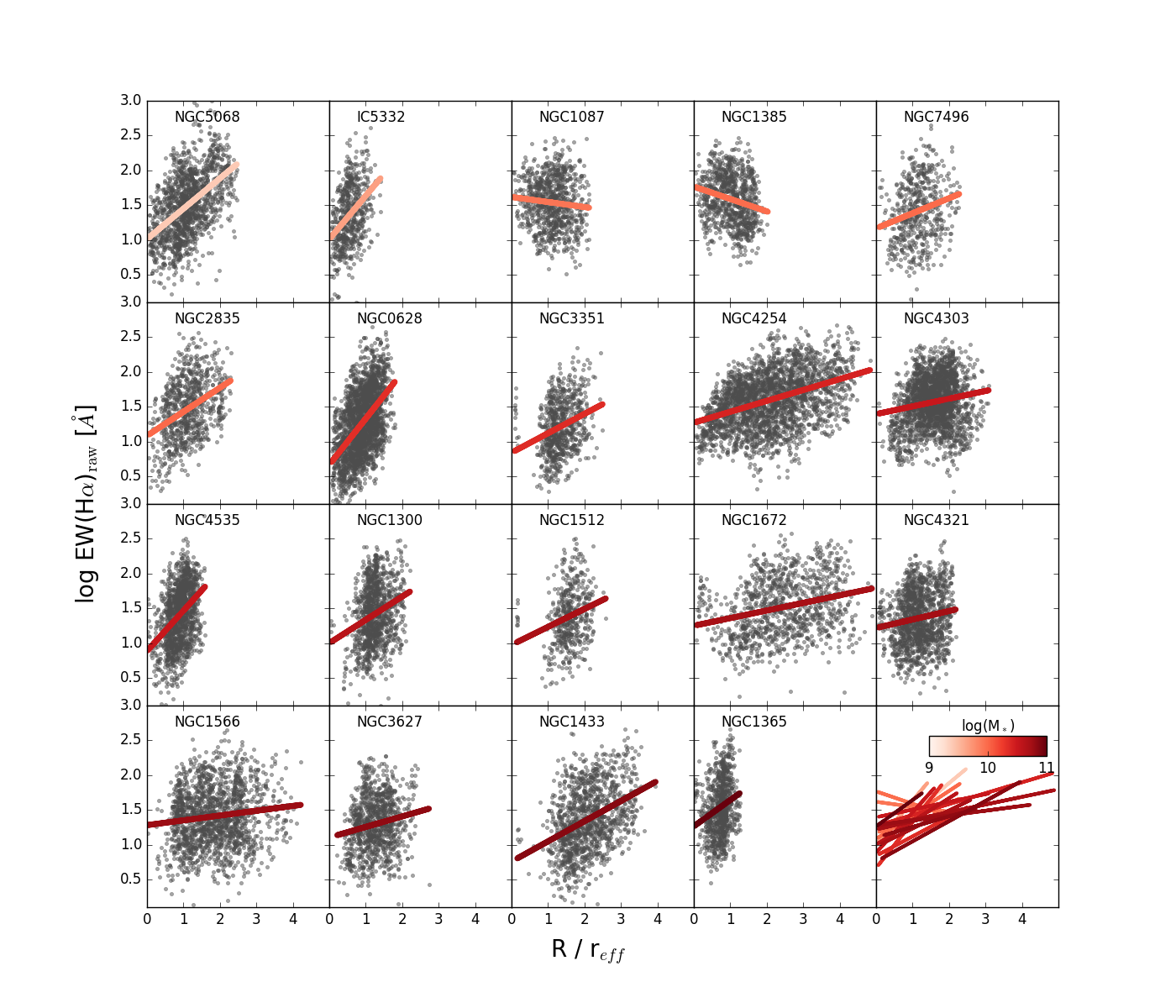}
    \caption{EW(Ha) radial gradients.}
    \label{fig:EW_gradients}
\end{figure*}

To give a more comprehensive look at the radial variations in nebular properties, we show the dust extinction derived from the Balmer decrement (A$_V$; Figure \ref{fig:ebv_gradient}), the extinction corrected \ha\ luminosity (Figure \ref{fig:LHa_gradient}), the ionization parameter (U; Figure \ref{fig:U_gradients}), and the equivalent width of \ha\ (EW \ha; Figure \ref{fig:EW_gradients}) as a function of effective radius (\reff) for all galaxies in our sample. In Table \ref{tab:rad_eff} we tabulate a representative value for each based on linear radial fits at 1 \reff. 

\begin{table}
\caption{Representative values of key physical parameters at 1 \reff    }
\label{tab:rad_eff}
\centering
\begin{tabular}{lrrrr}
\hline \hline
Galaxy & A$_V$ & log EW(\ha) & log L(\ha) & log U \\
 & [mag] & [\AA] & [erg s$^{-1}$] & \\
\hline
\hline
NGC0628 &  0.72  & 1.3 & 36.9 & -1.9  \\
NGC1087 &  0.80  & 1.5 & 37.6 & -1.6  \\
NGC1300 &  0.97  & 1.3 & 37.3 & -1.5  \\
NGC1365 &  0.69  & 1.6 & 37.6 & -1.5  \\
NGC1385 &  0.86  & 1.6 & 37.7 & -1.6  \\
NGC1433 &  0.69  & 1.0 & 37.2 & -1.6  \\
NGC1512 &  0.71  & 1.2 & 37.7 & -1.7  \\
NGC1566 &  0.86  & 1.3 & 37.8 & -1.5  \\
NGC1672 &  0.95  & 1.4 & 38.0 & -1.7  \\
NGC2835 &  0.45  & 1.4 & 37.2 & -1.7  \\
NGC3351 &  0.85  & 1.1 & 37.0 & -1.7  \\
NGC3627 &  1.18  & 1.3 & 37.9 & -2.1  \\
NGC4254 &  1.25  & 1.4 & 37.9 & -1.7  \\
NGC4303 &  0.96  & 1.5 & 38.0 & -1.7  \\
NGC4321 &  1.14  & 1.3 & 37.7 & -1.6  \\
NGC4535 &  0.87  & 1.5 & 37.1 & -1.7  \\
NGC5068 &  0.36  & 1.4 & 36.5 & -1.9  \\
NGC7496 &  0.74  & 1.4 & 37.4 & -1.6  \\
IC5332 &  0.25  & 1.6 & 36.4 & -1.6  \\
\hline
\end{tabular}
\end{table}

\section{Radial metallicity gradients for alternate prescriptions}
\label{appendix:altgrads}

\begin{figure*}
    \centering
    \includegraphics[width=6in]{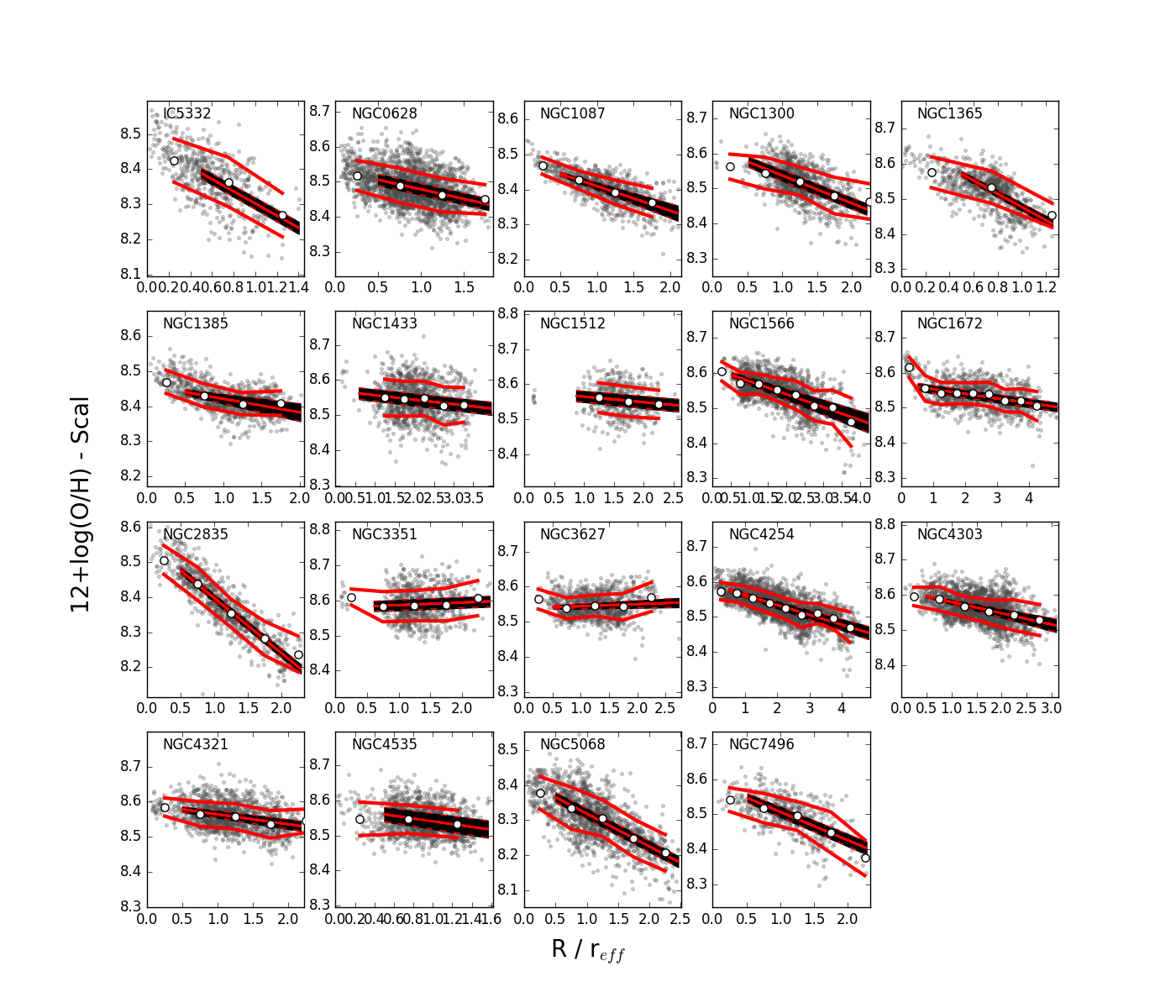}
    \caption{Metallicity radial gradients, using the Scal-PG16 prescription. All galaxies show the same dynamic range in 12+log(O/H). }
    \label{fig:zoom_scal}
\end{figure*}

\begin{figure*}
    \centering
    \includegraphics[width=6in]{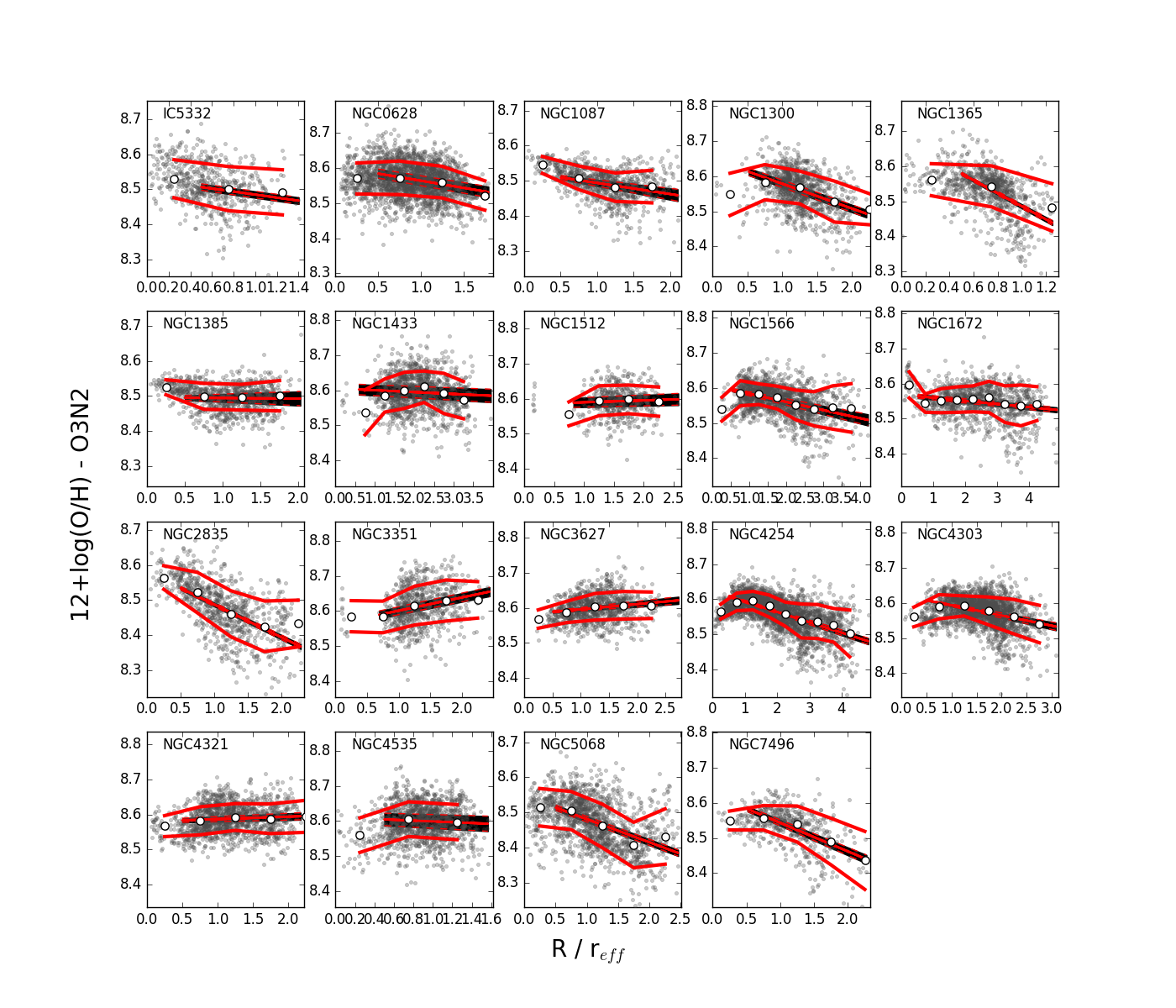}
    \caption{Metallicity radial gradients, using the O3N2-M13 prescription. All galaxies show the same dynamic range in 12+log(O/H). }
    \label{fig:zoom_O3N2_M13}
\end{figure*}

\begin{figure*}
    \centering
    \includegraphics[width=6in]{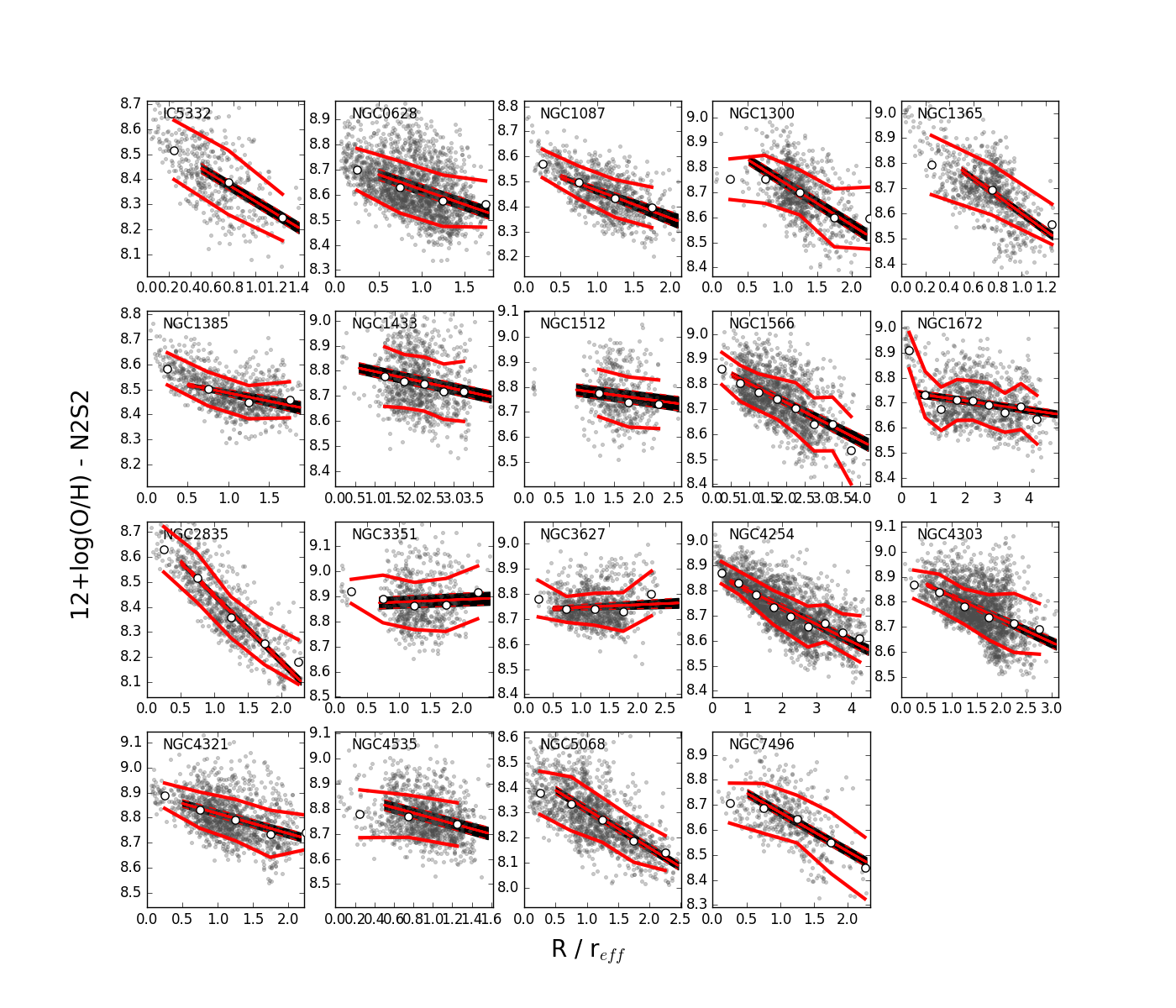}
    \caption{Metallicity radial gradients, using the N2S2-D16 prescription. All galaxies show the same dynamic range in 12+log(O/H). }
    \label{fig:zoom_N2S2}
\end{figure*}

A zoomed in view of the radial metallicity gradients for each galaxy is shown for metallicities calculated using the Scal-PG16 (Figure \ref{fig:zoom_scal}), O3N2-M13 (Figure \ref{fig:zoom_O3N2_M13}), and N2S2-D16 (Figure \ref{fig:zoom_N2S2}) prescriptions. Corresponding linear radial fits are provided for the Scal-PG16 (Table \ref{tab:metal_grad_scal}), O3N2-M13 (Table \ref{tab:metal_grad_O3N2_M13}), and N2S2-D16 (Table \ref{tab:metal_grad_N2S2}) prescriptions.

\begin{table}
\caption{Linear fit parameters for radial gradients in 12+log(O/H) using the O3N2 \citep{Marino2013} prescription, see Figure \ref{fig:zoom_N2S2}    }
\label{tab:metal_grad_O3N2_M13}
\centering
\begin{tabular}{lrrrr}
\hline \hline
Galaxy & intercept & slope [dex/r$_{eff}$] & value at r$_{eff}$ & $\sigma$(O/H)\\
\hline
\hline
IC5332 &  8.528 $\pm$ 0.010 & -0.043 $\pm$ 0.000 & 8.485 & 0.063 \\
NGC0628 &  8.605 $\pm$ 0.010 & -0.041 $\pm$ 0.002 & 8.564 & 0.046 \\
NGC1087 &  8.525 $\pm$ 0.002 & -0.032 $\pm$ 0.007 & 8.494 & 0.041 \\
NGC1300 &  8.649 $\pm$ 0.011 & -0.071 $\pm$ 0.000 & 8.579 & 0.049 \\
NGC1365 &  8.671 $\pm$ 0.003 & -0.186 $\pm$ 0.003 & 8.486 & 0.054 \\
NGC1385 &  8.495 $\pm$ 0.002 & -0.001 $\pm$ 0.009 & 8.494 & 0.039 \\
NGC1433 &  8.605 $\pm$ 0.013 & -0.005 $\pm$ 0.001 & 8.600 & 0.051 \\
NGC1512 &  8.584 $\pm$ 0.012 & 0.006 $\pm$ 0.002 & 8.590 & 0.042 \\
NGC1566 &  8.607 $\pm$ 0.003 & -0.023 $\pm$ 0.003 & 8.583 & 0.040 \\
NGC1672 &  8.568 $\pm$ 0.005 & -0.009 $\pm$ 0.000 & 8.559 & 0.045 \\
NGC2835 &  8.581 $\pm$ 0.004 & -0.093 $\pm$ 0.001 & 8.488 & 0.063 \\
NGC3351 &  8.563 $\pm$ 0.012 & 0.038 $\pm$ 0.001 & 8.601 & 0.055 \\
NGC3627 &  8.581 $\pm$ 0.002 & 0.015 $\pm$ 0.003 & 8.596 & 0.036 \\
NGC4254 &  8.623 $\pm$ 0.003 & -0.030 $\pm$ 0.002 & 8.594 & 0.039 \\
NGC4303 &  8.623 $\pm$ 0.000 & -0.029 $\pm$ 0.004 & 8.594 & 0.039 \\
NGC4321 &  8.580 $\pm$ 0.005 & 0.007 $\pm$ 0.002 & 8.587 & 0.040 \\
NGC4535 &  8.611 $\pm$ 0.014 & -0.011 $\pm$ 0.003 & 8.599 & 0.050 \\
NGC5068 &  8.547 $\pm$ 0.007 & -0.066 $\pm$ 0.001 & 8.481 & 0.062 \\
NGC7496 &  8.624 $\pm$ 0.005 & -0.081 $\pm$ 0.002 & 8.543 & 0.050 \\
\hline
\end{tabular}
\end{table}
\begin{table}
\caption{Linear fit parameters for radial gradients in 12+log(O/H) using the N2S2 \citep{Dopita2016} prescription, see Figure \ref{fig:zoom_N2S2}    }
\label{tab:metal_grad_N2S2}
\centering
\begin{tabular}{lrrrr}
\hline \hline
Galaxy & intercept & slope [dex/r$_{eff}$] & value at r$_{eff}$ & $\sigma$(O/H)\\
\hline
\hline
IC5332 &  8.575 $\pm$ 0.018 & -0.263 $\pm$ 0.003 & 8.312 & 0.120 \\
NGC0628 &  8.740 $\pm$ 0.020 & -0.119 $\pm$ 0.002 & 8.622 & 0.103 \\
NGC1087 &  8.574 $\pm$ 0.004 & -0.110 $\pm$ 0.012 & 8.464 & 0.072 \\
NGC1300 &  8.919 $\pm$ 0.017 & -0.175 $\pm$ 0.003 & 8.744 & 0.093 \\
NGC1365 &  8.947 $\pm$ 0.008 & -0.348 $\pm$ 0.005 & 8.599 & 0.091 \\
NGC1385 &  8.552 $\pm$ 0.000 & -0.066 $\pm$ 0.013 & 8.486 & 0.068 \\
NGC1433 &  8.831 $\pm$ 0.020 & -0.034 $\pm$ 0.001 & 8.797 & 0.111 \\
NGC1512 &  8.817 $\pm$ 0.022 & -0.032 $\pm$ 0.003 & 8.784 & 0.097 \\
NGC1566 &  8.877 $\pm$ 0.008 & -0.076 $\pm$ 0.004 & 8.802 & 0.088 \\
NGC1672 &  8.745 $\pm$ 0.014 & -0.019 $\pm$ 0.000 & 8.727 & 0.086 \\
NGC2835 &  8.712 $\pm$ 0.012 & -0.265 $\pm$ 0.003 & 8.447 & 0.083 \\
NGC3351 &  8.867 $\pm$ 0.024 & 0.010 $\pm$ 0.002 & 8.878 & 0.099 \\
NGC3627 &  8.737 $\pm$ 0.006 & 0.010 $\pm$ 0.005 & 8.747 & 0.067 \\
NGC4254 &  8.872 $\pm$ 0.008 & -0.068 $\pm$ 0.003 & 8.803 & 0.071 \\
NGC4303 &  8.916 $\pm$ 0.004 & -0.093 $\pm$ 0.005 & 8.823 & 0.087 \\
NGC4321 &  8.897 $\pm$ 0.014 & -0.080 $\pm$ 0.001 & 8.817 & 0.081 \\
NGC4535 &  8.868 $\pm$ 0.023 & -0.105 $\pm$ 0.004 & 8.763 & 0.084 \\
NGC5068 &  8.466 $\pm$ 0.015 & -0.153 $\pm$ 0.000 & 8.313 & 0.092 \\
NGC7496 &  8.825 $\pm$ 0.016 & -0.155 $\pm$ 0.002 & 8.669 & 0.101 \\
\hline
\end{tabular}
\end{table}

\bsp	
\label{lastpage}
\end{document}